\documentclass[%
 reprint,
 amsmath,amssymb,
 aps,
]{revtex4-2}

\usepackage{graphicx}
\usepackage{dcolumn}
\usepackage{bm}
\usepackage{amsmath}
\usepackage{float}
\usepackage[colorlinks,linkcolor=blue,citecolor=blue]{hyperref}

\begin{document}

\preprint{APS/123-QED}

\title{Anomalous spontaneous emission dynamics at chiral exceptional points}

\author{Yu-Wei Lu}
\affiliation{School of Physics and Optoelectronics, South China University of Technology, Guangzhou 510641, China}
\affiliation{School of Physics and Optoelectronic Engineering, Foshan University, Foshan 528000, China}
\author{Yanhui Zhao}
\affiliation{School of Physics and Optoelectronic Engineering, Ludong University, Yantai 264000, China}
\author{Runhua Li}%
\email[Corresponding Author: ]{rhli@scut.edu.cn}
\affiliation{School of Physics and Optoelectronics, South China University of Technology, Guangzhou 510641, China}
\author{Jingfeng Liu}
\affiliation{College of Electronic Engineering, South China Agricultural University, Guangzhou 510642, China}%


\begin{abstract}
An open quantum system operated at the spectral singularities where dimensionality reduces, known as exceptional points (EPs), demonstrates distinguishing behavior from the Hermitian counterpart. Here, we present an analytical description of local density of states (LDOS) for microcavity featuring chiral EPs, and unveil the anomalous spontaneous emission dynamics from a quantum emitter (QE) due to the non-Lorentzian response of EPs. Specifically, we reveal that a square Lorentzian term of LDOS contributed by chiral EPs can destructively interfere with the linear Lorentzian profile, resulting in the null Purcell enhancement to a QE with special transition frequency, which we call {\it{EP induced transparency}}. While for the case of constructive interference, the square Lorentzian term can narrow the linewidth of Rabi splitting even below that of bare components, and thus significantly suppresses the decay of Rabi oscillation. Interestingly, we further find that an open microcavity with chiral EPs supports atom-photon bound states for population trapping and decay suppression in long-time dynamics. As applications, we demonstrate the advantages of microcavity operated at chiral EPs in achieving high-fidelity entanglement generation and high-efficiency single-photon generation. Our work unveils the exotic cavity quantum electrodynamics unique to chiral EPs, which opens the door for controlling light-matter interaction at the quantum level through non-Hermiticity, and holds great potential in building high-performance quantum-optics devices.
\end{abstract}

\maketitle


\section{Introduction}

Exceptional points (EPs), the spectral degeneracies where two or more eigenvalues and the associated eigenstates simultaneously coalesce, are the central concept in non-Hermitian physics \cite{RN1,RN2,RN3,RN4,RN5}. A plethora of intriguing effects and exotic phenomena emerge around EPs due to their nontrivial topological properties and the dimensionality reduction, including ultrasensitive sensing \cite{RN6,RN7,RN8,RN9,RN10,RN11,lgl}, laser mode selection \cite{RN12,RN13}, chiral mode conversion \cite{RN14,RN15,RN17}, and unidirectional invisibility \cite{RN18,RN19}. Harnessing these peculiar features of non-Hermitian degeneracies for building novel devices with unprecedented performance has been experimentally demonstrated in various classical dissipative platforms, ranging from nanophotonics \cite{RN2,RN20,RN21,RN22,xrr}, acoustics \cite{RN27}, to macroscopic facilities, such as fiber network \cite{RN28}, electric circuits \cite{RN29} and heat diffusive system \cite{RN30}. In recent years, great efforts have been dedicated to accessing the quantum EPs by implementing the non-Hermiticity in quantum systems \cite{RN11,RN17,RN24,RN25,RN26,RN31,RN32,RN33,RN34,RN35,RN36,RN37,RN38,RN39}, and investigating the quantum states control through EPs \cite{RN17,RN31,RN32,RN33,RN38,RN40,RN41,RN42}. In this respect, pioneer works have shown the ability of EPs to tune the photon statistics \cite{RN32,RN33}, enhance the sensitivity of quantum sensing \cite{RN43,RN44}, and steer the evolution of single quantum system \cite{RN17,RN31,RN38,RN40,RN41,RN42,RN45,RN46,RJJ}. 

Despite these promising results, quantum effects of EPs beyond wave mechanics is still largely unexplored. Recently, the emission properties of a quantum emitter (QE) in electromagnetic environment supporting EPs attracts growing attention \cite{RN38,RN40,RN42,RN42,RN45}, since the modification of spontaneous emission (SE) exhibits a non-Lorentzian feature around EPs, contrast to the Lorentzian response in conventional single-mode cavity, which can lead to a greater enhancement of SE rate \cite{RN40,RN42,RN45}. It implies that the formation of EPs in nanophotonic structures significantly alters the local density of states (LDOS), which fully governs the interaction between a QE and arbitrary electromagnetic environment \cite{RN47,RN48,RN50}. Tailoring the LDOS of electromagnetic environment is crucial for optimizing the performance of many practical applications, ranging from traditional optoelectronic devices like lasers \cite{RN51,RN52} and solar cells \cite{RN53}, to advanced quantum technologies, such as quantum light sources \cite{RN54,RN55,RN56} and logical gate \cite{RN57,RN58}. Therefore, a LDOS theory that can capture the effects of EPs is of both fundamental and applied significance. The formalism of LDOS based on the classical Green function separates the EPs contribution from the usual Lorentzian term \cite{RN40,RN41,RN59}, while it leaves the origin of EPs unclear, i.e., information of the underlying cavity resonances (frequency and decay rate) is ambiguous, the coupling parameters of coalescent cavity modes that forms EPs are implicit, and thus the EPs condition cannot be given explicitly. As a consequence, there are at least two obvious limitations that hinder the exploration of quantum effects of EPs using macroscopic quantum electrodynamics (QED) based on the classical Green function: the quick and accurate design of nanophotonic structures possessing EPs remains challenging, and the nonlinear regime of cavity QED involving multiphoton interaction is exclusive. 

\begin{figure}[t]
\includegraphics[width=0.99\linewidth]{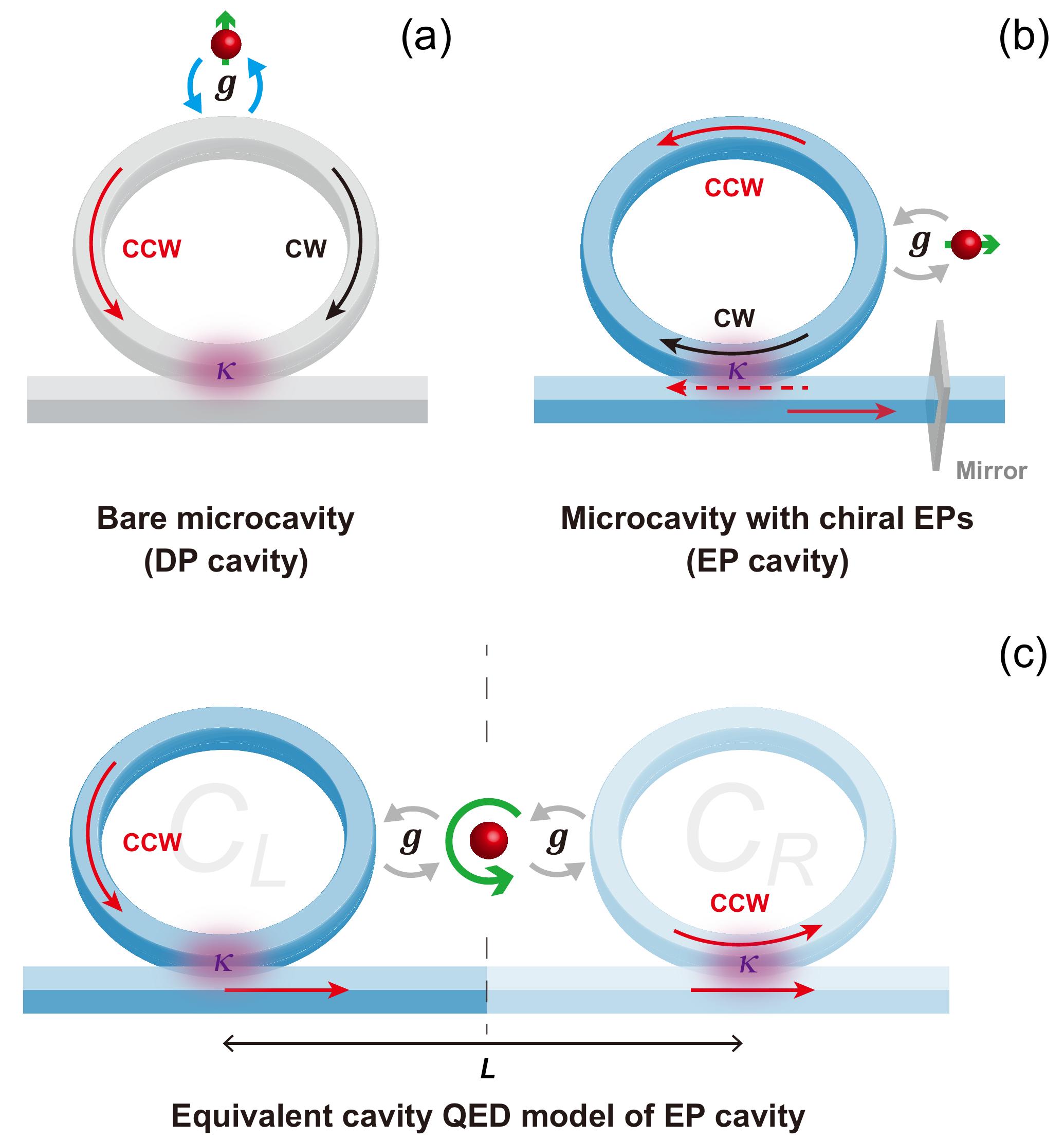}
\caption{\label{fig1} Schematic diagrams of cavity QED models. (a) A whispering-gallery-mode (WGM) cavity represented by microring resonator supports the degenerate clockwise (CW) and counterclockwise (CCW) modes. A linearly polarized quantum emitter (QE) is simultaneously coupled to both CW and CCW modes with equal coupling rate $g$. (b) A microring cavity coupled to a semi-infinite waveguide with a mirror at the one side can create chiral EPs due to the chiral coupling between the CW and CCW modes. (c) Conceptual illustration of the equivalent cavity QED model for the system in (b) with a perfect mirror, where the original CW mode is flipped to another CCW mode through mirror symmetry. Accordingly, the linearly polarized QE is replaced by a circularly polarized one. The distance between two waveguide-cavity junctions is $L$. }
\end{figure}

To overcome the aforementioned limitations of current LDOS theory for EPs and explore the quantum state control through EPs, in this work we propose an equivalent cavity QED model that can provide an intuitive and flexible quantum master equation approach for a class of novel microcavity with chiral EPs \cite{RN8,RN11,RN60,RN61,RN62,nc2022}, also known as exceptional surface, which is a collection of EPs in high-dimensional parameter space rather than a single EP. An analytical description of LDOS for such EP cavity is derived in Sec. \ref{sec2}, and parameterized by the resonance frequencies, decay rates and dissipative coupling rate of discrete cavity modes, as well as the coupling rates between the cavity and the QE. Therefore, it can be used to determine the parameters in quantum master equation from LDOS of a given photonic structure obtained by electromagnetic simulations, and establishes an explicit connection with cavity QED at EPs. Furthermore, the formulation of SE spectrum is also obtained to understand the peculiar quantum dynamics at EPs. Our LDOS theory not only explains the previously known properties of SE dynamics related to EPs only with the special parameters choice \cite{RN41,RN42}, but also allows the entire formalization and exploring more extraordinary effects of EPs in open quantum systems.

Based on the proposed LDOS theory, in Sec. \ref{sec3} we perform comprehensive study to show how the chiral EPs tailors the LDOS and affects the SE dynamics of a QE and compare to the conventional Lorentzian cavity. With the equivalent cavity QED model, we further find out the formation of a special kind of quantum state without decay at chiral EPs, and demonstrate its applications in quantum state control. Finally, we conclude in Sec. \ref{sec4}. 

\section{Model and theory}\label{sec2}

In this section, we build the quantum model to describe the interaction between a QE and the microcavity with chiral EP (we call EP cavity hereafter). In traveling wave optical resonators, such as the whispering-gallery-mode (WGM) microcavity, the chiral EPs emerges via two unidirectionally coupled cavity modes. Different strategies have been proposed to implement the unidirectional coupling between the clockwise (CW) and the counterclockwise (CCW) modes of WGM microcavity, by introducing two or more Rayleigh-type nano-scatters \cite{RN8,RN62} or coupling to a waveguide terminated by a mirror at the one side \cite{RN11,RN42,nc2022}. Despite different realizations, the underlying physical mechanism is essentially the same, where the unidirectional coupling is induced by auxiliary reservoir modes offered by either nano-scatters or waveguide. Here we employ the latter scheme for the sake of theoretical simplicity, but the results and conclusions presented in this work are suitable for other types of EP cavity.

Fig. \ref{fig1}(a) depicts a basic cavity QED system consisting of a QE and a WGM microcavity with infinite waveguide. The WGM microcavity has a decay rate $\kappa$ due to the coupling to waveguide. Since there is no coupling between the CW and CCW modes, the cavity has two degenerate modes, i.e., is operated at diabolic points (DPs) \cite{RN60,RN71}, and thus called DP cavity hereafter. The QE is linearly polarized, and coupled to a pair of degenerate CW and CCW modes with equal coupling rate $g$. The system is described by the two-mode Jaynes-Cummings (JC) model. While by coupling the WGM microcavity to a semi-infinite waveguide with a mirror at the right end, a unidirectional coupling from CCW mode to CW mode is created, as Fig. \ref{fig1}(b) illustrates. This photonic structure supports chiral EPs, and offers great tunability to SE rate of a QE. 

In order to develop an intuitive cavity QED model for this novel EP cavity, we transform the system to an equivalent one through mirror symmetry, as shown in Fig. \ref{fig1}(c). The original CW mode is flipped to a mirrored CCW mode, and the corresponding left-handed emission of QE is converted to the right-handed emission into the mirrored CCW mode as well. As a result, the linearly polarized QE becomes circularly polarized in the equivalent cavity QED model. Two CCW modes in the equivalent model constitute a cascaded system \cite{RN67,RN68,RN69}, the quantum dynamics of system is described by the extended cascaded quantum master equation $\left(\hbar=1\right)$
\begin{equation}\label{eq1}
\begin{gathered}
\dot{\rho}=-i\left[H_{0}+H_{I}, \rho\right]+\gamma \mathcal{L}\left[\sigma_{-}\right] \rho+\kappa \mathcal{L}\left[c_{L}\right] \rho+\kappa \mathcal{L}\left[c_{R}\right] \rho \\
+\kappa|r|\left(e^{i \varphi}\left[c_{L} \rho, c_{R}^{\dagger}\right]+e^{-i \varphi}\left[c_{R}, \rho c_{L}^{\dagger}\right]\right)
\end{gathered}
\end{equation}
with the free Hamiltonian 
\begin{equation}
H_{0}=\omega_{0} \sigma_{+} \sigma_{-}+\omega_{c} c_{L}^{\dagger} c_{L}+\omega_{c} c_{R}^{\dagger} c_{R}
\end{equation}
and the interaction Hamiltonian
\begin{equation}
H_{I}=g\left(e^{-i \phi} c_{L}^{\dagger} \sigma_{-}+h.c.\right)+g\left(e^{i \phi} c_{R}^{\dagger} \sigma_{-}+h.c.\right)
\end{equation}
where $\sigma_{-}$ is the lowering operator of QE with transition frequency $\omega_0$ and SE rate $\gamma$ in homogeneous medium. $\gamma=\mu^{2} \omega_{0}^{3} n_{b} /\left(3 \pi \hbar \varepsilon_{0} c^{3}\right)$, where $\varepsilon_{0}$ is the permittivity of vacuum, $\mu$ is the dipole moment of QE, and $n_b$ is the refractive index of background medium. $c_L / c_R$ is the bosonic annihilation operator for CCW mode in the left/right cavity with resonance frequency $\omega_c$. $\kappa$ is the dissipative coupling rate between the cavity modes and the waveguide. The intrinsic decay of two CCW modes is omitted, considering that the evanescent coupling to waveguide dominates the cavity dissipation. $\mathcal{L}[O] \rho=O \rho O^{\dagger}-\left\{O^{\dagger} O, \rho\right\} / 2$ is the Liouvillian superoperator for operator $O$. The second line of Eq. (\ref{eq1}) describes the unidirectional coupling from the left CCW mode to the right CCW mode, where $\left|r\right|$ is the reflectivity of mirror, and $\varphi=\beta L$ is the phase factor, with $\beta$ and $L$ being the propagation constant of waveguide and the distance between two cavities, respectively. In the interaction Hamiltonian, the phase factor $\phi$ originates from the traveling wave nature of WGM modes and depends on the azimuthal orientation of QE location \cite{RN70,RN71}. 

From Eq. (\ref{eq1}), we take the operator expectation values to obtain the equations of motion 
\begin{equation}\label{eq3}
\frac{d}{d t} \vec{p}=-i \mathbf{M} \vec{p}
\end{equation}
where $\vec{p}=\left(\left\langle c_{L}\right\rangle,\left\langle c_{R}\right\rangle,\left\langle\sigma_{-}\right\rangle\right)^{T}$ and the $3 \times 3$ matrix takes the form
\begin{equation}\label{eq4}
\mathbf{M}=\left(\begin{array}{ccc}
\omega_{c}-i \frac{\kappa}{2} & 0 & g e^{-i \phi} \\
-i \kappa|r| e^{i \varphi} & \omega_{c}-i \frac{\kappa}{2} & g e^{i \phi} \\
g e^{i \phi} & g e^{-i \phi} & \omega_{0}-i \frac{\gamma}{2}
\end{array}\right)
\end{equation}
The above equations do not include the time delay of itinerant photon. This is justified if the propagation time $L/v_g$, where $v_g$ is the group velocity of waveguide, is much smaller than the timescale of system evolution set by $\min \left[g^{-1}, \kappa^{-1}, \gamma^{-1}\right]$. In the following study, we consider the optical frequency and focus on the strong coupling regime, and thus the timescale of system evolution is determined by $g^{-1}$. For WGM cavity, $g$ is typically smaller than 1meV, and we evaluate that $g^{-1} \gg L / v_{g}$ for $L<20\mu\mathrm{m}$; in this case, we can assume that the photon propagates without time delay.

To study the quantum dynamics of a QE located in a realistic photonic structure, especially for the interactions involving multiphoton process, we need to determine the system parameters in quantum master equation from LDOS obtained by electromagnetic simulation. Therefore, an analytical LDOS theory is crucial to extract the system parameters. The normalized LDOS of electromagnetic environment, i.e., Purcell factor, is expressed as $F_P(\omega)=[J(\omega)+J_0 (\omega)]/J_0 (\omega)$, where $J(\omega)$ and $J_0 (\omega)=\gamma/2\pi$ are the spectral density of photonic structures and background medium (free space), respectively. For EP cavity shown in Fig. \ref{fig1}(b), the spectral density of cavity is given by $J(\omega)= \int_{-\infty}^{+\infty} d \tau e^{i \omega \tau} g^{2}\left\langle\left[c_{L}^{\dagger}(0)+e^{-i 2 \phi} c_{R}^{\dagger}(0)\right]\left[c_{L}(\tau)+e^{i 2 \phi} c_{R}(\tau)\right]\right\rangle$ \cite{PRLmb,PRBjm,np2022}, where the environmental correlation functions $\left\langle c_{L}^{\dagger}(0) c_{L}(\tau)\right\rangle$, $\left\langle c_{L}^{\dagger}(0) c_{R}(\tau)\right\rangle$ and $\left\langle c_{R}^{\dagger}(0) c_{R}(\tau)\right\rangle$ can be calculated by applying the quantum regression theorem $\left\langle O_{i}(t) O_{j}(t+\tau)\right\rangle=\operatorname{Tr}\left\{O_{j}(0) e^{\mathcal{L}^{\prime} \tau}\left[\rho(t) O_{i}(0)\right]\right\}$ \cite{RN72}, with $\mathcal{L}^{\prime} \rho^{\prime} = \dot{\rho}^{\prime}$, and $\dot{\rho}^{\prime}$ is given by
\begin{equation}
\begin{gathered}
\dot{\rho}^{\prime}=-i \omega_{c}\left[c_{L}^{\dagger} c_{L}+c_{R}^{\dagger} c_{R}, \rho\right]+\kappa \mathcal{L}\left[c_{L}\right] \rho+\kappa \mathcal{L}\left[c_{R}\right] \rho \\
+\kappa|r|\left(e^{i \varphi}\left[c_{L} \rho, c_{R}^{\dagger}\right]+e^{-i \varphi}\left[c_{R}, \rho c_{L}^{\dagger}\right]\right)
\end{gathered}
\end{equation}
From the above quantum master equation, we have
\begin{equation}
\frac{d}{d t}\vec{v_{LL}}=\left[\begin{array}{cc}
\omega_{c}-i \frac{\kappa}{2} & 0 \\
-i \kappa|r| e^{i \varphi} & \omega_{c}-i \frac{\kappa}{2}
\end{array}\right]\vec{v_{LL}}
\end{equation}
where $\vec{v_{LL}}=\left(\left\langle c_{L}^{\dagger}(0) c_{L}(\tau)\right\rangle, \left\langle c_{L}^{\dagger}(0) c_{R}(\tau)\right\rangle\right)^{T}$. Since the initial state of electromagnetic environment is the vacuum state, we can obtain the analytical expression of $\left\langle c_{L}^{\dagger}(0) c_{L}(\tau)\right\rangle$ with $\left\langle c_{L}^{\dagger}(0) c_{L}(0)\right\rangle = 1$; other correlation functions can be solved in similar fashion. We finally arrive at 
\begin{equation}\label{eq8}
J(\omega) = J_\mathrm{DP}(\omega) + J_\mathrm{EP}(\omega)
\end{equation}
with the DP and EP contributions
\begin{equation}
\begin{aligned}
J_{\mathrm{DP}}(\omega) &=-g^{2} \operatorname{Im}\left[\chi_{\mathrm{DP}}(\omega)\right] \\
J_{\mathrm{EP}}(\omega) &=-g^{2} \operatorname{Im}\left[\chi_{\mathrm{EP}}(\omega)\right]
\end{aligned}
\end{equation}
where $\chi_{\mathrm{DP}}(\omega)$ is the usual linear Lorentzian response given by $\left\langle c_{L}^{\dagger}(0) c_{L}(\tau)\right\rangle$ and $\left\langle c_{R}^{\dagger}(0) c_{R}(\tau)\right\rangle$
\begin{equation}\label{eq10}
\chi_{\mathrm{DP}}(\omega)=\frac{1}{\pi}\frac{2}{\left(\omega-\omega_{c}\right)+i \kappa / 2}
\end{equation}
and $\chi_{\mathrm{EP}}(\omega)$ is the EP term contributed by $\left\langle c_{L}^{\dagger}(0) c_{R}(\tau)\right\rangle$
\begin{equation}\label{eq11}
\chi_{\mathrm{EP}}(\omega)=\frac{1}{\pi}\frac{-i \kappa|r| e^{i \Delta \phi}}{\left[\left(\omega-\omega_{c}\right)+i \kappa / 2\right]^{2}}
\end{equation}
where $\Delta\phi=\varphi-2\phi$ is the phase difference. The factor 2 in the numerator of $\chi_{\mathrm{DP}}(\omega)$ represents that the QE is coupled to two cavity modes. Eq. (\ref{eq11}) clearly indicates the square Lorentzian profile of $\chi_{\mathrm{EP}}(\omega)$, a signature of second-order EPs. However, it also indicates that any loss of reflection amplitude, i.e., an imperfect mirror with $\left| r \right|<1$, will degrade the quantum effects of EPs. In the following study, we take $\left|r\right|=1$ unless specially noted. We highlight that our approach can also be applied to a more general case where the coupling between two cavities are bidirectional but asymmetrical, for example, introducing a Rayleigh-type nano-scatter in close proximity of microring \cite{RN8,RN19,RN63}. As a result, the system is drawn out of chiral EPs, and such configuration can be utilized to study the emergence of quantum effects of EPs. Furthermore, one can obtain the LDOS of target nanophotonic structure that owns desirable quantum effects of EPs for electromagnetic design. This is important for structure optimization, especially for these consisting of absorbing and dispersing medium like the plasmonic-photonic cavity \cite{RN73,RN74,RN75,RN76}, where the coupling rate between cavity modes is hard to evaluate due to their distinct features, contract to the EP cavity studied in this work. On the other hand, the quantum-optics properties of a given nanophotonic structure are predictable, by parametrizing the extended cascaded quantum master equation (Eq. (\ref{eq1})) using the simple curve fitting of LDOS obtained from electromagnetic simulations. Therefore, Eqs. (\ref{eq8})-(\ref{eq11}) build the bridge between the electromagnetic design of nanophotonic structures and the quantum state control by EPs. 

\begin{figure}[t]
\includegraphics[width=0.99\linewidth]{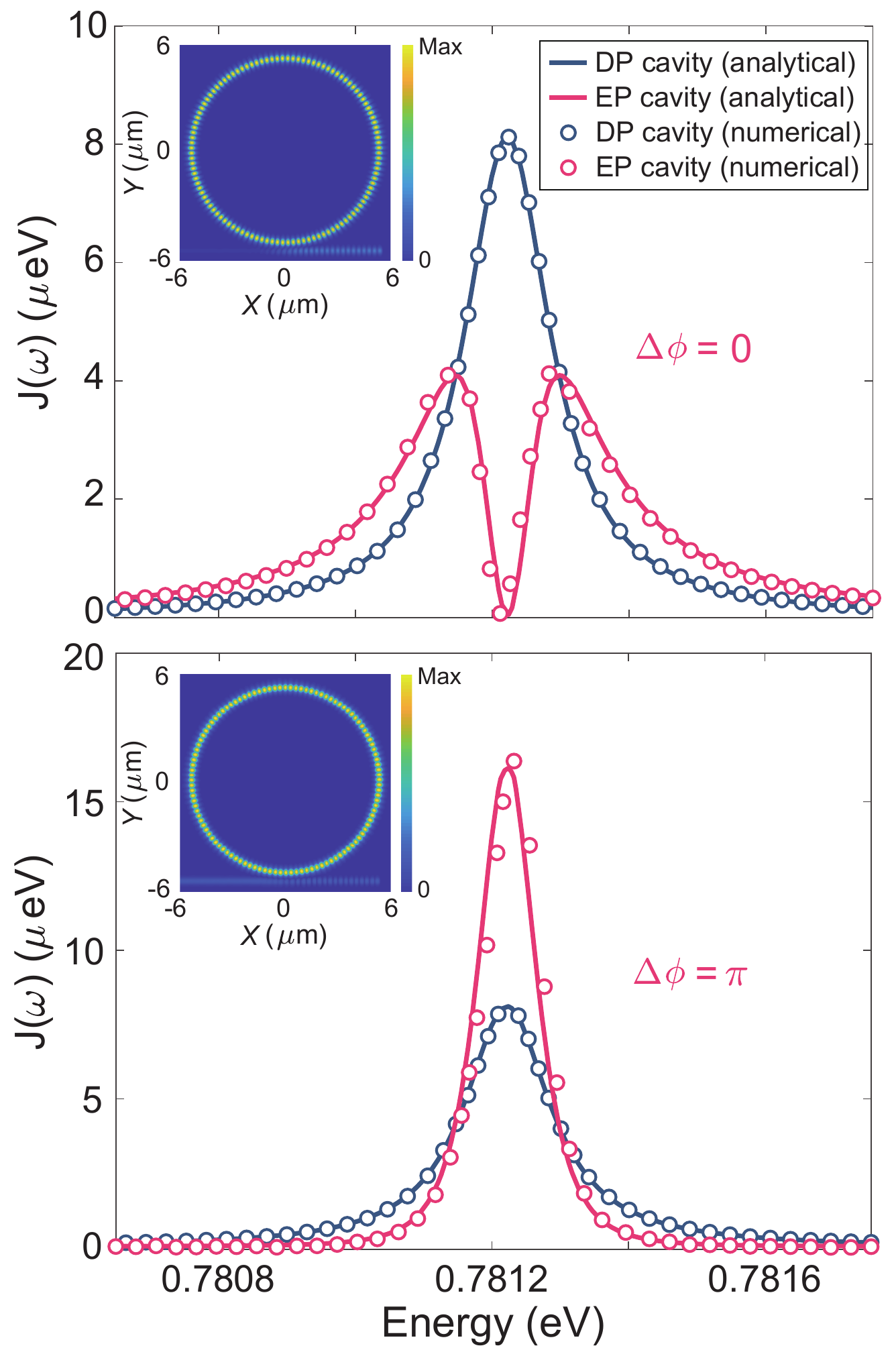}
\caption{\label{fig2} Comparison of analytical and numerical spectral density of EP cavity. (a) and (b) show the results of $\Delta\phi = 0$ and $\pi$, respectively. The design parameters of EP cavity are: outer radius $R=5\mu\mathrm{m}$, width $w=0.25\mu\mathrm{m}$, refractive index $n_c=3.47$, edge-to-edge separation to the waveguide $d=0.2\mu\mathrm{m}$. The width of waveguide is $d$, and the mirror is made of 100-nm thick silver. The whole structure is embedded in the background medium with refractive index of $n_b=1.44$. The insets show the corresponding electric field distribution at the resonance frequency of DP cavity $\omega_c=0.78122\mathrm{eV}$. }
\end{figure}

To validate our theory and as an example, in Fig. \ref{fig2} we compare the spectral density given by Eqs. (\ref{eq8})-(\ref{eq11}) with the electromagnetic simulations using a realistic EP cavity. The QE is $Z$-oriented (perpendicular to the cavity plane), and located at one-quarter of the perimeter from the waveguide-cavity junction ($X=0$) in the clockwise direction. For QE with a diploe moment of $\mu=60$ Debye, we can evaluate $(\omega_c,\kappa,g)=(0.78122\mathrm{eV},152.8\mu\mathrm{eV},24.9\mu\mathrm{eV})$ by curve fitting using Eq. (\ref{eq10}) from the corresponding DP cavity, i.e., EP cavity without the mirror in waveguide; subsequently, the spectral density of EP cavity can be analytically obtained from Eqs. (\ref{eq8})-(\ref{eq11}). Fig. \ref{fig2} shows the spectral density of EP cavity, where we can see the good accordance between the analytical expressions and the numerical simulation for $\Delta\phi=0$ and $\pi$. The results confirm the validness of our LDOS theory for EP cavity. With the obtained parameters, the quantum dynamics of system can be investigated by the extended cascaded quantum master equation. 

The exotic quantum dynamics at EPs can be well understood from the spectral properties of QE. The emission spectrum of QE is defined as $S(\omega)=(2 \pi)^{-1} \int_{0}^{\infty} d t_{1} \int_{0}^{\infty} d t_{2} e^{i \omega\left(t_{2}-t_{1}\right)}\left\langle \boldsymbol{E}^{-}\left(\boldsymbol{r}, t_{1}\right) \cdot \boldsymbol{E}^{+}\left(\boldsymbol{r}, t_{2}\right)\right\rangle$ \cite{RN72}, with $\boldsymbol{E}^{-}\left(\boldsymbol{r}, t_{1}\right)=\left[\boldsymbol{E}^{+}\left(\boldsymbol{r}, t_{1}\right)\right]^{\dagger}$ and $\boldsymbol{E}^{+}(\boldsymbol{r}, t)=e^{-i \omega_{0} t} \int_{0}^{t} d t^{\prime} \boldsymbol{G}\left(t, t^{\prime}\right) \sigma_{-}\left(t^{\prime}\right) e^{i \omega_{0} t^{\prime}}$, where $\boldsymbol{G}\left(t, t^{\prime}\right)$ is a kind of propagator dependent on the detection method. For assuming an initially excited QE and $\boldsymbol{r}=\boldsymbol{r}_{0}$, where $\boldsymbol{r}_{0}$ is the QE position, $S(\omega)$ becomes the SE spectrum, also called the polarization spectrum \cite{RN77,RN78}, which reflects the local dynamics of a QE, and is given by $S(\omega)=\int_{0}^{\infty} d t_{2} \int_{0}^{\infty} d t_{1} e^{i \omega\left(t_{2}-t_{1}\right)}\left\langle\sigma_{+}\left(t_{1}\right) \sigma_{-}\left(t_{2}\right)\right\rangle$. Denoting $\tau=t_2-t_1$ and taking the limit of $t_{1}=t \rightarrow \infty$, the SE spectrum can be expressed as $S(\omega)=\lim _{t \rightarrow \infty} \operatorname{Re}\left[\int_{0}^{\infty} d \tau\left(\sigma_{+}(t+\tau) \sigma_{-}(t)\right\rangle e^{i \omega \tau}\right]$, where $\left\langle\sigma_{+}(t+\tau) \sigma_{-}(t)\right\rangle$ can be calculated using the quantum regression theorem. Then the SE spectrum can be analytically obtained
\begin{equation}\label{eq12}
S(\omega)=\frac{1}{\pi} \frac{\gamma+\Gamma(\omega)}{\left[\omega-\omega_{0}-\Delta(\omega)\right]^{2}+\left[\frac{\gamma+\Gamma(\omega)}{2}\right]^{2}}
\end{equation}
with the photonic Lamb shift
\begin{equation}\label{eq13}
\Delta(\omega)=g^{2} \operatorname{Re}\left[\chi_{\mathrm{DP}}(\omega)+\chi_{\mathrm{EP}}(\omega)\right]
\end{equation}
and the local coupling strength
\begin{equation}\label{eq14}
\Gamma(\omega) =-2 g^{2} \operatorname{Im}\left[\chi_{\mathrm{DP}}(\omega)+\chi_{\mathrm{EP}}(\omega)\right]
\end{equation}
Note that the SE dynamics of QE can be retrieved from $\mathcal{F}[S(\omega)]$, the Fourier transform of SE spectrum. Eqs. (\ref{eq12})-(\ref{eq14}) indicate that LDOS is crucial for understanding the exotic behavior of quantum dynamics at EPs. The contributions of DP and EP are separated in Eqs. (\ref{eq13}) and (\ref{eq14}), which is beneficial to unravel how the emergence of EPs alters the quantum dynamics. 

\section{Results and discussion}\label{sec3}
\subsection{Exceptional point induced transparency}

\begin{figure}[t]
\includegraphics[width=0.99\linewidth]{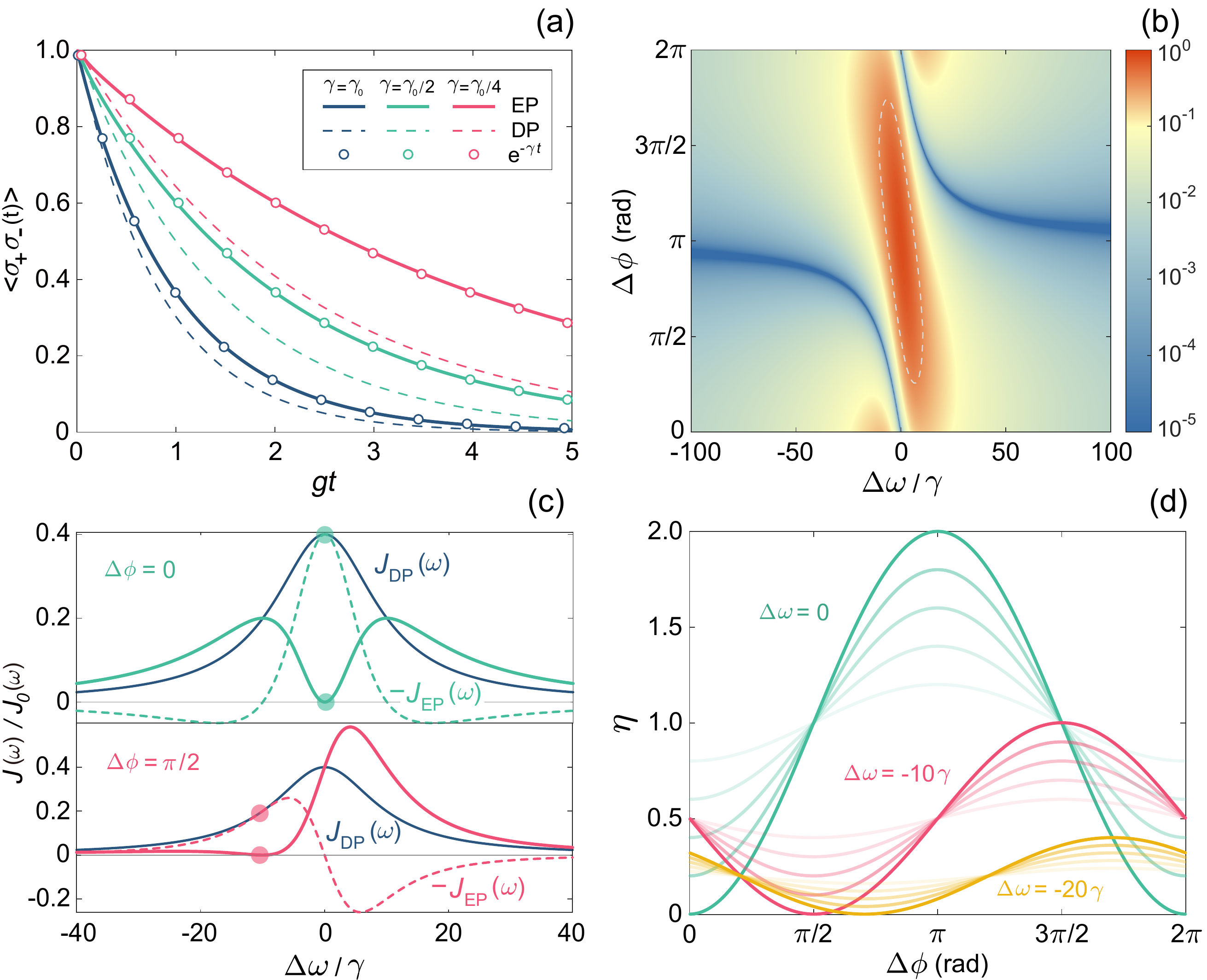}
\caption{\label{fig3} EP induced transparency and spectral density of EP cavity. (a) Comparison of the spontaneous emission dynamics of a QE resonantly and weakly coupled to the EP (solid lines) and DP (dashed lines) cavities with different SE rates $\gamma$. The QE dynamics in the free space is also shown for comparison (circles). The parameters are $g=\gamma_0$ and $g^2/\kappa\gamma=1/40$, where $\gamma_0$ represents the unit rate of QE decay. (b) Normalized spectral density of EP cavity $J(\omega)/J_0(\omega)$ versus frequency and phase difference $\Delta\phi$ with $g=\gamma$ and $\kappa=20\gamma$. The gray dashed line indicates the parameter region for Purcell factor enhancement $\eta>1$. (c) Decomposition of $J(\omega)$ of EP cavity versus frequency for $\Delta\phi=0$ and $\pi/2$. The DP and EP contributions are plotted by the solid blue lines and the dashed color lines, respectively. (d) $\eta$ of EP cavity as the function of $\Delta\phi$ for various $\Delta\omega$. $\left| r \right|$ declines from 1.0 to 0.2 in a linear fashion for curves with the same color from dark to light. The parameters of (c) and (d) are the same as (b). }
\end{figure}

We first consider the weak coupling regime, where the Purcell effect is expected to modify the SE rate of a QE. Note that the quality factor of modes in EP cavity is typically $10^4$ due to the dissipative coupling to waveguide, therefore, we will focus on the case of $\kappa \gg \gamma $ in the following discussion. Fig. \ref{fig3}(a) shows the SE dynamics of QEs with different $\gamma$, while the coupling rate $g$ and the cooperativity $C=8g^2/\kappa\gamma=0.2$ are set to be fixed. For resonant QE-cavity coupling, the maximum Purcell factor of DP cavity is equal to $(C+1)$, obtained by substituting Eq. (\ref{eq8}) into $F_P(\omega_c) = J(\omega_c) / J_0(\omega_c)+ 1$. Therefore, QEs experience the Purcell enhancement and the corresponding SE dynamics manifests a faster decay, as shown by the dashed lines in Fig. \ref{fig3}(a). On the contrary, the SE dynamics of QEs in EP cavity is counterintuitive, which shows good accordance with QEs in the free space as if the cavity is absent. Therefore, we call this intriguing phenomenon as {\it{EP induced transparency}}. The same effect has been reported in Ref. \cite{RN42} and explained as a consequence of decoupling between the QE and the cavity modes due to the formation of stand wave pattern of electric field at EPs. Different from the previous work, here we unravel from the perspective of LDOS that the EP induced transparency results from the precise cancellation of EP and DP contributions of $J\left(\omega\right)$ at special frequency point, giving arise to null Purcell enhancement for a QE with the same transition frequency. 

The frequency corresponding to null Purcell enhancement can be found from Eqs. (\ref{eq8})-(\ref{eq11}) by letting $J(\omega)=0$. The solution takes a simple form
\begin{equation}\label{eq15}
\Delta \omega=\Delta \omega_{m} \equiv-\frac{\kappa}{2} \tan \left(\frac{\Delta \phi}{2}\right)
\end{equation}
where $\Delta \omega=\omega-\omega_c$ is the frequency detuning. Fig. \ref{fig3}(b) displays the normalized LDOS of cavity $J(\omega)/J_0(\omega)$ versus $\Delta \omega$ and $\Delta\phi$, which can be easily tuned by varying the mirror location, i.e., the waveguide length $L$. We can see that the zero point of $J\left(\omega\right)$ goes away from cavity resonance as $\Delta\phi$ increase from 0 to $\pi$, and the opposite tendency is observed for $\Delta \phi \in[\pi, 2 \pi]$, where the zero point of frequency is larger than cavity resonance in this case. Especially, the null Purcell enhancement is achieved at the cavity resonance with $\Delta\phi=0$, which is just the circumstance studied in Ref. \cite{RN42}. Fig. \ref{fig3}(c) plots the decomposition of $J\left(\omega\right)$ for $\Delta\phi=0$ and $\pi/2$, where we plot $-J_\mathrm{EP} (\omega)$, i.e., we reverse the curve of EP term by taking a negative sign, to clearly display the cancellation of EP and DP contributions, see the shaded circles for indicating $\Delta\omega_m$. It shows that the EP term can be negative and weakens the Purcell effect. For $\Delta\phi=0$, the EP response features an even symmetry and a narrow linewidth, compared to a DP (Lorentzian) cavity. The upper panel of Fig. \ref{fig3}(c) shows $J(\omega)>0$ for $\left|\omega-\omega_c\right|>\kappa/2$, while $J(\omega)<0$ for $\left|\omega-\omega_c\right|<\kappa/2$. The resultant $J\left(\omega\right)$ is slightly enhanced at a wide frequency range far detuned from the cavity, but strongly suppressed around the cavity resonance. The EP response cancels the DP contribution at $\Delta\omega$, resulting in the vanishing Purcell effect for a QE resonantly coupled to EP cavity. The lower panel of Fig. \ref{fig3}(c) shows that the EP response exhibits a totally different profile for $\Delta\phi=\pi/2$, which becomes odd symmetry and changes its sign at cavity resonance. As a result, the disappeared Purcell effect occurs at the left side of cavity resonance $(\Delta\omega=-\kappa/2<0)$, while the enhanced Purcell effect is observed at the whole frequency range of $\omega>\omega_c$, leading to a strongly asymmetrical lineshape of $J\left(\omega\right)$. 

Fig. \ref{fig3}(d) plots the Purcell effect enhancement of EPs as the function of $\Delta\phi$ for various $\Delta\omega$ and $\left| r \right|$, which is defined as $\eta \equiv J(\omega_c)/J_\mathrm{DP}(\omega_c)=J_\mathrm{EP}(\omega_c)/J_\mathrm{DP}(\omega_c)+1$. It shows that EP cavity attains greatest tunabiliy to Purcell effect from chiral EPs. Particularly, with resonant QE-cavity coupling, a double increase of Purcell enhancement compared to a DP cavity can be realized with a prefect mirror and $\Delta\phi=\pi$, permitting the stronger light-matter interaction. For a QE detuned from cavity, the null Purcell enhancement still exists with $\left| r \right|=1$, but the maximum $\eta$ decreases. $\eta>1$ can be achieved inside the parameter region indicated by the gray dashed line in Fig. \ref{fig3}(b). Fig. \ref{fig3}(d) also shows that the maximum $\eta$ drops as the loss of reflection amplitude increases, but a mirror with $\left| r \right|>0.8$ still holds great tunability of Purcell effect, where, for example, the maximum $\eta$ is greater than $1.75$ at $\Delta\omega=0$. It is worth noting that $\left| r \right| \sim 0.98$ is achievable \cite{RN42,RN55}, and thus a practical mirror will not significantly weaken the ability of chiral EPs to tune the Purcell effect in experiment. 

\subsection{Decay suppression of Rabi oscillation at chiral EPs}

The above analysis reveals that the chiral EPs have the ability to significantly modify the SE process of a QE, from completely suppressed to enhanced Purcell effect, and thus a EP cavity can provide greater degrees of freedom to control the light-matter interaction than a DP cavity. We now go beyond the weak coupling regime and investigate the effects of chiral EPs on the coherent energy exchange between the QE and the cavity, i.e., the Rabi oscillation. 

\begin{figure}[t]
\includegraphics[width=0.99\linewidth]{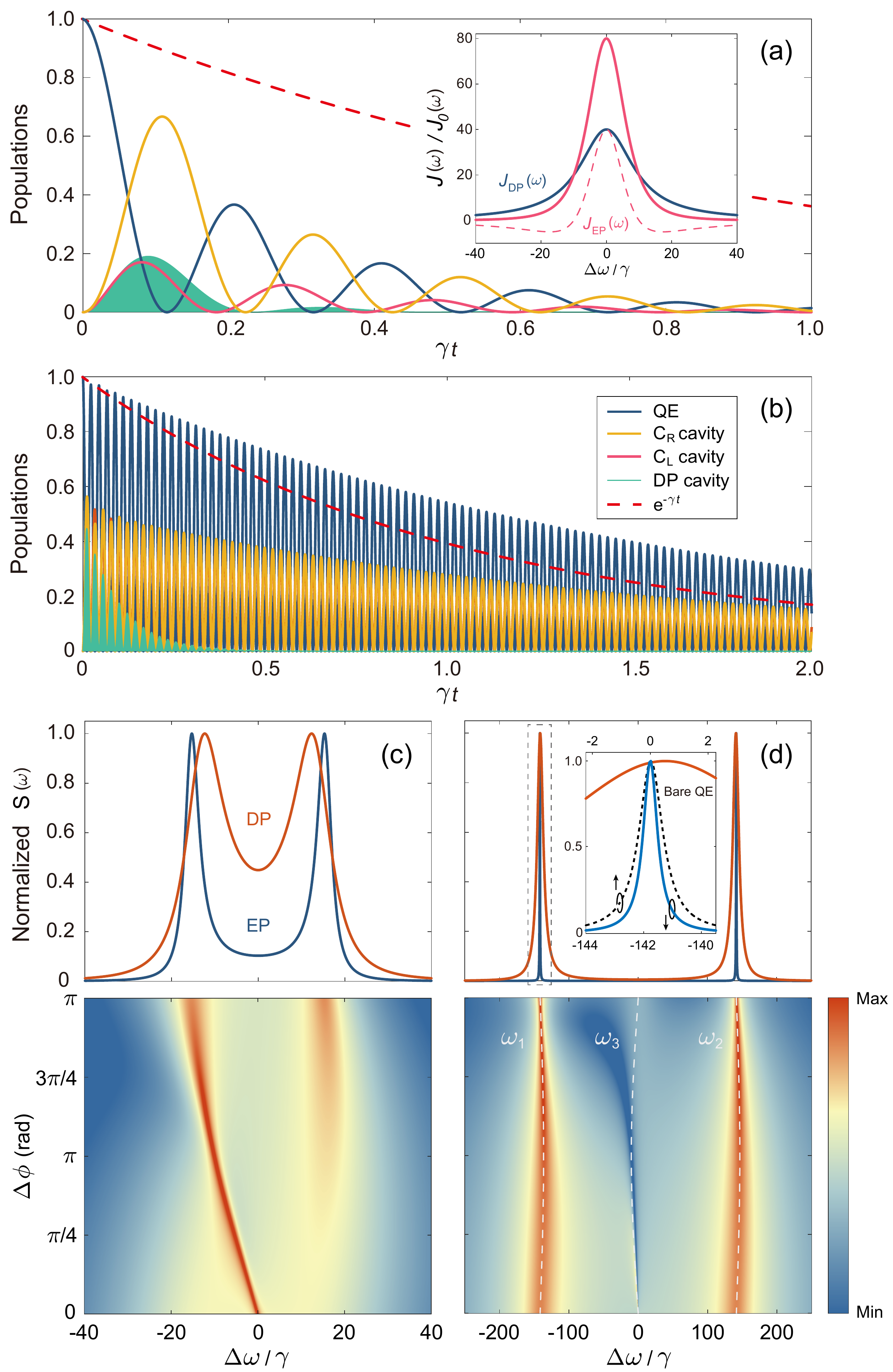}
\caption{\label{fig4} Decay suppression of Rabi oscillation in EP cavity. (a) Population dynamics for QE (blue line) and cavity modes of EP (yellow and pink lines) and DP (thin green line with shading) cavities. SE dynamics of a bare QE (in free space) with the same decay rate is also shown for comparison (dashed red line). The parameters are $g=10\gamma$, $\kappa=20\gamma$, $\omega_{0}=\omega_{c}$ and $\Delta\phi=\pi/2$. The inset plots the corresponding normalized LDOS of cavity. (b) is the same as (a) but for $g=100\gamma$. The upper panels of (c) and (d) plot the normalized SE spectra corresponding to (a) and (b) for EP (blue lines) and DP (red lines) cavities, respectively. While the lower panels show the SE spectrum versus $\Delta\phi$. The inset in the upper panel shows the peak of Rabi doublet with lower energy, where the SE spectrum of a bare QE is also presented for comparison (black dashed line). The white dashed lines in the lower panel track the eigenenergy levels given by Eqs. (\ref{eq16}) and (\ref{eq17}).  }
\end{figure}

\begin{figure*}[t]
\includegraphics[width=0.99\linewidth]{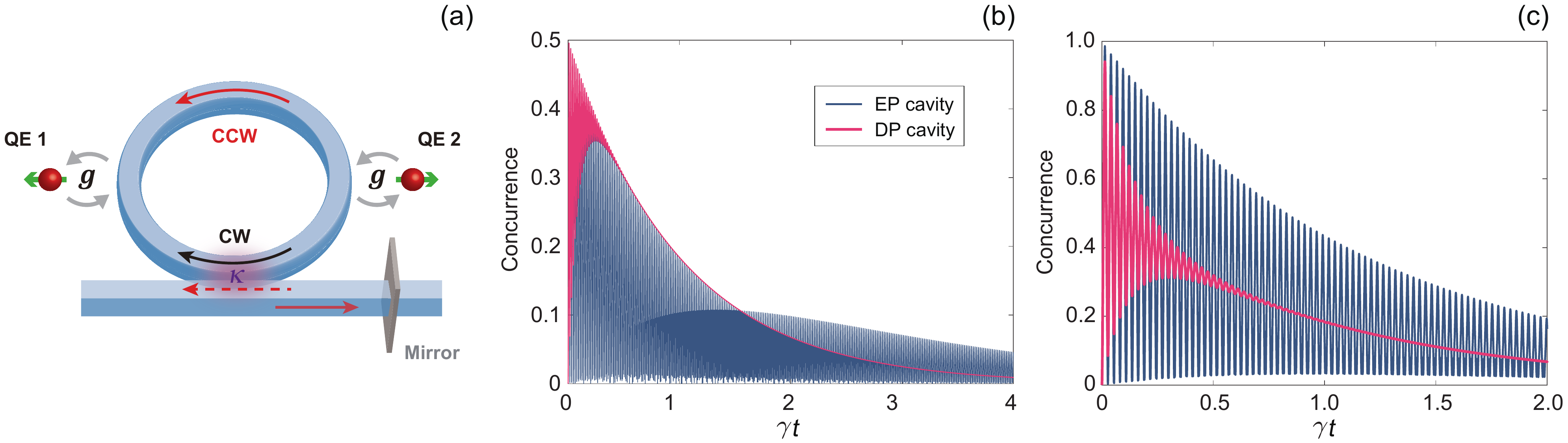}
\caption{\label{fig5} Enhanced entanglement generation in EP cavity. (a) Schematic of spontaneous entanglement generation between two identical qubits in EP cavity. (b) and (c) Dynamical concurrence between two qubits in DP (red lines) and EP cavities (blue lines) for $\Delta_{0c}=2.32g$ and $\Delta_{0c}=0$, respectively. Other parameters are $g=100\gamma$, $\kappa=20\gamma$, and $\Delta\phi=\pi$. The results are numerically calculated using QuTip \cite{RN79} with one qubit initially in the excited state while another in the ground state. }
\end{figure*}

Fig. \ref{fig4}(a) shows the time evolution of populations with an initially excited QE resonantly coupled to EP cavity, where the Rabi oscillation is evident. The parameters are $g=10\gamma$, $\kappa=20\gamma$, and $\Delta\phi=\pi$. It also shows that the maximum population of $c_L$ cavity is slightly lower than the DP cavity, due to the lack of direct energy blackflow from $c_R$ cavity. On the other hand, the Rabi oscillation of $c_L$ cavity sustains for a longer time and four cycles can be observed, indicating less energy dissipation in EP cavity. While the Rabi oscillation of DP cavity manifests a faster decay, and the population of second-cycle oscillation is much smaller than that of EP cavity and hard to recognize. Meanwhile, the maximum population of $c_R$ cavity is more than threefold compared to the DP cavity, and reaches $\sim 0.66$. However, though the higher population, the period of Rabi oscillation is not obviously changed. The SE spectra shown in the upper panel of Fig. \ref{fig4}(c) reveals that, it is because the splitting width is slightly enlarged, but the linewidth of Rabi peaks are greatly reduced. Eq. (\ref{eq12}) indicates that the reduction of linewidth and the resultant decay suppression of Rabi oscillation are attributed to the modification of LDOS by the EP term, see the inset of Fig. \ref{fig4}(a). It shows that for $\Delta\phi=\pi$, EP cavity exhibits the simultaneously enhanced LDOS and narrower linewidth compared to the DP cavity.

Fig. \ref{fig4}(b) shows the time evolution of populations for $g=100\gamma$, where the decay of Rabi oscillation is slower than not only the DP cavity but also a bare QE. This is counterintuitive, since the coupling to a lossy cavity will increase the dissipation rate due to the Purcell enhancement, and result in a faster decay. To gain insight into the problem, we obtain the approximate expressions of eigenenergies from the denominator of SE spectrum (Eq. (\ref{eq12})), which are given by
\begin{equation}\label{eq16}
\omega_{1,2} \approx \pm \sqrt{2} g+\frac{\kappa}{4} \sin (\Delta \phi)-i \frac{[\cos (\Delta \phi)+1] \kappa+\gamma}{4}
\end{equation}
\begin{equation}\label{eq17}
\omega_{3} \approx \frac{\kappa}{2} \sin (\Delta \phi)\left[\frac{\cos (\Delta \phi)}{C}-1\right]
 -i \frac{\kappa}{2}\left[\cos (\Delta \phi)-1\right]
\end{equation}
where we expand the eigenenergies up to second order with respect to $\kappa$ and $\gamma$. The white dashed lines in the lower panel of Fig. \ref{fig4}(d) track the energies of eigenenergy levels, where we find that two eigenenergies in Eq. (\ref{eq16}) correspond to the Rabi doublet in the SE spectrum shown in the upper panel of Fig. \ref{fig4}(d). Same as the case of $g=10\gamma$, the SE spectrum of $g=100\gamma$ exhibits the significant narrowing of Rabi doublet. However, Eq. (\ref{eq16}) indicates that the linewidth narrowing in this case is anomalous, since the minimum decay achieved at $\Delta\phi=\pi$ is $-2 \operatorname{Im}\left[\omega_{1,2}\right] \approx \gamma / 2$, a half of a bare QE, see the inset in the upper panel of Fig. \ref{fig4}(d), where the SE spectrum of a bare QE is shown for comparison. Eq. (\ref{eq16}) also indicates that the reduction of decay rate originates from the unidirectional coupling ($\kappa \cos(\Delta\phi)$ term) between two cavity modes for chiral EPs. The condition $\Delta\phi=\pi$ implies that the decay suppression of Rabi oscillation can be understood as a result of the backflow of partial leaky energy through the reflection of mirror, and thus unique to EP cavity. The lower panels of Fig. \ref{fig4}(c) and (d) display the SE spectra as the function of $\Delta\phi$ for $g=10\gamma$ and $100\gamma$. It is interesting to see that though both are under the strong coupling regime with the equal splitting seen in SE spectrum at $\Delta\phi=\pi$, the underlying evolution of Rabi splitting is distinguishing. The Rabi splitting in the case of $g=100\gamma$ are nearly unchanged as $\Delta\phi$ varies, while for $g=10\gamma$, the peak of Rabi splitting with lower energy evolves from cavity resonance, and the corresponding linewidth is much narrower than the peak with higher energy until $\Delta\phi$ approaches to $\pi$. 

\begin{figure*}[t]
\includegraphics[width=0.99\linewidth]{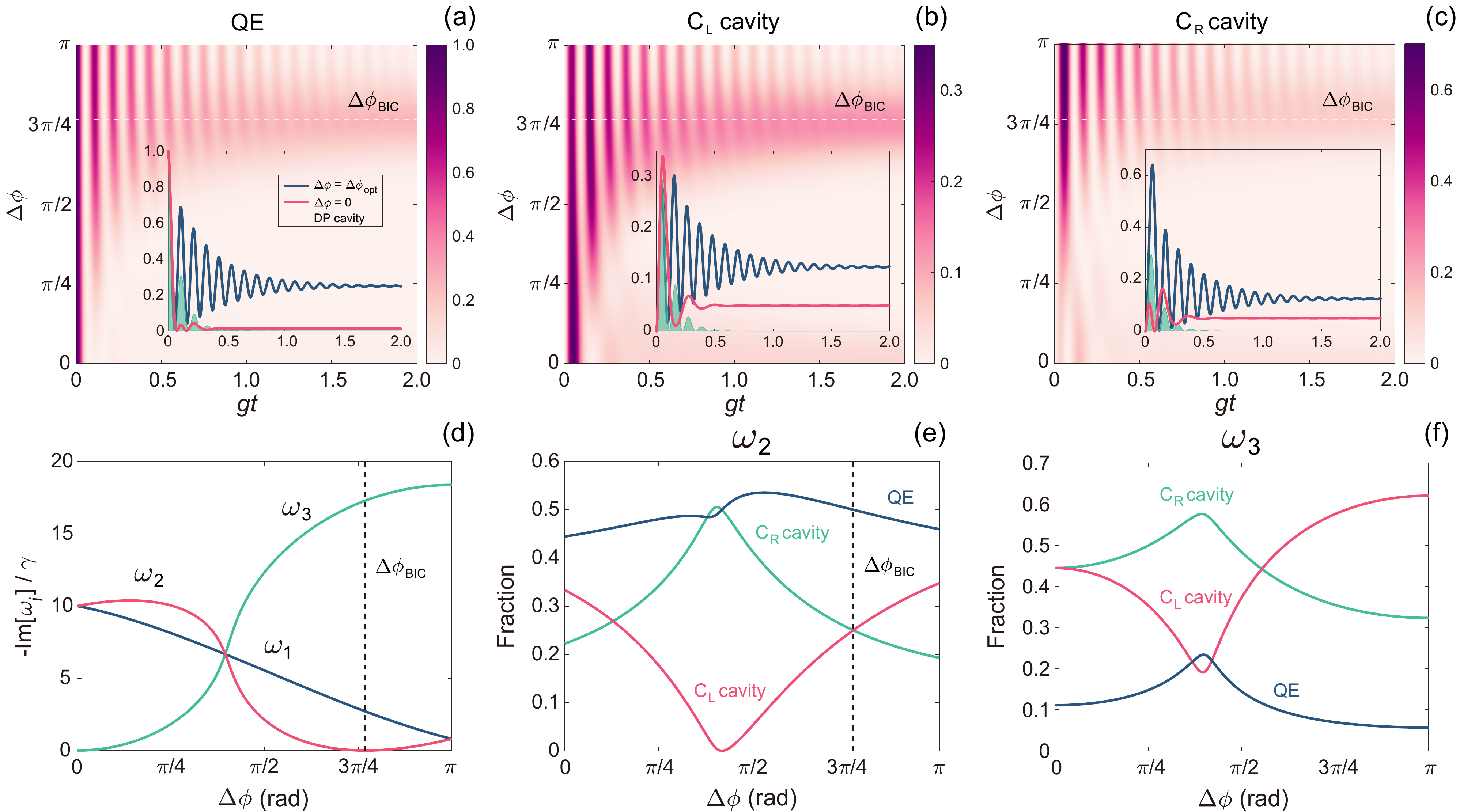}
\caption{\label{fig6} Atom-photon bound states in EP cavity. Time evolution of QE (a), $c_L$ cavity (b) and $c_R$ cavity (c) as the function of $\Delta\phi$ for $\Delta_\mathrm{0c}=0$, $g=20\gamma$ and $\kappa=20\gamma$. The insets show the time evolution for $\Delta\phi=0$ and $\Delta\phi_\mathrm{BIC}$. The results of DP cavity are also shown for comparison. The time evolution is obtained by numerically solving Eq.(\ref{eq3}). (d) Imaginary part of eigenvalues of Eq. (\ref{eq4}) versus $\Delta\phi$. The Hopfield coefficients of QE and two cavities contributed to eigenenergy levels $\omega_2$ and $\omega_3$ are shown in (e) and (f), respectively.  }
\end{figure*}

The above results indicate that EP cavity together with sufficiently strong QE-cavity interaction will have significant advantages in quantum-optics applications, such as the spontaneous entanglement generation (SEG) between qubits, benefiting from the decoherence suppression by chiral EPs. Here we demonstarte the EP enhanced two-qubit SEG in EP cavity, see Fig. \ref{fig5}(a) for schematic. Extending our model into the multi-QE case is straightforward 
\begin{equation}\label{eq18}
\begin{gathered}
\dot{\rho}_{M}=-i\left[H^{M}, \rho\right]+\gamma \sum_{i} \mathcal{L}\left[\sigma_{-}^{(i)}\right] \rho+\kappa \mathcal{L}\left[c_{L}\right] \rho+\kappa \mathcal{L}\left[c_{R}\right] \rho \\
+\kappa|r|\left(e^{i \varphi}\left[c_{L} \rho, c_{R}^{\dagger}\right]+e^{-i \varphi}\left[c_{R}, \rho c_{L}^{\dagger}\right]\right)
\end{gathered}
\end{equation}
with full Hamiltonian $H^M=H_0^M+H_I^M$, where the free Hamiltonian $H_0^M$ and the interaction Hamiltonian $H_I^M$ are given by
\begin{equation}
H_{0}^{M}=\omega_{0} \sum_{i} \sigma_{+}^{(i)} \sigma_{-}^{(i)}+\omega_{c} c_{L}^{\dagger} c_{L}+\omega_{c} c_{R}^{\dagger} c_{R}
\end{equation}
\begin{equation}
\begin{gathered}
H_{I}^{M}=g \sum_{i}\left(e^{-i \phi_{i}} c_{L}^{\dagger} \sigma_{-}^{(i)}+e^{i \phi_{i}} \sigma_{+}^{(i)} c_{L}\right)\\+g \sum_{i}\left(e^{i \phi_{i}} c_{R}^{\dagger} \sigma_{-}^{(i)}+e^{-i \phi_{i}} \sigma_{+}^{(i)} c_{R}\right)
\end{gathered}
\end{equation}
where we assume the identical qubits. With an initially excited qubit, the generated entanglement is quantified by the concurrence $C(t)$ between qubits, which in our system is given by the simple expression $C(t)=2\left| C_{eg} (t) C_{ge}^{*} (t)\right|$ \cite{RN86,np2022}, with $C_{eg} (t)$ and $C_{ge} (t)$ being the probability amplitudes of two single-excitation states that one qubit in the excited state while another in the ground state. 

In the case of resonant qubit-cavity coupling, the upper bound of maximum concurrence is 0.5. As shown in Fig. \ref{fig5}(b), the concurrence exponentially decays with a rate of $\gamma$ for a DP cavity; while in the EP cavity, the evolution of $C(t)$ manifests two distinguishing stages with different decay rates: the decay of $C(t)$ is identical as DP cavity for $t<1.6\gamma^{-1}$; after that, it is found to decay with a rate $\sim\kappa^3/32g^2$, obtained from the eigenenergies of Eq.(\ref{eq18}) in the single-excitation subspace \cite{np2022}. With a large coupling rate $g=100\gamma$, $\kappa^3/32g^2=1/40\gamma\ll\gamma$, and thus the decay of $C(t)$ for $t>1.6\gamma^{-1}$ is much slower than a DP cavity. It also indicates that for a larger $g$, the decay suppression of quantum entanglement will be more evident. Fig. \ref{fig5}(c) displays the $C(t)$ corresponding to the maximum concurrence $C_{\mathrm{max}}={\mathrm{max}}\left[C(t)\right]=0.9866$, achieved with the optimal qubit-cavity detuning $\Delta_{0c} \equiv \omega_0 - \omega_c=2.32g$. In this case, the concurrence decays much faster in a DP cavity than the EP cavity, exhibiting the great potential of EP cavity to preserve the generated entanglement.

\subsection{Population trapping at chiral EPs with atom-photon bound states}

\begin{figure*}[t]
\includegraphics[width=0.99\linewidth]{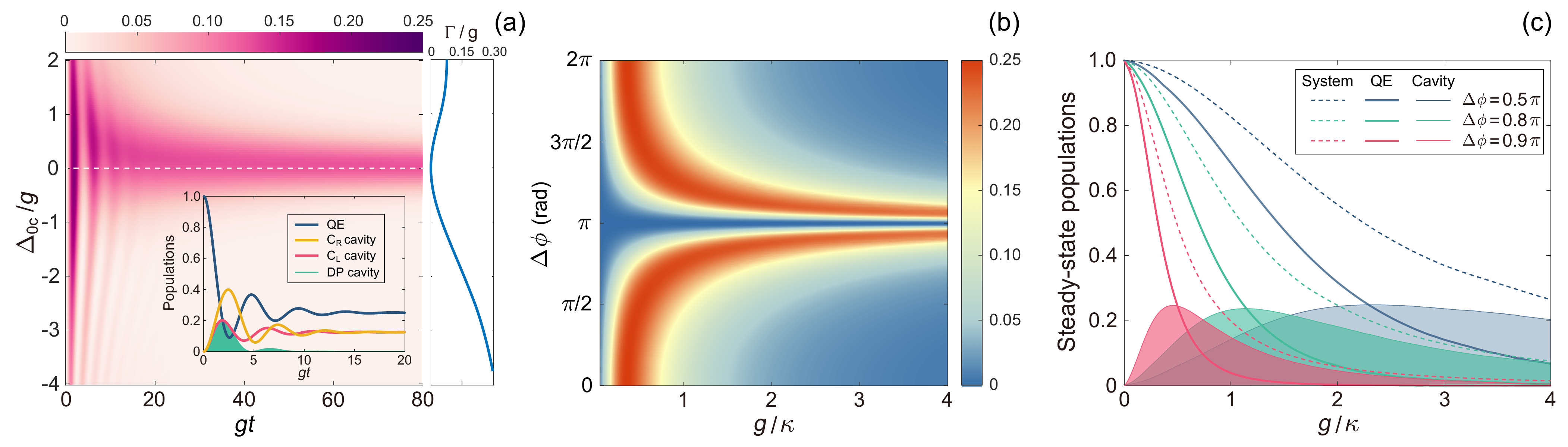}
\caption{\label{fig7} Steady-state population with atom-photon bound states at chiral EPs. (a) Time dynamics of cavity population ($c_L + c_R$) as the function of QE-cavity detuning $\Delta_{0c}$, with parameters $\Delta\phi=\pi/2$ and $g=10\gamma$. The inset shows the time dynamics at $\Delta_{0c}=0$. The right panel plots $\Gamma=-\operatorname{Im}\left[\omega_{\mathrm{BIC}}\right]$ as the function of $\Delta_{0c}$, where $\omega_{\mathrm{BIC}}$ is the eigenenergy corresponding to BIC. (b) Steady-state cavity population versus $g$ and $\Delta_{0c}$. (c) Steady-state populations of QE (solid lines), cavity (thin solid lines with shading) and the system (QE + cavity, dashed lines) as the function of $g$ for various $\Delta\phi$. In all figures, $\kappa=20\gamma$.}
\end{figure*}

Inspired by the above results that the spectral linewidth can be smaller than a bare QE, in the following we investigate the SE dynamics of an ideal QE at chiral EPs, i.e., $\gamma=0$. Fig. \ref{fig6}(a)-(c) plot the time evolution of QE and two cavity modes versus $\Delta\phi=0$ for resonant QE-cavity coupling, where we can observe the population trapping at $\Delta\phi=0$ and $0.77\pi$, while it cannot be achieved in a DP cavity, see the green shaded lines. We can also see that for $\Delta\phi=0$, the steady-state population of QE is almost zero and smaller than that of two cavity modes; however, the situation is reversed for $\Delta\phi=0.77\pi$, and the trapped population is also higher than the case of $\Delta\phi=0$. These phenomena indicate the formation of atom-photon bound states in the system, but the underlying physical mechanisms are different. For the former case ($\Delta\phi=0$), the destructive interference of two cavity modes through the reflection of mirror gives rise to stand wave inside the cavity for population trapping. While the mechanism of bound state for $\Delta\phi \neq 0$ is similar to the accidental bound states in the continuum (BIC) \cite{RN83,RN84,RN85}, also known as Friedrich-Wintgen BIC, which results from the destructive interference of two coupling pathways between QE and $c_R$ cavity, the direct coupling and the indirect coupling mediated by $c_L$ cavity. The explicit expression of non-zero $\Delta\phi$ for BIC can be obtained by finding purely real eigenvalues of characteristic matrix Eq. (\ref{eq4}), which is given by
\begin{equation}\label{eq21}
\Delta \phi_{\mathrm{BIC}}=2 \arccos \left(\frac{\kappa}{2 \sqrt{2} g}\right)
\end{equation}
In Fig. \ref{fig6}(d), we plot the imaginary part (decay) of eigenenergies in the single-excitation subspace as the function of $\Delta\phi$, where it shows that bound states for $\Delta\phi=0$ and $\Delta\phi_\mathrm{BIC}$ are achieved at different eigenenergy levels, $\omega_3$ and $\omega_2$, respectively. Fig. \ref{fig6}(e) shows the Hopfield coefficients (eigenvector components) of $\omega_2$ versus $\Delta\phi$, where we can see that the coefficients for cavity modes and QE are both equal to 0.5 at $\Delta\phi_\mathrm{BIC}$; by contrast, the Hopfield coefficient of QE for $\omega_2$ is less than 0.1 at $\Delta\phi=0$ and much smaller than cavity modes. It explains the different distributions of steady-state population for two bound states shown in Fig. \ref{fig6}(a)-(c). 

Since BIC exhibits better energy preservation at chiral EPs, we then consider a general case of arbitrary $\Delta\phi$, and find the QE-cavity detuning corresponding to BIC. From the denominator of SE spectrum (Eq. (\ref{eq12})), we obtain
\begin{equation}\label{eq22}
\Delta \omega_{\mathrm{BIC}}=\frac{2 g^{2}}{\kappa} \sin (\Delta \phi)+\Delta \omega_{m}
\end{equation}
We can see that Eq. (\ref{eq22}) encompasses the case of EP induced transparency, with $g \ll \kappa$ in the weak coupling regime. As $g$ increases, the contribution of the first term in Eq. (\ref{eq22}) becomes significant, the resultant $\Delta \omega_{\mathrm{BIC}}$ is no longer overlapped with the location of null Purcell enhancement for $\Delta \phi \neq 0$.

Fig. \ref{fig7}(a) shows the time evolution of cavity population versus QE-cavity detuning $\Delta_{0c}=\omega_0-\omega_c$, with parameters $\Delta\phi=\pi/2$, $g=10\gamma$, and $\kappa=20\gamma$. Eq. (\ref{eq22}), as well as the decay of eigenenergies, indicates that BIC is achieved at $\Delta_{0c}=0$, see the blue line in the right panel of Fig. \ref{fig7}(a). As a result, the decay of cavity population is obviously slower around $\Delta_{0c}=0$, and the populations of QE and two cavities are partially trapped after a few cycles of Rabi oscillation at $\Delta_{0c}=0$, as shown by the inset of Fig. \ref{fig7}(a). It shows that though the steady-state populations of two cavities are the same, $c_R$ cavity exhibits stronger Rabi oscillation due to the unidirectional energy transfer from $c_L$ cavity. The maximum population of $c_R$ cavity reaches 0.4, while that of $c_L$ cavity is about 0.2. 

\begin{figure*}[t]
\includegraphics[width=0.99\linewidth]{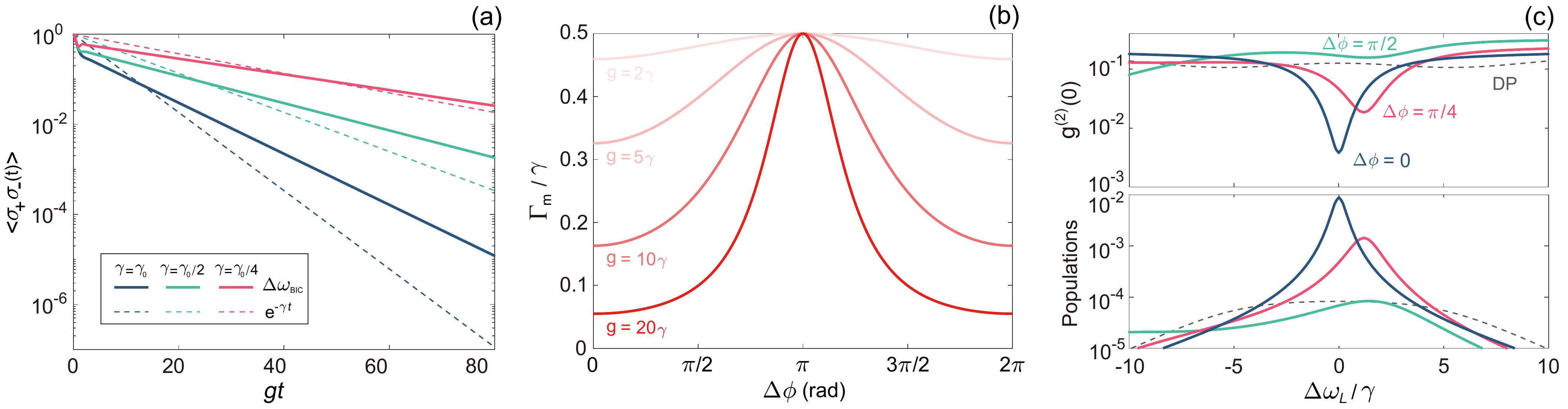}
\caption{\label{fig8} Decay suppression in long-time dynamics with atom-photon bound states at chiral EPs. (a) Logarithmic plot of QE dynamics with detuning $\Delta \omega_{\mathrm{BIC}}$ (solid lines) from cavity for various $\Delta\phi$. The dynamics of bare QEs are shown for comparison (dashed lines). The parameters are $g=5\gamma_0$ and $g^2/\kappa\gamma=5/4$, where $\gamma_0$ is the unit rate of QE decay. Note that the SE rate $\gamma$ is taken to be $\gamma_0$, $\gamma_0/2$, and $\gamma_0/4$ for $\Delta\phi=0$, $\pi/4$, and $\pi/2$, respectively, to distinguish different curves, but the corresponding $\Gamma_{m}$ is still the curve of $g=5\gamma$ in (b). (b) $\Gamma_{m}$ as the function of $\Delta\phi$ for various $g$. (c) $g^{(2)} (0)$ (upper panel) and population (lower panel) of $c_L$ cavity as the function of frequency detuning in the case of $c_R$ cavity drive. The results are obtained by implementing a driving Hamiltonian $H_{d}=\Omega\left(c_{R}^{\dagger}+c_{R}\right)$ in the extended cascaded quantum master equation (Eq. (\ref{eq1})) and numerically calculating using QuTip \cite{RN79}, with parameters $g=5\gamma$, $\Delta\phi=0$, $\Delta_{0c}=\Delta \omega_{\mathrm{BIC}}$ and driving strength $\Omega=0.2\gamma$. $\kappa=20\gamma$ for all figures. }
\end{figure*}

Fig. \ref{fig7}(b) displays the steady-state cavity population versus $\Delta\phi$ and $g$. It reveals that for a given $\Delta\phi$, there is an optimal $g$ for maximum population, denoted as $P_{c}^{o p t}(\Delta \phi)$. The maximal $P_{c}^{o p t}(\Delta \phi)$ is found to achieve at $\Delta\phi=0$, i.e., $P_{c}^{o p t}(0)$ is the upper bound of steady-state cavity population. $P_{c}^{o p t}(\Delta \phi)$ is evaluated to decrease by 0.03 as $\Delta\phi$ varies from 0 to $0.9\pi$, and thus is robust against the variation of $\Delta\phi$. For $\Delta\phi=0$, the steady-state populations can be analytically given from Eq. (\ref{eq3}), which are $P_{e}^{s s}=\kappa^{4} /\left(8 g^{2}+\kappa^{2}\right)^{2}$ for QE and $P_c^{ss}=8g^2 \kappa^2/(8g^2+\kappa^2 )^2$ for cavity. We can see that $P_e^{ss}$ reduces as $g$ increase, exhibiting distinguishing feature from $P_c^{ss}$. Furthermore, we can obtain the analytical expression of leaky energy $P_{\kappa}^{ss}=8g^2/(8g^2+{\kappa}^2 )$, which is monotonically increasing with respect to $g$. It implies that the optimal $g$ for maximum cavity population is the result of balance between the population transfer from QE and the system dissipation. Fig. \ref{fig7}(c) plots the steady-state populations versus $g$ for various $\Delta\phi$. It shows that for $\Delta\phi>0.9\pi$, trapping with high population can be achieved for cavity at a wide range of $g$. 

Finally, we investigate how the atom-photon bound states modify the SE dynamics at chiral EPs by taking into account the SE rate of QE to the free space, i.e., $\gamma \neq 0$. Fig. \ref{fig8}(a) compares the SE dynamics of initial excited QEs in the free space and in EP cavity with optimal detuning $\Delta \omega_{\mathrm{BIC}}$ from the cavity. Remarkably, we can see from the slope of decay curves that BIC at chiral EPs can offer different degrees of suppression on the decoherence process, according to the value of $\Delta\phi$. As a result, the population of QEs in EP cavity can be larger than the bare QEs for $t>50g^{-1}$.

The damping rate of long-time dynamics depends on the minimum decay of eigenenergies, i.e., $\Gamma_{m}=\min \left[-\operatorname{Im}\left[\omega_{i}\right]\right]$. Fig. \ref{fig8}(b) plots $\Gamma_{m}$ for various $g$, where it shows that $\Gamma_{m}$ reaches the minimum at $\Delta\phi=0$, and increases as $\Delta\phi$ varies from 0 to $\pi$. Consistent with the time dynamics shown in Fig. \ref{fig8}(a), $\Gamma_{m}$ of EP cavity can be smaller than $\gamma/2$, which is not possible for a DP cavity with $\kappa \geq \gamma$, since $\Gamma_{m}$ of a standard JC system is $\gamma/2$ for weak coupling and greater than $\gamma/2$ for strong coupling. Fig. \ref{fig8}(b) also shows that enhancing the QE-cavity interaction is beneficial to reduce the decay rate, and a tenfold reduction $(\Gamma_{m} \approx \gamma / 20)$ can be achieved with a moderate QE-cavity coupling $g=20\gamma$ at $\Delta\phi=0$. While for arbitrary $g$, $\Gamma_{m} \rightarrow \gamma / 2$ as $\Delta\phi$ approaches to $\pi$, implying that QE is decoupled from cavity at $\Delta\phi=\pi$. 

As applications, the significantly reduced decay of eigenenergies makes the EP cavity-QE system advantageous for single-photon generation exploiting the photon blockade \cite{RN80,RN81}, where the photon antibunching takes place at the frequency of one of eigenenergy levels in the single-excitation subspace with a coherent input. Therefore, single-photon blockade can attain efficiency improvement at chiral EPs. In the case of $c_R$ cavity drive, Fig. \ref{fig8}(c) compares the single-photon purity $g^{(2)}(0)=\left\langle c_{L}^{\dagger} c_{L}^{\dagger} c_{L} c_{L}\right\rangle / n_{L}^{2}$ and population $n_{L}=\left\langle c_{L}^{\dagger} c_{L}\right\rangle$ of DP and EP cavities, with parameters $\Delta\phi=0$, $g=5\gamma$ and $\kappa=20\gamma$. We evaluate $g \approx g_c=\sqrt{(\kappa^2+\gamma^2)}/4$, where $g_c$ is the critical coupling rate of strong coupling for a DP cavity \cite{RN82}. It implies that the strong antibunching is absent in DP cavity, since the system just reaches the strong coupling regime. As shown in Fig. \ref{fig8}(c), $g^{(2)} (0)$ curve of DP cavity is flat, and the minimum $g^{(2)} (0)$ is $\sim 0.1$. With the resonant QE-cavity coupling, which is also the optimal configuration detuning according to Eq. (\ref{eq22}), the $c_R$ cavity demonstrates a great enhancement of single-photon purity by over an order of magnitude at $\Delta\phi=0$, with the minimal $g^{(2)} (0) \sim 0.005$, accompanied by a hundredfold improvement of population. While for $\Delta\phi=\pi/4$, the single-photon purity and population are both improved by about an order of magnitude at chiral EPs. Therefore, EP cavity shows great potential in building high-efficiency single-photon source.

\section{Conclusion}\label{sec4}

In this work, we study the spontaneous emission of a quantum emitter located in microcavity supporting chiral EPs, and find the intriguing phenomena of EP induced transparency and the anomalously small decay rate of transient and long-time dynamics. An analytical description of LDOS and SE spectrum is presented to unveil that the non-Lorentzian response and atom-photon bound states are responsible for these peculiar quantum dynamics at chiral EPs. The results demonstrate the striking ability of chiral EPs to suppress the decoherence process occurred in both weak- and strong-coupling regimes. Therefore, we envision that EPs can offer the robustness against the dephasing for various quantum-optics applications, including but not limited to high-efficiency single-photon sources \cite{RN87,RN88}, nonlinear interaction at the single-photon level \cite{RN89}, and high-fidelity entanglement generation and transport \cite{RN90,RN91}. Besides the nanophotonic structures, the quantum effects of chiral EPs demonstrated in this work can also be implemented in other kind of platforms, such as superconducting \cite{RN23,RN92}, cavity optomechanics \cite{RN93,RN94}, and open cavity magnonic systems \cite{RN24,RN95}. We believe that our work can provide insights on the effects of EPs in diverse quantum systems and is instructive for harnessing the non-Hermiticity for building novel quantum devices. 

\begin{acknowledgments}
We are grateful to Haishu Tan and Yongyao Li for helpful discussions and supports. This work is supported by the Postdoctor Startup Project of Foshan and the High-level Talent Scientific Research Startup Project of Foshan University (CGZ07001).
\end{acknowledgments}

%
%
%
%
%
%
%

\nocite{*}

\bibliography{eqed}

\begin{thebibliography}{97}%
\makeatletter
\providecommand \@ifxundefined [1]{%
 \@ifx{#1\undefined}
}%
\providecommand \@ifnum [1]{%
 \ifnum #1\expandafter \@firstoftwo
 \else \expandafter \@secondoftwo
 \fi
}%
\providecommand \@ifx [1]{%
 \ifx #1\expandafter \@firstoftwo
 \else \expandafter \@secondoftwo
 \fi
}%
\providecommand \natexlab [1]{#1}%
\providecommand \enquote  [1]{``#1''}%
\providecommand \bibnamefont  [1]{#1}%
\providecommand \bibfnamefont [1]{#1}%
\providecommand \citenamefont [1]{#1}%
\providecommand \href@noop [0]{\@secondoftwo}%
\providecommand \href [0]{\begingroup \@sanitize@url \@href}%
\providecommand \@href[1]{\@@startlink{#1}\@@href}%
\providecommand \@@href[1]{\endgroup#1\@@endlink}%
\providecommand \@sanitize@url [0]{\catcode `\\12\catcode `\$12\catcode
  `\&12\catcode `\#12\catcode `\^12\catcode `\_12\catcode `\%12\relax}%
\providecommand \@@startlink[1]{}%
\providecommand \@@endlink[0]{}%
\providecommand \url  [0]{\begingroup\@sanitize@url \@url }%
\providecommand \@url [1]{\endgroup\@href {#1}{\urlprefix }}%
\providecommand \urlprefix  [0]{URL }%
\providecommand \Eprint [0]{\href }%
\providecommand \doibase [0]{https://doi.org/}%
\providecommand \selectlanguage [0]{\@gobble}%
\providecommand \bibinfo  [0]{\@secondoftwo}%
\providecommand \bibfield  [0]{\@secondoftwo}%
\providecommand \translation [1]{[#1]}%
\providecommand \BibitemOpen [0]{}%
\providecommand \bibitemStop [0]{}%
\providecommand \bibitemNoStop [0]{.\EOS\space}%
\providecommand \EOS [0]{\spacefactor3000\relax}%
\providecommand \BibitemShut  [1]{\csname bibitem#1\endcsname}%
\let\auto@bib@innerbib\@empty
\bibitem [{\citenamefont {Bender}\ and\ \citenamefont {Boettcher}(1998)}]{RN1}%
  \BibitemOpen
  \bibfield  {author} {\bibinfo {author} {\bibfnamefont {C.~M.}\ \bibnamefont
  {Bender}}\ and\ \bibinfo {author} {\bibfnamefont {S.}~\bibnamefont
  {Boettcher}},\ }\bibfield  {title} {\bibinfo {title} {Real spectra in
  non-hermitian hamiltonians having $\mathcal{PT}$ symmetry},\ }\href
  {https://doi.org/10.1103/PhysRevLett.80.5243} {\bibfield  {journal} {\bibinfo
   {journal} {Physical Review Letters}\ }\textbf {\bibinfo {volume} {80}},\
  \bibinfo {pages} {5243} (\bibinfo {year} {1998})}\BibitemShut {NoStop}%
\bibitem [{\citenamefont {Miri}\ and\ \citenamefont {Alù}(2019)}]{RN2}%
  \BibitemOpen
  \bibfield  {author} {\bibinfo {author} {\bibfnamefont {M.-A.}\ \bibnamefont
  {Miri}}\ and\ \bibinfo {author} {\bibfnamefont {A.}~\bibnamefont {Alù}},\
  }\bibfield  {title} {\bibinfo {title} {Exceptional points in optics and
  photonics},\ }\href {https://doi.org/10.1126/science.aar7709 %J Science}
  {\bibfield  {journal} {\bibinfo  {journal} {Science}\ }\textbf {\bibinfo
  {volume} {363}},\ \bibinfo {pages} {eaar7709} (\bibinfo {year}
  {2019})}\BibitemShut {NoStop}%
\bibitem [{\citenamefont {{\"O}zdemir}\ \emph {et~al.}(2019)\citenamefont
  {{\"O}zdemir}, \citenamefont {Rotter}, \citenamefont {Nori},\ and\
  \citenamefont {Yang}}]{RN3}%
  \BibitemOpen
  \bibfield  {author} {\bibinfo {author} {\bibfnamefont {S.~K.}\ \bibnamefont
  {{\"O}zdemir}}, \bibinfo {author} {\bibfnamefont {S.}~\bibnamefont {Rotter}},
  \bibinfo {author} {\bibfnamefont {F.}~\bibnamefont {Nori}},\ and\ \bibinfo
  {author} {\bibfnamefont {L.}~\bibnamefont {Yang}},\ }\bibfield  {title}
  {\bibinfo {title} {Parity–time symmetry and exceptional points in
  photonics},\ }\href {https://doi.org/10.1038/s41563-019-0304-9} {\bibfield
  {journal} {\bibinfo  {journal} {Nature Materials}\ }\textbf {\bibinfo
  {volume} {18}},\ \bibinfo {pages} {783} (\bibinfo {year} {2019})}\BibitemShut
  {NoStop}%
\bibitem [{\citenamefont {El-Ganainy}\ \emph {et~al.}(2018)\citenamefont
  {El-Ganainy}, \citenamefont {Makris}, \citenamefont {Khajavikhan},
  \citenamefont {Musslimani}, \citenamefont {Rotter},\ and\ \citenamefont
  {Christodoulides}}]{RN4}%
  \BibitemOpen
  \bibfield  {author} {\bibinfo {author} {\bibfnamefont {R.}~\bibnamefont
  {El-Ganainy}}, \bibinfo {author} {\bibfnamefont {K.~G.}\ \bibnamefont
  {Makris}}, \bibinfo {author} {\bibfnamefont {M.}~\bibnamefont {Khajavikhan}},
  \bibinfo {author} {\bibfnamefont {Z.~H.}\ \bibnamefont {Musslimani}},
  \bibinfo {author} {\bibfnamefont {S.}~\bibnamefont {Rotter}},\ and\ \bibinfo
  {author} {\bibfnamefont {D.~N.}\ \bibnamefont {Christodoulides}},\ }\bibfield
   {title} {\bibinfo {title} {Non-hermitian physics and $\mathcal{PT}$
  symmetry},\ }\href {https://doi.org/10.1038/nphys4323} {\bibfield  {journal}
  {\bibinfo  {journal} {Nature Physics}\ }\textbf {\bibinfo {volume} {14}},\
  \bibinfo {pages} {11} (\bibinfo {year} {2018})}\BibitemShut {NoStop}%
\bibitem [{\citenamefont {Bergholtz}\ \emph {et~al.}(2021)\citenamefont
  {Bergholtz}, \citenamefont {Budich},\ and\ \citenamefont {Kunst}}]{RN5}%
  \BibitemOpen
  \bibfield  {author} {\bibinfo {author} {\bibfnamefont {E.~J.}\ \bibnamefont
  {Bergholtz}}, \bibinfo {author} {\bibfnamefont {J.~C.}\ \bibnamefont
  {Budich}},\ and\ \bibinfo {author} {\bibfnamefont {F.~K.}\ \bibnamefont
  {Kunst}},\ }\bibfield  {title} {\bibinfo {title} {Exceptional topology of
  non-hermitian systems},\ }\href
  {https://doi.org/10.1103/RevModPhys.93.015005} {\bibfield  {journal}
  {\bibinfo  {journal} {Reviews of Modern Physics}\ }\textbf {\bibinfo {volume}
  {93}},\ \bibinfo {pages} {015005} (\bibinfo {year} {2021})}\BibitemShut
  {NoStop}%
\bibitem [{\citenamefont {Hodaei}\ \emph {et~al.}(2017)\citenamefont {Hodaei},
  \citenamefont {Hassan}, \citenamefont {Wittek}, \citenamefont
  {Garcia-Gracia}, \citenamefont {El-Ganainy}, \citenamefont
  {Christodoulides},\ and\ \citenamefont {Khajavikhan}}]{RN6}%
  \BibitemOpen
  \bibfield  {author} {\bibinfo {author} {\bibfnamefont {H.}~\bibnamefont
  {Hodaei}}, \bibinfo {author} {\bibfnamefont {A.~U.}\ \bibnamefont {Hassan}},
  \bibinfo {author} {\bibfnamefont {S.}~\bibnamefont {Wittek}}, \bibinfo
  {author} {\bibfnamefont {H.}~\bibnamefont {Garcia-Gracia}}, \bibinfo {author}
  {\bibfnamefont {R.}~\bibnamefont {El-Ganainy}}, \bibinfo {author}
  {\bibfnamefont {D.~N.}\ \bibnamefont {Christodoulides}},\ and\ \bibinfo
  {author} {\bibfnamefont {M.}~\bibnamefont {Khajavikhan}},\ }\bibfield
  {title} {\bibinfo {title} {Enhanced sensitivity at higher-order exceptional
  points},\ }\href {https://doi.org/10.1038/nature23280} {\bibfield  {journal}
  {\bibinfo  {journal} {Nature}\ }\textbf {\bibinfo {volume} {548}},\ \bibinfo
  {pages} {187} (\bibinfo {year} {2017})}\BibitemShut {NoStop}%
\bibitem [{\citenamefont {Chen}\ \emph {et~al.}(2017)\citenamefont {Chen},
  \citenamefont {Kaya~{\"O}zdemir}, \citenamefont {Zhao}, \citenamefont
  {Wiersig},\ and\ \citenamefont {Yang}}]{RN7}%
  \BibitemOpen
  \bibfield  {author} {\bibinfo {author} {\bibfnamefont {W.}~\bibnamefont
  {Chen}}, \bibinfo {author} {\bibfnamefont {S.}~\bibnamefont
  {Kaya~{\"O}zdemir}}, \bibinfo {author} {\bibfnamefont {G.}~\bibnamefont
  {Zhao}}, \bibinfo {author} {\bibfnamefont {J.}~\bibnamefont {Wiersig}},\ and\
  \bibinfo {author} {\bibfnamefont {L.}~\bibnamefont {Yang}},\ }\bibfield
  {title} {\bibinfo {title} {Exceptional points enhance sensing in an optical
  microcavity},\ }\href {https://doi.org/10.1038/nature23281} {\bibfield
  {journal} {\bibinfo  {journal} {Nature}\ }\textbf {\bibinfo {volume} {548}},\
  \bibinfo {pages} {192} (\bibinfo {year} {2017})}\BibitemShut {NoStop}%
\bibitem [{\citenamefont {Wiersig}(2014)}]{RN8}%
  \BibitemOpen
  \bibfield  {author} {\bibinfo {author} {\bibfnamefont {J.}~\bibnamefont
  {Wiersig}},\ }\bibfield  {title} {\bibinfo {title} {Enhancing the sensitivity
  of frequency and energy splitting detection by using exceptional points:
  Application to microcavity sensors for single-particle detection},\ }\href
  {https://doi.org/10.1103/PhysRevLett.112.203901} {\bibfield  {journal}
  {\bibinfo  {journal} {Physical Review Letters}\ }\textbf {\bibinfo {volume}
  {112}},\ \bibinfo {pages} {203901} (\bibinfo {year} {2014})}\BibitemShut
  {NoStop}%
\bibitem [{\citenamefont {Lai}\ \emph {et~al.}(2019)\citenamefont {Lai},
  \citenamefont {Lu}, \citenamefont {Suh}, \citenamefont {Yuan},\ and\
  \citenamefont {Vahala}}]{RN9}%
  \BibitemOpen
  \bibfield  {author} {\bibinfo {author} {\bibfnamefont {Y.-H.}\ \bibnamefont
  {Lai}}, \bibinfo {author} {\bibfnamefont {Y.-K.}\ \bibnamefont {Lu}},
  \bibinfo {author} {\bibfnamefont {M.-G.}\ \bibnamefont {Suh}}, \bibinfo
  {author} {\bibfnamefont {Z.}~\bibnamefont {Yuan}},\ and\ \bibinfo {author}
  {\bibfnamefont {K.}~\bibnamefont {Vahala}},\ }\bibfield  {title} {\bibinfo
  {title} {Observation of the exceptional-point-enhanced sagnac effect},\
  }\href {https://doi.org/10.1038/s41586-019-1777-z} {\bibfield  {journal}
  {\bibinfo  {journal} {Nature}\ }\textbf {\bibinfo {volume} {576}},\ \bibinfo
  {pages} {65} (\bibinfo {year} {2019})}\BibitemShut {NoStop}%
\bibitem [{\citenamefont {Hokmabadi}\ \emph {et~al.}(2019)\citenamefont
  {Hokmabadi}, \citenamefont {Schumer}, \citenamefont {Christodoulides},\ and\
  \citenamefont {Khajavikhan}}]{RN10}%
  \BibitemOpen
  \bibfield  {author} {\bibinfo {author} {\bibfnamefont {M.~P.}\ \bibnamefont
  {Hokmabadi}}, \bibinfo {author} {\bibfnamefont {A.}~\bibnamefont {Schumer}},
  \bibinfo {author} {\bibfnamefont {D.~N.}\ \bibnamefont {Christodoulides}},\
  and\ \bibinfo {author} {\bibfnamefont {M.}~\bibnamefont {Khajavikhan}},\
  }\bibfield  {title} {\bibinfo {title} {Non-hermitian ring laser gyroscopes
  with enhanced sagnac sensitivity},\ }\href
  {https://doi.org/10.1038/s41586-019-1780-4} {\bibfield  {journal} {\bibinfo
  {journal} {Nature}\ }\textbf {\bibinfo {volume} {576}},\ \bibinfo {pages}
  {70} (\bibinfo {year} {2019})}\BibitemShut {NoStop}%
\bibitem [{\citenamefont {Zhong}\ \emph {et~al.}(2019)\citenamefont {Zhong},
  \citenamefont {Ren}, \citenamefont {Khajavikhan}, \citenamefont
  {Christodoulides}, \citenamefont {{\"O}zdemir},\ and\ \citenamefont
  {El-Ganainy}}]{RN11}%
  \BibitemOpen
  \bibfield  {author} {\bibinfo {author} {\bibfnamefont {Q.}~\bibnamefont
  {Zhong}}, \bibinfo {author} {\bibfnamefont {J.}~\bibnamefont {Ren}}, \bibinfo
  {author} {\bibfnamefont {M.}~\bibnamefont {Khajavikhan}}, \bibinfo {author}
  {\bibfnamefont {D.~N.}\ \bibnamefont {Christodoulides}}, \bibinfo {author}
  {\bibfnamefont {S.~K.}\ \bibnamefont {{\"O}zdemir}},\ and\ \bibinfo {author}
  {\bibfnamefont {R.}~\bibnamefont {El-Ganainy}},\ }\bibfield  {title}
  {\bibinfo {title} {Sensing with exceptional surfaces in order to combine
  sensitivity with robustness},\ }\href
  {https://doi.org/10.1103/PhysRevLett.122.153902} {\bibfield  {journal}
  {\bibinfo  {journal} {Physical Review Letters}\ }\textbf {\bibinfo {volume}
  {122}},\ \bibinfo {pages} {153902} (\bibinfo {year} {2019})}\BibitemShut
  {NoStop}%
\bibitem [{\citenamefont {Qin}\ \emph {et~al.}(2021)\citenamefont {Qin},
  \citenamefont {Xie}, \citenamefont {Zhang}, \citenamefont {Hu}, \citenamefont
  {Wang}, \citenamefont {Li}, \citenamefont {Xu}, \citenamefont {Lei},
  \citenamefont {Ruan},\ and\ \citenamefont {Long}}]{lgl}%
  \BibitemOpen
  \bibfield  {author} {\bibinfo {author} {\bibfnamefont {G.-Q.}\ \bibnamefont
  {Qin}}, \bibinfo {author} {\bibfnamefont {R.-R.}\ \bibnamefont {Xie}},
  \bibinfo {author} {\bibfnamefont {H.}~\bibnamefont {Zhang}}, \bibinfo
  {author} {\bibfnamefont {Y.-Q.}\ \bibnamefont {Hu}}, \bibinfo {author}
  {\bibfnamefont {M.}~\bibnamefont {Wang}}, \bibinfo {author} {\bibfnamefont
  {G.-Q.}\ \bibnamefont {Li}}, \bibinfo {author} {\bibfnamefont
  {H.}~\bibnamefont {Xu}}, \bibinfo {author} {\bibfnamefont {F.}~\bibnamefont
  {Lei}}, \bibinfo {author} {\bibfnamefont {D.}~\bibnamefont {Ruan}},\ and\
  \bibinfo {author} {\bibfnamefont {G.-L.}\ \bibnamefont {Long}},\ }\bibfield
  {title} {\bibinfo {title} {Experimental realization of sensitivity
  enhancement and suppression with exceptional surfaces},\ }\href
  {https://doi.org/https://doi.org/10.1002/lpor.202000569} {\bibfield
  {journal} {\bibinfo  {journal} {Laser \& Photonics Reviews}\ }\textbf
  {\bibinfo {volume} {15}},\ \bibinfo {pages} {2000569} (\bibinfo {year}
  {2021})}\BibitemShut {NoStop}%
\bibitem [{\citenamefont {Feng}\ \emph {et~al.}(2014)\citenamefont {Feng},
  \citenamefont {Wong}, \citenamefont {Ma}, \citenamefont {Wang},\ and\
  \citenamefont {Zhang}}]{RN12}%
  \BibitemOpen
  \bibfield  {author} {\bibinfo {author} {\bibfnamefont {L.}~\bibnamefont
  {Feng}}, \bibinfo {author} {\bibfnamefont {Z.~J.}\ \bibnamefont {Wong}},
  \bibinfo {author} {\bibfnamefont {R.~M.}\ \bibnamefont {Ma}}, \bibinfo
  {author} {\bibfnamefont {Y.}~\bibnamefont {Wang}},\ and\ \bibinfo {author}
  {\bibfnamefont {X.}~\bibnamefont {Zhang}},\ }\bibfield  {title} {\bibinfo
  {title} {Single-mode laser by parity-time symmetry breaking},\ }\href
  {https://doi.org/10.1126/science.1258479} {\bibfield  {journal} {\bibinfo
  {journal} {Science}\ }\textbf {\bibinfo {volume} {346}},\ \bibinfo {pages}
  {972} (\bibinfo {year} {2014})}\BibitemShut {NoStop}%
\bibitem [{\citenamefont {Hodaei}\ \emph {et~al.}(2014)\citenamefont {Hodaei},
  \citenamefont {Miri}, \citenamefont {Heinrich}, \citenamefont
  {Christodoulides},\ and\ \citenamefont {Khajavikhan}}]{RN13}%
  \BibitemOpen
  \bibfield  {author} {\bibinfo {author} {\bibfnamefont {H.}~\bibnamefont
  {Hodaei}}, \bibinfo {author} {\bibfnamefont {M.~A.}\ \bibnamefont {Miri}},
  \bibinfo {author} {\bibfnamefont {M.}~\bibnamefont {Heinrich}}, \bibinfo
  {author} {\bibfnamefont {D.~N.}\ \bibnamefont {Christodoulides}},\ and\
  \bibinfo {author} {\bibfnamefont {M.}~\bibnamefont {Khajavikhan}},\
  }\bibfield  {title} {\bibinfo {title} {Parity-time-symmetric microring
  lasers},\ }\href {https://doi.org/10.1126/science.1258480} {\bibfield
  {journal} {\bibinfo  {journal} {Science}\ }\textbf {\bibinfo {volume}
  {346}},\ \bibinfo {pages} {975} (\bibinfo {year} {2014})}\BibitemShut
  {NoStop}%
\bibitem [{\citenamefont {Xu}\ \emph {et~al.}(2016)\citenamefont {Xu},
  \citenamefont {Mason}, \citenamefont {Jiang},\ and\ \citenamefont
  {Harris}}]{RN14}%
  \BibitemOpen
  \bibfield  {author} {\bibinfo {author} {\bibfnamefont {H.}~\bibnamefont
  {Xu}}, \bibinfo {author} {\bibfnamefont {D.}~\bibnamefont {Mason}}, \bibinfo
  {author} {\bibfnamefont {L.}~\bibnamefont {Jiang}},\ and\ \bibinfo {author}
  {\bibfnamefont {J.~G.~E.}\ \bibnamefont {Harris}},\ }\bibfield  {title}
  {\bibinfo {title} {Topological energy transfer in an optomechanical system
  with exceptional points},\ }\href {https://doi.org/10.1038/nature18604}
  {\bibfield  {journal} {\bibinfo  {journal} {Nature}\ }\textbf {\bibinfo
  {volume} {537}},\ \bibinfo {pages} {80} (\bibinfo {year} {2016})}\BibitemShut
  {NoStop}%
\bibitem [{\citenamefont {Doppler}\ \emph {et~al.}(2016)\citenamefont
  {Doppler}, \citenamefont {Mailybaev}, \citenamefont {Böhm}, \citenamefont
  {Kuhl}, \citenamefont {Girschik}, \citenamefont {Libisch}, \citenamefont
  {Milburn}, \citenamefont {Rabl}, \citenamefont {Moiseyev},\ and\
  \citenamefont {Rotter}}]{RN15}%
  \BibitemOpen
  \bibfield  {author} {\bibinfo {author} {\bibfnamefont {J.}~\bibnamefont
  {Doppler}}, \bibinfo {author} {\bibfnamefont {A.~A.}\ \bibnamefont
  {Mailybaev}}, \bibinfo {author} {\bibfnamefont {J.}~\bibnamefont {Böhm}},
  \bibinfo {author} {\bibfnamefont {U.}~\bibnamefont {Kuhl}}, \bibinfo {author}
  {\bibfnamefont {A.}~\bibnamefont {Girschik}}, \bibinfo {author}
  {\bibfnamefont {F.}~\bibnamefont {Libisch}}, \bibinfo {author} {\bibfnamefont
  {T.~J.}\ \bibnamefont {Milburn}}, \bibinfo {author} {\bibfnamefont
  {P.}~\bibnamefont {Rabl}}, \bibinfo {author} {\bibfnamefont {N.}~\bibnamefont
  {Moiseyev}},\ and\ \bibinfo {author} {\bibfnamefont {S.}~\bibnamefont
  {Rotter}},\ }\bibfield  {title} {\bibinfo {title} {Dynamically encircling an
  exceptional point for asymmetric mode switching},\ }\href
  {https://doi.org/10.1038/nature18605} {\bibfield  {journal} {\bibinfo
  {journal} {Nature}\ }\textbf {\bibinfo {volume} {537}},\ \bibinfo {pages}
  {76} (\bibinfo {year} {2016})}\BibitemShut {NoStop}%
\bibitem [{\citenamefont {Chen}\ \emph {et~al.}(2020)\citenamefont {Chen},
  \citenamefont {Liu}, \citenamefont {Luan}, \citenamefont {Liu}, \citenamefont
  {Wang}, \citenamefont {Zhu}, \citenamefont {Li}, \citenamefont {Gu},
  \citenamefont {Liang}, \citenamefont {Gao}, \citenamefont {Lu}, \citenamefont
  {Ge}, \citenamefont {Zhang}, \citenamefont {Zhu},\ and\ \citenamefont
  {Ma}}]{RN17}%
  \BibitemOpen
  \bibfield  {author} {\bibinfo {author} {\bibfnamefont {H.-Z.}\ \bibnamefont
  {Chen}}, \bibinfo {author} {\bibfnamefont {T.}~\bibnamefont {Liu}}, \bibinfo
  {author} {\bibfnamefont {H.-Y.}\ \bibnamefont {Luan}}, \bibinfo {author}
  {\bibfnamefont {R.-J.}\ \bibnamefont {Liu}}, \bibinfo {author} {\bibfnamefont
  {X.-Y.}\ \bibnamefont {Wang}}, \bibinfo {author} {\bibfnamefont {X.-F.}\
  \bibnamefont {Zhu}}, \bibinfo {author} {\bibfnamefont {Y.-B.}\ \bibnamefont
  {Li}}, \bibinfo {author} {\bibfnamefont {Z.-M.}\ \bibnamefont {Gu}}, \bibinfo
  {author} {\bibfnamefont {S.-J.}\ \bibnamefont {Liang}}, \bibinfo {author}
  {\bibfnamefont {H.}~\bibnamefont {Gao}}, \bibinfo {author} {\bibfnamefont
  {L.}~\bibnamefont {Lu}}, \bibinfo {author} {\bibfnamefont {L.}~\bibnamefont
  {Ge}}, \bibinfo {author} {\bibfnamefont {S.}~\bibnamefont {Zhang}}, \bibinfo
  {author} {\bibfnamefont {J.}~\bibnamefont {Zhu}},\ and\ \bibinfo {author}
  {\bibfnamefont {R.-M.}\ \bibnamefont {Ma}},\ }\bibfield  {title} {\bibinfo
  {title} {Revealing the missing dimension at an exceptional point},\ }\href
  {https://doi.org/10.1038/s41567-020-0807-y} {\bibfield  {journal} {\bibinfo
  {journal} {Nature Physics}\ }\textbf {\bibinfo {volume} {16}},\ \bibinfo
  {pages} {571} (\bibinfo {year} {2020})}\BibitemShut {NoStop}%
\bibitem [{\citenamefont {Lin}\ \emph {et~al.}(2011)\citenamefont {Lin},
  \citenamefont {Ramezani}, \citenamefont {Eichelkraut}, \citenamefont
  {Kottos}, \citenamefont {Cao},\ and\ \citenamefont {Christodoulides}}]{RN18}%
  \BibitemOpen
  \bibfield  {author} {\bibinfo {author} {\bibfnamefont {Z.}~\bibnamefont
  {Lin}}, \bibinfo {author} {\bibfnamefont {H.}~\bibnamefont {Ramezani}},
  \bibinfo {author} {\bibfnamefont {T.}~\bibnamefont {Eichelkraut}}, \bibinfo
  {author} {\bibfnamefont {T.}~\bibnamefont {Kottos}}, \bibinfo {author}
  {\bibfnamefont {H.}~\bibnamefont {Cao}},\ and\ \bibinfo {author}
  {\bibfnamefont {D.~N.}\ \bibnamefont {Christodoulides}},\ }\bibfield  {title}
  {\bibinfo {title} {Unidirectional invisibility induced by
  $\mathcal{PT}$-symmetric periodic structures},\ }\href
  {https://doi.org/10.1103/PhysRevLett.106.213901} {\bibfield  {journal}
  {\bibinfo  {journal} {Physical Review Letters}\ }\textbf {\bibinfo {volume}
  {106}},\ \bibinfo {pages} {213901} (\bibinfo {year} {2011})}\BibitemShut
  {NoStop}%
\bibitem [{\citenamefont {Peng}\ \emph {et~al.}(2014)\citenamefont {Peng},
  \citenamefont {{\"O}zdemir}, \citenamefont {Lei}, \citenamefont {Monifi},
  \citenamefont {Gianfreda}, \citenamefont {Long}, \citenamefont {Fan},
  \citenamefont {Nori}, \citenamefont {Bender},\ and\ \citenamefont
  {Yang}}]{RN19}%
  \BibitemOpen
  \bibfield  {author} {\bibinfo {author} {\bibfnamefont {B.}~\bibnamefont
  {Peng}}, \bibinfo {author} {\bibfnamefont {S.~K.}\ \bibnamefont
  {{\"O}zdemir}}, \bibinfo {author} {\bibfnamefont {F.}~\bibnamefont {Lei}},
  \bibinfo {author} {\bibfnamefont {F.}~\bibnamefont {Monifi}}, \bibinfo
  {author} {\bibfnamefont {M.}~\bibnamefont {Gianfreda}}, \bibinfo {author}
  {\bibfnamefont {G.~L.}\ \bibnamefont {Long}}, \bibinfo {author}
  {\bibfnamefont {S.}~\bibnamefont {Fan}}, \bibinfo {author} {\bibfnamefont
  {F.}~\bibnamefont {Nori}}, \bibinfo {author} {\bibfnamefont {C.~M.}\
  \bibnamefont {Bender}},\ and\ \bibinfo {author} {\bibfnamefont
  {L.}~\bibnamefont {Yang}},\ }\bibfield  {title} {\bibinfo {title}
  {Parity–time-symmetric whispering-gallery microcavities},\ }\href
  {https://doi.org/10.1038/nphys2927} {\bibfield  {journal} {\bibinfo
  {journal} {Nature Physics}\ }\textbf {\bibinfo {volume} {10}},\ \bibinfo
  {pages} {394} (\bibinfo {year} {2014})}\BibitemShut {NoStop}%
\bibitem [{\citenamefont {Parto}\ \emph {et~al.}(2020)\citenamefont {Parto},
  \citenamefont {Liu}, \citenamefont {Bahari}, \citenamefont {Khajavikhan},\
  and\ \citenamefont {Christodoulides}}]{RN20}%
  \BibitemOpen
  \bibfield  {author} {\bibinfo {author} {\bibfnamefont {M.}~\bibnamefont
  {Parto}}, \bibinfo {author} {\bibfnamefont {Y.~G.~N.}\ \bibnamefont {Liu}},
  \bibinfo {author} {\bibfnamefont {B.}~\bibnamefont {Bahari}}, \bibinfo
  {author} {\bibfnamefont {M.}~\bibnamefont {Khajavikhan}},\ and\ \bibinfo
  {author} {\bibfnamefont {D.~N.}\ \bibnamefont {Christodoulides}},\ }\bibfield
   {title} {\bibinfo {title} {Non-hermitian and topological photonics: optics
  at an exceptional point},\ }\href
  {https://doi.org/https://doi.org/10.1515/nanoph-2020-0434} {\bibfield
  {journal} {\bibinfo  {journal} {Nanophotonics}\ ,\ \bibinfo {pages} {403}}
  (\bibinfo {year} {2020})}\BibitemShut {NoStop}%
\bibitem [{\citenamefont {Wang}\ \emph {et~al.}(2021)\citenamefont {Wang},
  \citenamefont {Sweeney}, \citenamefont {Stone},\ and\ \citenamefont
  {Yang}}]{RN21}%
  \BibitemOpen
  \bibfield  {author} {\bibinfo {author} {\bibfnamefont {C.}~\bibnamefont
  {Wang}}, \bibinfo {author} {\bibfnamefont {W.~R.}\ \bibnamefont {Sweeney}},
  \bibinfo {author} {\bibfnamefont {A.~D.}\ \bibnamefont {Stone}},\ and\
  \bibinfo {author} {\bibfnamefont {L.}~\bibnamefont {Yang}},\ }\bibfield
  {title} {\bibinfo {title} {Coherent perfect absorption at an exceptional
  point},\ }\href {https://doi.org/doi:10.1126/science.abj1028} {\bibfield
  {journal} {\bibinfo  {journal} {Science}\ }\textbf {\bibinfo {volume}
  {373}},\ \bibinfo {pages} {1261} (\bibinfo {year} {2021})}\BibitemShut
  {NoStop}%
\bibitem [{\citenamefont {Jiang}\ \emph {et~al.}(2022)\citenamefont {Jiang},
  \citenamefont {Zhang}, \citenamefont {Lu}, \citenamefont {Ye}, \citenamefont
  {Lin}, \citenamefont {Tang}, \citenamefont {Xue}, \citenamefont {Li},
  \citenamefont {Xu},\ and\ \citenamefont {Gong}}]{RN22}%
  \BibitemOpen
  \bibfield  {author} {\bibinfo {author} {\bibfnamefont {H.}~\bibnamefont
  {Jiang}}, \bibinfo {author} {\bibfnamefont {W.}~\bibnamefont {Zhang}},
  \bibinfo {author} {\bibfnamefont {G.}~\bibnamefont {Lu}}, \bibinfo {author}
  {\bibfnamefont {L.}~\bibnamefont {Ye}}, \bibinfo {author} {\bibfnamefont
  {H.}~\bibnamefont {Lin}}, \bibinfo {author} {\bibfnamefont {J.}~\bibnamefont
  {Tang}}, \bibinfo {author} {\bibfnamefont {Z.}~\bibnamefont {Xue}}, \bibinfo
  {author} {\bibfnamefont {Z.}~\bibnamefont {Li}}, \bibinfo {author}
  {\bibfnamefont {H.}~\bibnamefont {Xu}},\ and\ \bibinfo {author}
  {\bibfnamefont {Q.}~\bibnamefont {Gong}},\ }\bibfield  {title} {\bibinfo
  {title} {Exceptional points and enhanced nanoscale sensing with a
  plasmon-exciton hybrid system},\ }\href {https://doi.org/10.1364/PRJ.445855}
  {\bibfield  {journal} {\bibinfo  {journal} {Photonics Research}\ }\textbf
  {\bibinfo {volume} {10}},\ \bibinfo {pages} {557} (\bibinfo {year}
  {2022})}\BibitemShut {NoStop}%
\bibitem [{\citenamefont {Xie}\ \emph {et~al.}(2021)\citenamefont {Xie},
  \citenamefont {Qin}, \citenamefont {Zhang}, \citenamefont {Wang},
  \citenamefont {Li}, \citenamefont {Ruan},\ and\ \citenamefont {Long}}]{xrr}%
  \BibitemOpen
  \bibfield  {author} {\bibinfo {author} {\bibfnamefont {R.-R.}\ \bibnamefont
  {Xie}}, \bibinfo {author} {\bibfnamefont {G.-Q.}\ \bibnamefont {Qin}},
  \bibinfo {author} {\bibfnamefont {H.}~\bibnamefont {Zhang}}, \bibinfo
  {author} {\bibfnamefont {M.}~\bibnamefont {Wang}}, \bibinfo {author}
  {\bibfnamefont {G.-Q.}\ \bibnamefont {Li}}, \bibinfo {author} {\bibfnamefont
  {D.}~\bibnamefont {Ruan}},\ and\ \bibinfo {author} {\bibfnamefont {G.-L.}\
  \bibnamefont {Long}},\ }\bibfield  {title} {\bibinfo {title}
  {Phase-controlled dual-wavelength resonance in a self-coupling
  whispering-gallery-mode microcavity},\ }\href
  {https://doi.org/10.1364/OL.416973} {\bibfield  {journal} {\bibinfo
  {journal} {Optics Letters}\ }\textbf {\bibinfo {volume} {46}},\ \bibinfo
  {pages} {773} (\bibinfo {year} {2021})}\BibitemShut {NoStop}%
\bibitem [{\citenamefont {Tang}\ \emph {et~al.}(2020)\citenamefont {Tang},
  \citenamefont {Jiang}, \citenamefont {Ding}, \citenamefont {Xiao},
  \citenamefont {Zhang}, \citenamefont {Chan},\ and\ \citenamefont
  {Ma}}]{RN27}%
  \BibitemOpen
  \bibfield  {author} {\bibinfo {author} {\bibfnamefont {W.}~\bibnamefont
  {Tang}}, \bibinfo {author} {\bibfnamefont {X.}~\bibnamefont {Jiang}},
  \bibinfo {author} {\bibfnamefont {K.}~\bibnamefont {Ding}}, \bibinfo {author}
  {\bibfnamefont {Y.-X.}\ \bibnamefont {Xiao}}, \bibinfo {author}
  {\bibfnamefont {Z.-Q.}\ \bibnamefont {Zhang}}, \bibinfo {author}
  {\bibfnamefont {C.~T.}\ \bibnamefont {Chan}},\ and\ \bibinfo {author}
  {\bibfnamefont {G.}~\bibnamefont {Ma}},\ }\bibfield  {title} {\bibinfo
  {title} {Exceptional nexus with a hybrid topological invariant},\ }\href
  {https://doi.org/10.1126/science.abd8872 %J Science} {\bibfield  {journal}
  {\bibinfo  {journal} {Science}\ }\textbf {\bibinfo {volume} {370}},\ \bibinfo
  {pages} {1077} (\bibinfo {year} {2020})}\BibitemShut {NoStop}%
\bibitem [{\citenamefont {Jahromi}\ \emph {et~al.}(2017)\citenamefont
  {Jahromi}, \citenamefont {Hassan}, \citenamefont {Christodoulides},\ and\
  \citenamefont {Abouraddy}}]{RN28}%
  \BibitemOpen
  \bibfield  {author} {\bibinfo {author} {\bibfnamefont {A.~K.}\ \bibnamefont
  {Jahromi}}, \bibinfo {author} {\bibfnamefont {A.~U.}\ \bibnamefont {Hassan}},
  \bibinfo {author} {\bibfnamefont {D.~N.}\ \bibnamefont {Christodoulides}},\
  and\ \bibinfo {author} {\bibfnamefont {A.~F.}\ \bibnamefont {Abouraddy}},\
  }\bibfield  {title} {\bibinfo {title} {Statistical parity-time-symmetric
  lasing in an optical fibre network},\ }\href
  {https://doi.org/10.1038/s41467-017-00958-x} {\bibfield  {journal} {\bibinfo
  {journal} {Nature Communications}\ }\textbf {\bibinfo {volume} {8}},\
  \bibinfo {pages} {1359} (\bibinfo {year} {2017})}\BibitemShut {NoStop}%
\bibitem [{\citenamefont {Assawaworrarit}\ \emph {et~al.}(2017)\citenamefont
  {Assawaworrarit}, \citenamefont {Yu},\ and\ \citenamefont {Fan}}]{RN29}%
  \BibitemOpen
  \bibfield  {author} {\bibinfo {author} {\bibfnamefont {S.}~\bibnamefont
  {Assawaworrarit}}, \bibinfo {author} {\bibfnamefont {X.}~\bibnamefont {Yu}},\
  and\ \bibinfo {author} {\bibfnamefont {S.}~\bibnamefont {Fan}},\ }\bibfield
  {title} {\bibinfo {title} {Robust wireless power transfer using a nonlinear
  parity-time-symmetric circuit},\ }\href {https://doi.org/10.1038/nature22404}
  {\bibfield  {journal} {\bibinfo  {journal} {Nature}\ }\textbf {\bibinfo
  {volume} {546}},\ \bibinfo {pages} {387} (\bibinfo {year}
  {2017})}\BibitemShut {NoStop}%
\bibitem [{\citenamefont {Li}\ \emph {et~al.}(2019)\citenamefont {Li},
  \citenamefont {Peng}, \citenamefont {Han}, \citenamefont {Miri},
  \citenamefont {Li}, \citenamefont {Xiao}, \citenamefont {Zhu}, \citenamefont
  {Zhao}, \citenamefont {Alù}, \citenamefont {Fan},\ and\ \citenamefont
  {Qiu}}]{RN30}%
  \BibitemOpen
  \bibfield  {author} {\bibinfo {author} {\bibfnamefont {Y.}~\bibnamefont
  {Li}}, \bibinfo {author} {\bibfnamefont {Y.-G.}\ \bibnamefont {Peng}},
  \bibinfo {author} {\bibfnamefont {L.}~\bibnamefont {Han}}, \bibinfo {author}
  {\bibfnamefont {M.-A.}\ \bibnamefont {Miri}}, \bibinfo {author}
  {\bibfnamefont {W.}~\bibnamefont {Li}}, \bibinfo {author} {\bibfnamefont
  {M.}~\bibnamefont {Xiao}}, \bibinfo {author} {\bibfnamefont {X.-F.}\
  \bibnamefont {Zhu}}, \bibinfo {author} {\bibfnamefont {J.}~\bibnamefont
  {Zhao}}, \bibinfo {author} {\bibfnamefont {A.}~\bibnamefont {Alù}}, \bibinfo
  {author} {\bibfnamefont {S.}~\bibnamefont {Fan}},\ and\ \bibinfo {author}
  {\bibfnamefont {C.-W.}\ \bibnamefont {Qiu}},\ }\bibfield  {title} {\bibinfo
  {title} {Anti-parity-time symmetry in diffusive systems},\ }\href
  {https://doi.org/doi:10.1126/science.aaw6259} {\bibfield  {journal} {\bibinfo
   {journal} {Science}\ }\textbf {\bibinfo {volume} {364}},\ \bibinfo {pages}
  {170} (\bibinfo {year} {2019})}\BibitemShut {NoStop}%
\bibitem [{\citenamefont {Zhang}\ \emph
  {et~al.}(2019{\natexlab{a}})\citenamefont {Zhang}, \citenamefont {Ding},
  \citenamefont {Zhou}, \citenamefont {Xu},\ and\ \citenamefont {Jin}}]{RN24}%
  \BibitemOpen
  \bibfield  {author} {\bibinfo {author} {\bibfnamefont {X.}~\bibnamefont
  {Zhang}}, \bibinfo {author} {\bibfnamefont {K.}~\bibnamefont {Ding}},
  \bibinfo {author} {\bibfnamefont {X.}~\bibnamefont {Zhou}}, \bibinfo {author}
  {\bibfnamefont {J.}~\bibnamefont {Xu}},\ and\ \bibinfo {author}
  {\bibfnamefont {D.}~\bibnamefont {Jin}},\ }\bibfield  {title} {\bibinfo
  {title} {Experimental observation of an exceptional surface in synthetic
  dimensions with magnon polaritons},\ }\href
  {https://doi.org/10.1103/PhysRevLett.123.237202} {\bibfield  {journal}
  {\bibinfo  {journal} {Physical Review Letters}\ }\textbf {\bibinfo {volume}
  {123}},\ \bibinfo {pages} {237202} (\bibinfo {year}
  {2019}{\natexlab{a}})}\BibitemShut {NoStop}%
\bibitem [{\citenamefont {Zhang}\ \emph {et~al.}(2017)\citenamefont {Zhang},
  \citenamefont {Luo}, \citenamefont {Wang}, \citenamefont {Li},\ and\
  \citenamefont {You}}]{RN25}%
  \BibitemOpen
  \bibfield  {author} {\bibinfo {author} {\bibfnamefont {D.}~\bibnamefont
  {Zhang}}, \bibinfo {author} {\bibfnamefont {X.-Q.}\ \bibnamefont {Luo}},
  \bibinfo {author} {\bibfnamefont {Y.-P.}\ \bibnamefont {Wang}}, \bibinfo
  {author} {\bibfnamefont {T.-F.}\ \bibnamefont {Li}},\ and\ \bibinfo {author}
  {\bibfnamefont {J.~Q.}\ \bibnamefont {You}},\ }\bibfield  {title} {\bibinfo
  {title} {Observation of the exceptional point in cavity magnon-polaritons},\
  }\href {https://doi.org/10.1038/s41467-017-01634-w} {\bibfield  {journal}
  {\bibinfo  {journal} {Nature Communications}\ }\textbf {\bibinfo {volume}
  {8}},\ \bibinfo {pages} {1368} (\bibinfo {year} {2017})}\BibitemShut
  {NoStop}%
\bibitem [{\citenamefont {Zhao}\ \emph {et~al.}(2020)\citenamefont {Zhao},
  \citenamefont {Liu}, \citenamefont {Wu}, \citenamefont {Duan}, \citenamefont
  {Liu},\ and\ \citenamefont {Du}}]{RN26}%
  \BibitemOpen
  \bibfield  {author} {\bibinfo {author} {\bibfnamefont {J.}~\bibnamefont
  {Zhao}}, \bibinfo {author} {\bibfnamefont {Y.}~\bibnamefont {Liu}}, \bibinfo
  {author} {\bibfnamefont {L.}~\bibnamefont {Wu}}, \bibinfo {author}
  {\bibfnamefont {C.-K.}\ \bibnamefont {Duan}}, \bibinfo {author}
  {\bibfnamefont {Y.-x.}\ \bibnamefont {Liu}},\ and\ \bibinfo {author}
  {\bibfnamefont {J.}~\bibnamefont {Du}},\ }\bibfield  {title} {\bibinfo
  {title} {Observation of anti-pt-symmetry phase transition in the
  magnon-cavity-magnon coupled system},\ }\href
  {https://doi.org/10.1103/PhysRevApplied.13.014053} {\bibfield  {journal}
  {\bibinfo  {journal} {Physical Review Applied}\ }\textbf {\bibinfo {volume}
  {13}},\ \bibinfo {pages} {014053} (\bibinfo {year} {2020})}\BibitemShut
  {NoStop}%
\bibitem [{\citenamefont {Wu}\ \emph {et~al.}(2019)\citenamefont {Wu},
  \citenamefont {Liu}, \citenamefont {Geng}, \citenamefont {Song},
  \citenamefont {Ye}, \citenamefont {Duan}, \citenamefont {Rong},\ and\
  \citenamefont {Du}}]{RN31}%
  \BibitemOpen
  \bibfield  {author} {\bibinfo {author} {\bibfnamefont {Y.}~\bibnamefont
  {Wu}}, \bibinfo {author} {\bibfnamefont {W.}~\bibnamefont {Liu}}, \bibinfo
  {author} {\bibfnamefont {J.}~\bibnamefont {Geng}}, \bibinfo {author}
  {\bibfnamefont {X.}~\bibnamefont {Song}}, \bibinfo {author} {\bibfnamefont
  {X.}~\bibnamefont {Ye}}, \bibinfo {author} {\bibfnamefont {C.~K.}\
  \bibnamefont {Duan}}, \bibinfo {author} {\bibfnamefont {X.}~\bibnamefont
  {Rong}},\ and\ \bibinfo {author} {\bibfnamefont {J.}~\bibnamefont {Du}},\
  }\bibfield  {title} {\bibinfo {title} {Observation of parity-time symmetry
  breaking in a single-spin system},\ }\href
  {https://doi.org/10.1126/science.aaw8205} {\bibfield  {journal} {\bibinfo
  {journal} {Science}\ }\textbf {\bibinfo {volume} {364}},\ \bibinfo {pages}
  {878} (\bibinfo {year} {2019})}\BibitemShut {NoStop}%
\bibitem [{\citenamefont {Klauck}\ \emph {et~al.}(2019)\citenamefont {Klauck},
  \citenamefont {Teuber}, \citenamefont {Ornigotti}, \citenamefont {Heinrich},
  \citenamefont {Scheel},\ and\ \citenamefont {Szameit}}]{RN32}%
  \BibitemOpen
  \bibfield  {author} {\bibinfo {author} {\bibfnamefont {F.}~\bibnamefont
  {Klauck}}, \bibinfo {author} {\bibfnamefont {L.}~\bibnamefont {Teuber}},
  \bibinfo {author} {\bibfnamefont {M.}~\bibnamefont {Ornigotti}}, \bibinfo
  {author} {\bibfnamefont {M.}~\bibnamefont {Heinrich}}, \bibinfo {author}
  {\bibfnamefont {S.}~\bibnamefont {Scheel}},\ and\ \bibinfo {author}
  {\bibfnamefont {A.}~\bibnamefont {Szameit}},\ }\bibfield  {title} {\bibinfo
  {title} {Observation of pt-symmetric quantum interference},\ }\href
  {https://doi.org/10.1038/s41566-019-0517-0} {\bibfield  {journal} {\bibinfo
  {journal} {Nature Photonics}\ }\textbf {\bibinfo {volume} {13}},\ \bibinfo
  {pages} {883} (\bibinfo {year} {2019})}\BibitemShut {NoStop}%
\bibitem [{\citenamefont {Huang}\ \emph {et~al.}(2020)\citenamefont {Huang},
  \citenamefont {{\"O}zdemir}, \citenamefont {Liao}, \citenamefont {Minganti},
  \citenamefont {Kuang}, \citenamefont {Nori},\ and\ \citenamefont
  {Jing}}]{RN33}%
  \BibitemOpen
  \bibfield  {author} {\bibinfo {author} {\bibfnamefont {R.}~\bibnamefont
  {Huang}}, \bibinfo {author} {\bibfnamefont {S.~K.}\ \bibnamefont
  {{\"O}zdemir}}, \bibinfo {author} {\bibfnamefont {J.-Q.}\ \bibnamefont
  {Liao}}, \bibinfo {author} {\bibfnamefont {F.}~\bibnamefont {Minganti}},
  \bibinfo {author} {\bibfnamefont {L.-M.}\ \bibnamefont {Kuang}}, \bibinfo
  {author} {\bibfnamefont {F.}~\bibnamefont {Nori}},\ and\ \bibinfo {author}
  {\bibfnamefont {H.}~\bibnamefont {Jing}},\ }\bibfield  {title} {\bibinfo
  {title} {Exceptional photon blockade},\ }\href
  {https://doi.org/10.48550/arXiv.2001.09492} {\bibfield  {journal} {\bibinfo
  {journal} {arXiv}\ ,\ \bibinfo {pages} {2001.09492}} (\bibinfo {year}
  {2020})}\BibitemShut {NoStop}%
\bibitem [{\citenamefont {Naghiloo}\ \emph {et~al.}(2019)\citenamefont
  {Naghiloo}, \citenamefont {Abbasi}, \citenamefont {Joglekar},\ and\
  \citenamefont {Murch}}]{RN34}%
  \BibitemOpen
  \bibfield  {author} {\bibinfo {author} {\bibfnamefont {M.}~\bibnamefont
  {Naghiloo}}, \bibinfo {author} {\bibfnamefont {M.}~\bibnamefont {Abbasi}},
  \bibinfo {author} {\bibfnamefont {Y.~N.}\ \bibnamefont {Joglekar}},\ and\
  \bibinfo {author} {\bibfnamefont {K.~W.}\ \bibnamefont {Murch}},\ }\bibfield
  {title} {\bibinfo {title} {Quantum state tomography across the exceptional
  point in a single dissipative qubit},\ }\href
  {https://doi.org/10.1038/s41567-019-0652-z} {\bibfield  {journal} {\bibinfo
  {journal} {Nature Physics}\ }\textbf {\bibinfo {volume} {15}},\ \bibinfo
  {pages} {1232} (\bibinfo {year} {2019})}\BibitemShut {NoStop}%
\bibitem [{\citenamefont {Choi}\ \emph {et~al.}(2010)\citenamefont {Choi},
  \citenamefont {Kang}, \citenamefont {Lim}, \citenamefont {Kim}, \citenamefont
  {Kim}, \citenamefont {Lee},\ and\ \citenamefont {An}}]{RN35}%
  \BibitemOpen
  \bibfield  {author} {\bibinfo {author} {\bibfnamefont {Y.}~\bibnamefont
  {Choi}}, \bibinfo {author} {\bibfnamefont {S.}~\bibnamefont {Kang}}, \bibinfo
  {author} {\bibfnamefont {S.}~\bibnamefont {Lim}}, \bibinfo {author}
  {\bibfnamefont {W.}~\bibnamefont {Kim}}, \bibinfo {author} {\bibfnamefont
  {J.~R.}\ \bibnamefont {Kim}}, \bibinfo {author} {\bibfnamefont {J.~H.}\
  \bibnamefont {Lee}},\ and\ \bibinfo {author} {\bibfnamefont {K.}~\bibnamefont
  {An}},\ }\bibfield  {title} {\bibinfo {title} {Quasieigenstate coalescence in
  an atom-cavity quantum composite},\ }\href
  {https://doi.org/10.1103/PhysRevLett.104.153601} {\bibfield  {journal}
  {\bibinfo  {journal} {Physical Review Letters}\ }\textbf {\bibinfo {volume}
  {104}},\ \bibinfo {pages} {153601} (\bibinfo {year} {2010})}\BibitemShut
  {NoStop}%
\bibitem [{\citenamefont {L{\"u}}\ \emph {et~al.}(2017)\citenamefont {L{\"u}},
  \citenamefont {{\"O}zdemir}, \citenamefont {Kuang}, \citenamefont {Nori},\
  and\ \citenamefont {Jing}}]{RN36}%
  \BibitemOpen
  \bibfield  {author} {\bibinfo {author} {\bibfnamefont {H.}~\bibnamefont
  {L{\"u}}}, \bibinfo {author} {\bibfnamefont {S.}~\bibnamefont {{\"O}zdemir}},
  \bibinfo {author} {\bibfnamefont {L.~M.}\ \bibnamefont {Kuang}}, \bibinfo
  {author} {\bibfnamefont {F.}~\bibnamefont {Nori}},\ and\ \bibinfo {author}
  {\bibfnamefont {H.}~\bibnamefont {Jing}},\ }\bibfield  {title} {\bibinfo
  {title} {Exceptional points in random-defect phonon lasers},\ }\href
  {https://doi.org/10.1103/PhysRevApplied.8.044020} {\bibfield  {journal}
  {\bibinfo  {journal} {Physical Review Applied}\ }\textbf {\bibinfo {volume}
  {8}},\ \bibinfo {pages} {044020} (\bibinfo {year} {2017})}\BibitemShut
  {NoStop}%
\bibitem [{\citenamefont {Zhang}\ \emph {et~al.}(2018)\citenamefont {Zhang},
  \citenamefont {Peng}, \citenamefont {{\"O}zdemir}, \citenamefont {Pichler},
  \citenamefont {Krimer}, \citenamefont {Zhao}, \citenamefont {Nori},
  \citenamefont {Liu}, \citenamefont {Rotter},\ and\ \citenamefont
  {Yang}}]{RN37}%
  \BibitemOpen
  \bibfield  {author} {\bibinfo {author} {\bibfnamefont {J.}~\bibnamefont
  {Zhang}}, \bibinfo {author} {\bibfnamefont {B.}~\bibnamefont {Peng}},
  \bibinfo {author} {\bibfnamefont {S.~K.}\ \bibnamefont {{\"O}zdemir}},
  \bibinfo {author} {\bibfnamefont {K.}~\bibnamefont {Pichler}}, \bibinfo
  {author} {\bibfnamefont {D.~O.}\ \bibnamefont {Krimer}}, \bibinfo {author}
  {\bibfnamefont {G.}~\bibnamefont {Zhao}}, \bibinfo {author} {\bibfnamefont
  {F.}~\bibnamefont {Nori}}, \bibinfo {author} {\bibfnamefont {Y.-x.}\
  \bibnamefont {Liu}}, \bibinfo {author} {\bibfnamefont {S.}~\bibnamefont
  {Rotter}},\ and\ \bibinfo {author} {\bibfnamefont {L.}~\bibnamefont {Yang}},\
  }\bibfield  {title} {\bibinfo {title} {A phonon laser operating at an
  exceptional point},\ }\href {https://doi.org/10.1038/s41566-018-0213-5}
  {\bibfield  {journal} {\bibinfo  {journal} {Nature Photonics}\ }\textbf
  {\bibinfo {volume} {12}},\ \bibinfo {pages} {479} (\bibinfo {year}
  {2018})}\BibitemShut {NoStop}%
\bibitem [{\citenamefont {Garmon}\ \emph {et~al.}(2021)\citenamefont {Garmon},
  \citenamefont {Ordonez},\ and\ \citenamefont {Hatano}}]{RN38}%
  \BibitemOpen
  \bibfield  {author} {\bibinfo {author} {\bibfnamefont {S.}~\bibnamefont
  {Garmon}}, \bibinfo {author} {\bibfnamefont {G.}~\bibnamefont {Ordonez}},\
  and\ \bibinfo {author} {\bibfnamefont {N.}~\bibnamefont {Hatano}},\
  }\bibfield  {title} {\bibinfo {title} {Anomalous-order exceptional point and
  non-markovian purcell effect at threshold in one-dimensional continuum
  systems},\ }\href {https://doi.org/10.1103/PhysRevResearch.3.033029}
  {\bibfield  {journal} {\bibinfo  {journal} {Physical Review Research}\
  }\textbf {\bibinfo {volume} {3}},\ \bibinfo {pages} {033029} (\bibinfo {year}
  {2021})}\BibitemShut {NoStop}%
\bibitem [{\citenamefont {Purkayastha}\ \emph {et~al.}(2020)\citenamefont
  {Purkayastha}, \citenamefont {Kulkarni},\ and\ \citenamefont
  {Joglekar}}]{RN39}%
  \BibitemOpen
  \bibfield  {author} {\bibinfo {author} {\bibfnamefont {A.}~\bibnamefont
  {Purkayastha}}, \bibinfo {author} {\bibfnamefont {M.}~\bibnamefont
  {Kulkarni}},\ and\ \bibinfo {author} {\bibfnamefont {Y.~N.}\ \bibnamefont
  {Joglekar}},\ }\bibfield  {title} {\bibinfo {title} {Emergent $\mathcal{PT}$
  symmetry in a double-quantum-dot circuit qed setup},\ }\href
  {https://doi.org/10.1103/PhysRevResearch.2.043075} {\bibfield  {journal}
  {\bibinfo  {journal} {Physical Review Research}\ }\textbf {\bibinfo {volume}
  {2}},\ \bibinfo {pages} {043075} (\bibinfo {year} {2020})}\BibitemShut
  {NoStop}%
\bibitem [{\citenamefont {Pick}\ \emph {et~al.}(2017)\citenamefont {Pick},
  \citenamefont {Zhen}, \citenamefont {Miller}, \citenamefont {Hsu},
  \citenamefont {Hernandez}, \citenamefont {Rodriguez}, \citenamefont
  {Soljacic},\ and\ \citenamefont {Johnson}}]{RN40}%
  \BibitemOpen
  \bibfield  {author} {\bibinfo {author} {\bibfnamefont {A.}~\bibnamefont
  {Pick}}, \bibinfo {author} {\bibfnamefont {B.}~\bibnamefont {Zhen}}, \bibinfo
  {author} {\bibfnamefont {O.~D.}\ \bibnamefont {Miller}}, \bibinfo {author}
  {\bibfnamefont {C.~W.}\ \bibnamefont {Hsu}}, \bibinfo {author} {\bibfnamefont
  {F.}~\bibnamefont {Hernandez}}, \bibinfo {author} {\bibfnamefont {A.~W.}\
  \bibnamefont {Rodriguez}}, \bibinfo {author} {\bibfnamefont {M.}~\bibnamefont
  {Soljacic}},\ and\ \bibinfo {author} {\bibfnamefont {S.~G.}\ \bibnamefont
  {Johnson}},\ }\bibfield  {title} {\bibinfo {title} {General theory of
  spontaneous emission near exceptional points},\ }\href
  {https://doi.org/10.1364/OE.25.012325} {\bibfield  {journal} {\bibinfo
  {journal} {Optics Express}\ }\textbf {\bibinfo {volume} {25}},\ \bibinfo
  {pages} {12325} (\bibinfo {year} {2017})}\BibitemShut {NoStop}%
\bibitem [{\citenamefont {Khanbekyan}\ and\ \citenamefont
  {Wiersig}(2020)}]{RN41}%
  \BibitemOpen
  \bibfield  {author} {\bibinfo {author} {\bibfnamefont {M.}~\bibnamefont
  {Khanbekyan}}\ and\ \bibinfo {author} {\bibfnamefont {J.}~\bibnamefont
  {Wiersig}},\ }\bibfield  {title} {\bibinfo {title} {Decay suppression of
  spontaneous emission of a single emitter in a high-$q$ cavity at exceptional
  points},\ }\href {https://doi.org/10.1103/PhysRevResearch.2.023375}
  {\bibfield  {journal} {\bibinfo  {journal} {Physical Review Research}\
  }\textbf {\bibinfo {volume} {2}},\ \bibinfo {pages} {023375} (\bibinfo {year}
  {2020})}\BibitemShut {NoStop}%
\bibitem [{\citenamefont {Zhong}\ \emph {et~al.}(2021)\citenamefont {Zhong},
  \citenamefont {Hashemi}, \citenamefont {{\"O}zdemir},\ and\ \citenamefont
  {El-Ganainy}}]{RN42}%
  \BibitemOpen
  \bibfield  {author} {\bibinfo {author} {\bibfnamefont {Q.}~\bibnamefont
  {Zhong}}, \bibinfo {author} {\bibfnamefont {A.}~\bibnamefont {Hashemi}},
  \bibinfo {author} {\bibfnamefont {S.~K.}\ \bibnamefont {{\"O}zdemir}},\ and\
  \bibinfo {author} {\bibfnamefont {R.}~\bibnamefont {El-Ganainy}},\ }\bibfield
   {title} {\bibinfo {title} {Control of spontaneous emission dynamics in
  microcavities with chiral exceptional surfaces},\ }\href
  {https://doi.org/10.1103/PhysRevResearch.3.013220} {\bibfield  {journal}
  {\bibinfo  {journal} {Physical Review Research}\ }\textbf {\bibinfo {volume}
  {3}},\ \bibinfo {pages} {013220} (\bibinfo {year} {2021})}\BibitemShut
  {NoStop}%
\bibitem [{\citenamefont {Zhang}\ \emph
  {et~al.}(2019{\natexlab{b}})\citenamefont {Zhang}, \citenamefont {Sweeney},
  \citenamefont {Hsu}, \citenamefont {Yang}, \citenamefont {Stone},\ and\
  \citenamefont {Jiang}}]{RN43}%
  \BibitemOpen
  \bibfield  {author} {\bibinfo {author} {\bibfnamefont {M.}~\bibnamefont
  {Zhang}}, \bibinfo {author} {\bibfnamefont {W.}~\bibnamefont {Sweeney}},
  \bibinfo {author} {\bibfnamefont {C.~W.}\ \bibnamefont {Hsu}}, \bibinfo
  {author} {\bibfnamefont {L.}~\bibnamefont {Yang}}, \bibinfo {author}
  {\bibfnamefont {A.}~\bibnamefont {Stone}},\ and\ \bibinfo {author}
  {\bibfnamefont {L.}~\bibnamefont {Jiang}},\ }\bibfield  {title} {\bibinfo
  {title} {Quantum noise theory of exceptional point amplifying sensors},\
  }\href {https://doi.org/10.1103/PhysRevLett.123.180501} {\bibfield  {journal}
  {\bibinfo  {journal} {Physical Review Letters}\ }\textbf {\bibinfo {volume}
  {123}},\ \bibinfo {pages} {180501} (\bibinfo {year}
  {2019}{\natexlab{b}})}\BibitemShut {NoStop}%
\bibitem [{\citenamefont {Chu}\ \emph {et~al.}(2020)\citenamefont {Chu},
  \citenamefont {Liu}, \citenamefont {Liu},\ and\ \citenamefont {Cai}}]{RN44}%
  \BibitemOpen
  \bibfield  {author} {\bibinfo {author} {\bibfnamefont {Y.}~\bibnamefont
  {Chu}}, \bibinfo {author} {\bibfnamefont {Y.}~\bibnamefont {Liu}}, \bibinfo
  {author} {\bibfnamefont {H.}~\bibnamefont {Liu}},\ and\ \bibinfo {author}
  {\bibfnamefont {J.}~\bibnamefont {Cai}},\ }\bibfield  {title} {\bibinfo
  {title} {Quantum sensing with a single-qubit pseudo-hermitian system},\
  }\href {https://doi.org/10.1103/PhysRevLett.124.020501} {\bibfield  {journal}
  {\bibinfo  {journal} {Physical Review Letters}\ }\textbf {\bibinfo {volume}
  {124}},\ \bibinfo {pages} {020501} (\bibinfo {year} {2020})}\BibitemShut
  {NoStop}%
\bibitem [{\citenamefont {Lin}\ \emph {et~al.}(2016)\citenamefont {Lin},
  \citenamefont {Pick}, \citenamefont {Lončar},\ and\ \citenamefont
  {Rodriguez}}]{RN45}%
  \BibitemOpen
  \bibfield  {author} {\bibinfo {author} {\bibfnamefont {Z.}~\bibnamefont
  {Lin}}, \bibinfo {author} {\bibfnamefont {A.}~\bibnamefont {Pick}}, \bibinfo
  {author} {\bibfnamefont {M.}~\bibnamefont {Lončar}},\ and\ \bibinfo {author}
  {\bibfnamefont {A.~W.}\ \bibnamefont {Rodriguez}},\ }\bibfield  {title}
  {\bibinfo {title} {Enhanced spontaneous emission at third-order dirac
  exceptional points in inverse-designed photonic crystals},\ }\href
  {https://doi.org/10.1103/PhysRevLett.117.107402} {\bibfield  {journal}
  {\bibinfo  {journal} {Physical Review Letters}\ }\textbf {\bibinfo {volume}
  {117}},\ \bibinfo {pages} {107402} (\bibinfo {year} {2016})}\BibitemShut
  {NoStop}%
\bibitem [{\citenamefont {Lu}\ \emph {et~al.}(2018)\citenamefont {Lu},
  \citenamefont {Peng}, \citenamefont {Cao}, \citenamefont {Xu}, \citenamefont
  {Wiersig}, \citenamefont {Gong},\ and\ \citenamefont {Xiao}}]{RN46}%
  \BibitemOpen
  \bibfield  {author} {\bibinfo {author} {\bibfnamefont {Y.-K.}\ \bibnamefont
  {Lu}}, \bibinfo {author} {\bibfnamefont {P.}~\bibnamefont {Peng}}, \bibinfo
  {author} {\bibfnamefont {Q.-T.}\ \bibnamefont {Cao}}, \bibinfo {author}
  {\bibfnamefont {D.}~\bibnamefont {Xu}}, \bibinfo {author} {\bibfnamefont
  {J.}~\bibnamefont {Wiersig}}, \bibinfo {author} {\bibfnamefont
  {Q.}~\bibnamefont {Gong}},\ and\ \bibinfo {author} {\bibfnamefont {Y.-F.}\
  \bibnamefont {Xiao}},\ }\bibfield  {title} {\bibinfo {title} {Spontaneous
  t-symmetry breaking and exceptional points in cavity quantum electrodynamics
  systems},\ }\href {https://doi.org/10.1016/j.scib.2018.07.020} {\bibfield
  {journal} {\bibinfo  {journal} {Science Bulletin}\ }\textbf {\bibinfo
  {volume} {63}},\ \bibinfo {pages} {1096} (\bibinfo {year}
  {2018})}\BibitemShut {NoStop}%
\bibitem [{\citenamefont {Ren}\ \emph {et~al.}(2022)\citenamefont {Ren},
  \citenamefont {Franke},\ and\ \citenamefont {Hughes}}]{RJJ}%
  \BibitemOpen
  \bibfield  {author} {\bibinfo {author} {\bibfnamefont {J.}~\bibnamefont
  {Ren}}, \bibinfo {author} {\bibfnamefont {S.}~\bibnamefont {Franke}},\ and\
  \bibinfo {author} {\bibfnamefont {S.}~\bibnamefont {Hughes}},\ }\bibfield
  {title} {\bibinfo {title} {Quasinormal mode theory of chiral power flow from
  linearly polarized dipole emitters coupled to index-modulated microring
  resonators close to an exceptional point},\ }\bibfield  {journal} {\bibinfo
  {journal} {ACS Photonics}\ }\href
  {https://doi.org/10.1021/acsphotonics.1c01848} {10.1021/acsphotonics.1c01848}
  (\bibinfo {year} {2022})\BibitemShut {NoStop}%
\bibitem [{\citenamefont {Pelton}(2015)}]{RN47}%
  \BibitemOpen
  \bibfield  {author} {\bibinfo {author} {\bibfnamefont {M.}~\bibnamefont
  {Pelton}},\ }\bibfield  {title} {\bibinfo {title} {Modified spontaneous
  emission in nanophotonic structures},\ }\href
  {https://doi.org/10.1038/nphoton.2015.103} {\bibfield  {journal} {\bibinfo
  {journal} {Nature Photonics}\ }\textbf {\bibinfo {volume} {9}},\ \bibinfo
  {pages} {427} (\bibinfo {year} {2015})}\BibitemShut {NoStop}%
\bibitem [{\citenamefont {Novotny}\ and\ \citenamefont {Hecht}(2012)}]{RN48}%
  \BibitemOpen
  \bibfield  {author} {\bibinfo {author} {\bibfnamefont {L.}~\bibnamefont
  {Novotny}}\ and\ \bibinfo {author} {\bibfnamefont {B.}~\bibnamefont
  {Hecht}},\ }\href@noop {} {\emph {\bibinfo {title} {Principles of
  nano-optics}}},\ \bibinfo {edition} {2nd}\ ed.\ (\bibinfo  {publisher}
  {Cambridge University Press},\ \bibinfo {year} {2012})\BibitemShut {NoStop}%
\bibitem [{\citenamefont {Liu}\ \emph {et~al.}(2012)\citenamefont {Liu},
  \citenamefont {Jiang}, \citenamefont {Jin}, \citenamefont {Wang},
  \citenamefont {Gan}, \citenamefont {Jia},\ and\ \citenamefont {Gu}}]{RN50}%
  \BibitemOpen
  \bibfield  {author} {\bibinfo {author} {\bibfnamefont {J.-F.}\ \bibnamefont
  {Liu}}, \bibinfo {author} {\bibfnamefont {H.-X.}\ \bibnamefont {Jiang}},
  \bibinfo {author} {\bibfnamefont {C.-J.}\ \bibnamefont {Jin}}, \bibinfo
  {author} {\bibfnamefont {X.-H.}\ \bibnamefont {Wang}}, \bibinfo {author}
  {\bibfnamefont {Z.-S.}\ \bibnamefont {Gan}}, \bibinfo {author} {\bibfnamefont
  {B.-H.}\ \bibnamefont {Jia}},\ and\ \bibinfo {author} {\bibfnamefont
  {M.}~\bibnamefont {Gu}},\ }\bibfield  {title} {\bibinfo {title}
  {Orientation-dependent local density of states in three-dimensional photonic
  crystals},\ }\href {https://doi.org/10.1103/PhysRevA.85.015802} {\bibfield
  {journal} {\bibinfo  {journal} {Physical Review A}\ }\textbf {\bibinfo
  {volume} {85}},\ \bibinfo {pages} {015802} (\bibinfo {year}
  {2012})}\BibitemShut {NoStop}%
\bibitem [{\citenamefont {Khurgin}\ and\ \citenamefont {Sun}(2014)}]{RN51}%
  \BibitemOpen
  \bibfield  {author} {\bibinfo {author} {\bibfnamefont {J.~B.}\ \bibnamefont
  {Khurgin}}\ and\ \bibinfo {author} {\bibfnamefont {G.}~\bibnamefont {Sun}},\
  }\bibfield  {title} {\bibinfo {title} {Comparative analysis of spasers,
  vertical-cavity surface-emitting lasers and surface-plasmon-emitting
  diodes},\ }\href {https://doi.org/10.1038/nphoton.2014.94} {\bibfield
  {journal} {\bibinfo  {journal} {Nature Photonics}\ }\textbf {\bibinfo
  {volume} {8}},\ \bibinfo {pages} {468} (\bibinfo {year} {2014})}\BibitemShut
  {NoStop}%
\bibitem [{\citenamefont {Andre}\ \emph {et~al.}(2019)\citenamefont {Andre},
  \citenamefont {Protsenko}, \citenamefont {Uskov}, \citenamefont {Mork},\ and\
  \citenamefont {Wubs}}]{RN52}%
  \BibitemOpen
  \bibfield  {author} {\bibinfo {author} {\bibfnamefont {E.~C.}\ \bibnamefont
  {Andre}}, \bibinfo {author} {\bibfnamefont {I.~E.}\ \bibnamefont
  {Protsenko}}, \bibinfo {author} {\bibfnamefont {A.~V.}\ \bibnamefont
  {Uskov}}, \bibinfo {author} {\bibfnamefont {J.}~\bibnamefont {Mork}},\ and\
  \bibinfo {author} {\bibfnamefont {M.}~\bibnamefont {Wubs}},\ }\bibfield
  {title} {\bibinfo {title} {On collective rabi splitting in nanolasers and
  nano-leds},\ }\href {https://doi.org/10.1364/OL.44.001415} {\bibfield
  {journal} {\bibinfo  {journal} {Optics Letters}\ }\textbf {\bibinfo {volume}
  {44}},\ \bibinfo {pages} {1415} (\bibinfo {year} {2019})}\BibitemShut
  {NoStop}%
\bibitem [{\citenamefont {Lenert}\ \emph {et~al.}(2014)\citenamefont {Lenert},
  \citenamefont {Bierman}, \citenamefont {Nam}, \citenamefont {Chan},
  \citenamefont {Celanović}, \citenamefont {Soljačić},\ and\ \citenamefont
  {Wang}}]{RN53}%
  \BibitemOpen
  \bibfield  {author} {\bibinfo {author} {\bibfnamefont {A.}~\bibnamefont
  {Lenert}}, \bibinfo {author} {\bibfnamefont {D.~M.}\ \bibnamefont {Bierman}},
  \bibinfo {author} {\bibfnamefont {Y.}~\bibnamefont {Nam}}, \bibinfo {author}
  {\bibfnamefont {W.~R.}\ \bibnamefont {Chan}}, \bibinfo {author}
  {\bibfnamefont {I.}~\bibnamefont {Celanović}}, \bibinfo {author}
  {\bibfnamefont {M.}~\bibnamefont {Soljačić}},\ and\ \bibinfo {author}
  {\bibfnamefont {E.~N.}\ \bibnamefont {Wang}},\ }\bibfield  {title} {\bibinfo
  {title} {A nanophotonic solar thermophotovoltaic device},\ }\href
  {https://doi.org/10.1038/nnano.2013.286} {\bibfield  {journal} {\bibinfo
  {journal} {Nature Nanotechnology}\ }\textbf {\bibinfo {volume} {9}},\
  \bibinfo {pages} {126} (\bibinfo {year} {2014})}\BibitemShut {NoStop}%
\bibitem [{\citenamefont {Bozhevolnyi}\ and\ \citenamefont
  {Khurgin}(2016)}]{RN54}%
  \BibitemOpen
  \bibfield  {author} {\bibinfo {author} {\bibfnamefont {S.~I.}\ \bibnamefont
  {Bozhevolnyi}}\ and\ \bibinfo {author} {\bibfnamefont {J.~B.}\ \bibnamefont
  {Khurgin}},\ }\bibfield  {title} {\bibinfo {title} {Fundamental limitations
  in spontaneous emission rate of single-photon sources},\ }\href
  {https://doi.org/10.1364/optica.3.001418} {\bibfield  {journal} {\bibinfo
  {journal} {Optica}\ }\textbf {\bibinfo {volume} {3}},\ \bibinfo {pages}
  {1418} (\bibinfo {year} {2016})}\BibitemShut {NoStop}%
\bibitem [{\citenamefont {Liu}\ \emph {et~al.}(2019)\citenamefont {Liu},
  \citenamefont {Su}, \citenamefont {Wei}, \citenamefont {Yao}, \citenamefont
  {Silva}, \citenamefont {Yu}, \citenamefont {Iles-Smith}, \citenamefont
  {Srinivasan}, \citenamefont {Rastelli}, \citenamefont {Li},\ and\
  \citenamefont {Wang}}]{RN55}%
  \BibitemOpen
  \bibfield  {author} {\bibinfo {author} {\bibfnamefont {J.}~\bibnamefont
  {Liu}}, \bibinfo {author} {\bibfnamefont {R.}~\bibnamefont {Su}}, \bibinfo
  {author} {\bibfnamefont {Y.}~\bibnamefont {Wei}}, \bibinfo {author}
  {\bibfnamefont {B.}~\bibnamefont {Yao}}, \bibinfo {author} {\bibfnamefont
  {S.~F. C.~d.}\ \bibnamefont {Silva}}, \bibinfo {author} {\bibfnamefont
  {Y.}~\bibnamefont {Yu}}, \bibinfo {author} {\bibfnamefont {J.}~\bibnamefont
  {Iles-Smith}}, \bibinfo {author} {\bibfnamefont {K.}~\bibnamefont
  {Srinivasan}}, \bibinfo {author} {\bibfnamefont {A.}~\bibnamefont
  {Rastelli}}, \bibinfo {author} {\bibfnamefont {J.}~\bibnamefont {Li}},\ and\
  \bibinfo {author} {\bibfnamefont {X.}~\bibnamefont {Wang}},\ }\bibfield
  {title} {\bibinfo {title} {A solid-state source of strongly entangled photon
  pairs with high brightness and indistinguishability},\ }\href
  {https://doi.org/10.1038/s41565-019-0435-9} {\bibfield  {journal} {\bibinfo
  {journal} {Nature Nanotechnology}\ }\textbf {\bibinfo {volume} {14}},\
  \bibinfo {pages} {586} (\bibinfo {year} {2019})}\BibitemShut {NoStop}%
\bibitem [{\citenamefont {Liu}\ \emph {et~al.}(2018)\citenamefont {Liu},
  \citenamefont {Brash}, \citenamefont {O’Hara}, \citenamefont {Martins},
  \citenamefont {Phillips}, \citenamefont {Coles}, \citenamefont {Royall},
  \citenamefont {Clarke}, \citenamefont {Bentham}, \citenamefont {Prtljaga},
  \citenamefont {Itskevich}, \citenamefont {Wilson}, \citenamefont {Skolnick},\
  and\ \citenamefont {Fox}}]{RN56}%
  \BibitemOpen
  \bibfield  {author} {\bibinfo {author} {\bibfnamefont {F.}~\bibnamefont
  {Liu}}, \bibinfo {author} {\bibfnamefont {A.~J.}\ \bibnamefont {Brash}},
  \bibinfo {author} {\bibfnamefont {J.}~\bibnamefont {O’Hara}}, \bibinfo
  {author} {\bibfnamefont {L.~M. P.~P.}\ \bibnamefont {Martins}}, \bibinfo
  {author} {\bibfnamefont {C.~L.}\ \bibnamefont {Phillips}}, \bibinfo {author}
  {\bibfnamefont {R.~J.}\ \bibnamefont {Coles}}, \bibinfo {author}
  {\bibfnamefont {B.}~\bibnamefont {Royall}}, \bibinfo {author} {\bibfnamefont
  {E.}~\bibnamefont {Clarke}}, \bibinfo {author} {\bibfnamefont
  {C.}~\bibnamefont {Bentham}}, \bibinfo {author} {\bibfnamefont
  {N.}~\bibnamefont {Prtljaga}}, \bibinfo {author} {\bibfnamefont {I.~E.}\
  \bibnamefont {Itskevich}}, \bibinfo {author} {\bibfnamefont {L.~R.}\
  \bibnamefont {Wilson}}, \bibinfo {author} {\bibfnamefont {M.~S.}\
  \bibnamefont {Skolnick}},\ and\ \bibinfo {author} {\bibfnamefont {A.~M.}\
  \bibnamefont {Fox}},\ }\bibfield  {title} {\bibinfo {title} {High purcell
  factor generation of indistinguishable on-chip single photons},\ }\href
  {https://doi.org/10.1038/s41565-018-0188-x} {\bibfield  {journal} {\bibinfo
  {journal} {Nature Nanotechnology}\ }\textbf {\bibinfo {volume} {13}},\
  \bibinfo {pages} {835} (\bibinfo {year} {2018})}\BibitemShut {NoStop}%
\bibitem [{\citenamefont {Söllner}\ \emph {et~al.}(2015)\citenamefont
  {Söllner}, \citenamefont {Mahmoodian}, \citenamefont {Hansen}, \citenamefont
  {Midolo}, \citenamefont {Javadi}, \citenamefont {Kiršanskė}, \citenamefont
  {Pregnolato}, \citenamefont {El-Ella}, \citenamefont {Lee}, \citenamefont
  {Song}, \citenamefont {Stobbe},\ and\ \citenamefont {Lodahl}}]{RN57}%
  \BibitemOpen
  \bibfield  {author} {\bibinfo {author} {\bibfnamefont {I.}~\bibnamefont
  {Söllner}}, \bibinfo {author} {\bibfnamefont {S.}~\bibnamefont
  {Mahmoodian}}, \bibinfo {author} {\bibfnamefont {S.~L.}\ \bibnamefont
  {Hansen}}, \bibinfo {author} {\bibfnamefont {L.}~\bibnamefont {Midolo}},
  \bibinfo {author} {\bibfnamefont {A.}~\bibnamefont {Javadi}}, \bibinfo
  {author} {\bibfnamefont {G.}~\bibnamefont {Kiršanskė}}, \bibinfo {author}
  {\bibfnamefont {T.}~\bibnamefont {Pregnolato}}, \bibinfo {author}
  {\bibfnamefont {H.}~\bibnamefont {El-Ella}}, \bibinfo {author} {\bibfnamefont
  {E.~H.}\ \bibnamefont {Lee}}, \bibinfo {author} {\bibfnamefont {J.~D.}\
  \bibnamefont {Song}}, \bibinfo {author} {\bibfnamefont {S.}~\bibnamefont
  {Stobbe}},\ and\ \bibinfo {author} {\bibfnamefont {P.}~\bibnamefont
  {Lodahl}},\ }\bibfield  {title} {\bibinfo {title} {Deterministic
  photon–emitter coupling in chiral photonic circuits},\ }\href
  {https://doi.org/10.1038/nnano.2015.159} {\bibfield  {journal} {\bibinfo
  {journal} {Nature Nanotechnology}\ }\textbf {\bibinfo {volume} {10}},\
  \bibinfo {pages} {775} (\bibinfo {year} {2015})}\BibitemShut {NoStop}%
\bibitem [{\citenamefont {Lodahl}\ \emph {et~al.}(2017)\citenamefont {Lodahl},
  \citenamefont {Mahmoodian}, \citenamefont {Stobbe}, \citenamefont
  {Rauschenbeutel}, \citenamefont {Schneeweiss}, \citenamefont {Volz},
  \citenamefont {Pichler},\ and\ \citenamefont {Zoller}}]{RN58}%
  \BibitemOpen
  \bibfield  {author} {\bibinfo {author} {\bibfnamefont {P.}~\bibnamefont
  {Lodahl}}, \bibinfo {author} {\bibfnamefont {S.}~\bibnamefont {Mahmoodian}},
  \bibinfo {author} {\bibfnamefont {S.}~\bibnamefont {Stobbe}}, \bibinfo
  {author} {\bibfnamefont {A.}~\bibnamefont {Rauschenbeutel}}, \bibinfo
  {author} {\bibfnamefont {P.}~\bibnamefont {Schneeweiss}}, \bibinfo {author}
  {\bibfnamefont {J.}~\bibnamefont {Volz}}, \bibinfo {author} {\bibfnamefont
  {H.}~\bibnamefont {Pichler}},\ and\ \bibinfo {author} {\bibfnamefont
  {P.}~\bibnamefont {Zoller}},\ }\bibfield  {title} {\bibinfo {title} {Chiral
  quantum optics},\ }\href {https://doi.org/10.1038/nature21037} {\bibfield
  {journal} {\bibinfo  {journal} {Nature}\ }\textbf {\bibinfo {volume} {541}},\
  \bibinfo {pages} {473} (\bibinfo {year} {2017})}\BibitemShut {NoStop}%
\bibitem [{\citenamefont {Heiss}(2015)}]{RN59}%
  \BibitemOpen
  \bibfield  {author} {\bibinfo {author} {\bibfnamefont {W.~D.}\ \bibnamefont
  {Heiss}},\ }\bibfield  {title} {\bibinfo {title} {Green’s functions at
  exceptional points},\ }\href {https://doi.org/10.1007/s10773-014-2428-7}
  {\bibfield  {journal} {\bibinfo  {journal} {International Journal of
  Theoretical Physics}\ }\textbf {\bibinfo {volume} {54}},\ \bibinfo {pages}
  {3954} (\bibinfo {year} {2015})}\BibitemShut {NoStop}%
\bibitem [{\citenamefont {Wiersig}(2020)}]{RN60}%
  \BibitemOpen
  \bibfield  {author} {\bibinfo {author} {\bibfnamefont {J.}~\bibnamefont
  {Wiersig}},\ }\bibfield  {title} {\bibinfo {title} {Review of exceptional
  point-based sensors},\ }\href {https://doi.org/10.1364/PRJ.396115} {\bibfield
   {journal} {\bibinfo  {journal} {Photonics Research}\ }\textbf {\bibinfo
  {volume} {8}},\ \bibinfo {pages} {1457} (\bibinfo {year} {2020})}\BibitemShut
  {NoStop}%
\bibitem [{\citenamefont {Peng}\ \emph {et~al.}(2016)\citenamefont {Peng},
  \citenamefont {{\"O}zdemir}, \citenamefont {Liertzer}, \citenamefont {Chen},
  \citenamefont {Kramer}, \citenamefont {Yilmaz}, \citenamefont {Wiersig},
  \citenamefont {Rotter},\ and\ \citenamefont {Yang}}]{RN61}%
  \BibitemOpen
  \bibfield  {author} {\bibinfo {author} {\bibfnamefont {B.}~\bibnamefont
  {Peng}}, \bibinfo {author} {\bibfnamefont {S.~K.}\ \bibnamefont
  {{\"O}zdemir}}, \bibinfo {author} {\bibfnamefont {M.}~\bibnamefont
  {Liertzer}}, \bibinfo {author} {\bibfnamefont {W.}~\bibnamefont {Chen}},
  \bibinfo {author} {\bibfnamefont {J.}~\bibnamefont {Kramer}}, \bibinfo
  {author} {\bibfnamefont {H.}~\bibnamefont {Yilmaz}}, \bibinfo {author}
  {\bibfnamefont {J.}~\bibnamefont {Wiersig}}, \bibinfo {author} {\bibfnamefont
  {S.}~\bibnamefont {Rotter}},\ and\ \bibinfo {author} {\bibfnamefont
  {L.}~\bibnamefont {Yang}},\ }\bibfield  {title} {\bibinfo {title} {Chiral
  modes and directional lasing at exceptional points},\ }\href
  {https://doi.org/10.1073/pnas.1603318113} {\bibfield  {journal} {\bibinfo
  {journal} {Proceedings of the National Academy of Sciences USA}\ }\textbf
  {\bibinfo {volume} {113}},\ \bibinfo {pages} {6845} (\bibinfo {year}
  {2016})}\BibitemShut {NoStop}%
\bibitem [{\citenamefont {Hashemi}\ \emph {et~al.}(2021)\citenamefont
  {Hashemi}, \citenamefont {Rezaei}, \citenamefont {{\"O}zdemir},\ and\
  \citenamefont {El-Ganainy}}]{RN62}%
  \BibitemOpen
  \bibfield  {author} {\bibinfo {author} {\bibfnamefont {A.}~\bibnamefont
  {Hashemi}}, \bibinfo {author} {\bibfnamefont {S.~M.}\ \bibnamefont {Rezaei}},
  \bibinfo {author} {\bibfnamefont {S.~K.}\ \bibnamefont {{\"O}zdemir}},\ and\
  \bibinfo {author} {\bibfnamefont {R.}~\bibnamefont {El-Ganainy}},\ }\bibfield
   {title} {\bibinfo {title} {New perspective on chiral exceptional points with
  application to discrete photonics},\ }\href
  {https://doi.org/10.1063/5.0045459} {\bibfield  {journal} {\bibinfo
  {journal} {APL Photonics}\ }\textbf {\bibinfo {volume} {6}},\ \bibinfo
  {pages} {040803} (\bibinfo {year} {2021})}\BibitemShut {NoStop}%
\bibitem [{\citenamefont {Hashemi}\ \emph {et~al.}(2022)\citenamefont
  {Hashemi}, \citenamefont {Busch}, \citenamefont {Christodoulides},
  \citenamefont {Ozdemir},\ and\ \citenamefont {El-Ganainy}}]{nc2022}%
  \BibitemOpen
  \bibfield  {author} {\bibinfo {author} {\bibfnamefont {A.}~\bibnamefont
  {Hashemi}}, \bibinfo {author} {\bibfnamefont {K.}~\bibnamefont {Busch}},
  \bibinfo {author} {\bibfnamefont {D.~N.}\ \bibnamefont {Christodoulides}},
  \bibinfo {author} {\bibfnamefont {S.~K.}\ \bibnamefont {Ozdemir}},\ and\
  \bibinfo {author} {\bibfnamefont {R.}~\bibnamefont {El-Ganainy}},\ }\bibfield
   {title} {\bibinfo {title} {Linear response theory of open systems with
  exceptional points},\ }\href {https://doi.org/10.1038/s41467-022-30715-8}
  {\bibfield  {journal} {\bibinfo  {journal} {Nature Communications}\ }\textbf
  {\bibinfo {volume} {13}},\ \bibinfo {pages} {3281} (\bibinfo {year}
  {2022})}\BibitemShut {NoStop}%
\bibitem [{\citenamefont {Yang}\ \emph {et~al.}(2021)\citenamefont {Yang},
  \citenamefont {Shi}, \citenamefont {Xie}, \citenamefont {Wu}, \citenamefont
  {Xiao}, \citenamefont {Song}, \citenamefont {Dang}, \citenamefont {Sun},
  \citenamefont {Yang}, \citenamefont {wang}, \citenamefont {Ge}, \citenamefont
  {Li}, \citenamefont {Zuo}, \citenamefont {Jin},\ and\ \citenamefont
  {Xu}}]{RN71}%
  \BibitemOpen
  \bibfield  {author} {\bibinfo {author} {\bibfnamefont {J.}~\bibnamefont
  {Yang}}, \bibinfo {author} {\bibfnamefont {S.}~\bibnamefont {Shi}}, \bibinfo
  {author} {\bibfnamefont {X.}~\bibnamefont {Xie}}, \bibinfo {author}
  {\bibfnamefont {S.}~\bibnamefont {Wu}}, \bibinfo {author} {\bibfnamefont
  {S.}~\bibnamefont {Xiao}}, \bibinfo {author} {\bibfnamefont {F.}~\bibnamefont
  {Song}}, \bibinfo {author} {\bibfnamefont {J.}~\bibnamefont {Dang}}, \bibinfo
  {author} {\bibfnamefont {S.}~\bibnamefont {Sun}}, \bibinfo {author}
  {\bibfnamefont {L.}~\bibnamefont {Yang}}, \bibinfo {author} {\bibfnamefont
  {Y.}~\bibnamefont {wang}}, \bibinfo {author} {\bibfnamefont {Z.-Y.}\
  \bibnamefont {Ge}}, \bibinfo {author} {\bibfnamefont {B.-B.}\ \bibnamefont
  {Li}}, \bibinfo {author} {\bibfnamefont {Z.}~\bibnamefont {Zuo}}, \bibinfo
  {author} {\bibfnamefont {K.}~\bibnamefont {Jin}},\ and\ \bibinfo {author}
  {\bibfnamefont {X.}~\bibnamefont {Xu}},\ }\bibfield  {title} {\bibinfo
  {title} {Enhanced emission from a single quantum dot in a microdisk at a
  deterministic diabolical point},\ }\href {https://doi.org/10.1364/OE.419740}
  {\bibfield  {journal} {\bibinfo  {journal} {Optics Express}\ }\textbf
  {\bibinfo {volume} {29}},\ \bibinfo {pages} {14231} (\bibinfo {year}
  {2021})}\BibitemShut {NoStop}%
\bibitem [{\citenamefont {Gardiner}(1993)}]{RN67}%
  \BibitemOpen
  \bibfield  {author} {\bibinfo {author} {\bibfnamefont {C.~W.}\ \bibnamefont
  {Gardiner}},\ }\bibfield  {title} {\bibinfo {title} {Driving a quantum system
  with the output field from another driven quantum system},\ }\href
  {https://doi.org/10.1103/PhysRevLett.70.2269} {\bibfield  {journal} {\bibinfo
   {journal} {Physical Review Letters}\ }\textbf {\bibinfo {volume} {70}},\
  \bibinfo {pages} {2269} (\bibinfo {year} {1993})}\BibitemShut {NoStop}%
\bibitem [{\citenamefont {Carmichael}(1993)}]{RN68}%
  \BibitemOpen
  \bibfield  {author} {\bibinfo {author} {\bibfnamefont {H.~J.}\ \bibnamefont
  {Carmichael}},\ }\bibfield  {title} {\bibinfo {title} {Quantum trajectory
  theory for cascaded open systems},\ }\href
  {https://doi.org/10.1103/PhysRevLett.70.2273} {\bibfield  {journal} {\bibinfo
   {journal} {Physical Review Letters}\ }\textbf {\bibinfo {volume} {70}},\
  \bibinfo {pages} {2273} (\bibinfo {year} {1993})}\BibitemShut {NoStop}%
\bibitem [{\citenamefont {Downing}\ \emph {et~al.}(2019)\citenamefont
  {Downing}, \citenamefont {Carreno}, \citenamefont {Laussy}, \citenamefont
  {del Valle},\ and\ \citenamefont {Fernandez-Dominguez}}]{RN69}%
  \BibitemOpen
  \bibfield  {author} {\bibinfo {author} {\bibfnamefont {C.~A.}\ \bibnamefont
  {Downing}}, \bibinfo {author} {\bibfnamefont {J.~C.~L.}\ \bibnamefont
  {Carreno}}, \bibinfo {author} {\bibfnamefont {F.~P.}\ \bibnamefont {Laussy}},
  \bibinfo {author} {\bibfnamefont {E.}~\bibnamefont {del Valle}},\ and\
  \bibinfo {author} {\bibfnamefont {A.~I.}\ \bibnamefont
  {Fernandez-Dominguez}},\ }\bibfield  {title} {\bibinfo {title} {Quasichiral
  interactions between quantum emitters at the nanoscale},\ }\href
  {https://doi.org/ARTN 057401 10.1103/PhysRevLett.122.057401} {\bibfield
  {journal} {\bibinfo  {journal} {Physical Review Letters}\ }\textbf {\bibinfo
  {volume} {122}},\ \bibinfo {pages} {057401} (\bibinfo {year}
  {2019})}\BibitemShut {NoStop}%
\bibitem [{\citenamefont {Srinivasan}\ and\ \citenamefont
  {Painter}(2007)}]{RN70}%
  \BibitemOpen
  \bibfield  {author} {\bibinfo {author} {\bibfnamefont {K.}~\bibnamefont
  {Srinivasan}}\ and\ \bibinfo {author} {\bibfnamefont {O.}~\bibnamefont
  {Painter}},\ }\bibfield  {title} {\bibinfo {title} {Mode coupling and
  cavity-quantum-dot interactions in a fiber-coupled microdisk cavity},\ }\href
  {https://doi.org/10.1103/PhysRevA.75.023814} {\bibfield  {journal} {\bibinfo
  {journal} {Physical Review A}\ }\textbf {\bibinfo {volume} {75}},\ \bibinfo
  {pages} {023814} (\bibinfo {year} {2007})}\BibitemShut {NoStop}%
\bibitem [{\citenamefont {Tamascelli}\ \emph {et~al.}(2018)\citenamefont
  {Tamascelli}, \citenamefont {Smirne}, \citenamefont {Huelga},\ and\
  \citenamefont {Plenio}}]{PRLmb}%
  \BibitemOpen
  \bibfield  {author} {\bibinfo {author} {\bibfnamefont {D.}~\bibnamefont
  {Tamascelli}}, \bibinfo {author} {\bibfnamefont {A.}~\bibnamefont {Smirne}},
  \bibinfo {author} {\bibfnamefont {S.~F.}\ \bibnamefont {Huelga}},\ and\
  \bibinfo {author} {\bibfnamefont {M.~B.}\ \bibnamefont {Plenio}},\ }\bibfield
   {title} {\bibinfo {title} {Nonperturbative treatment of non-markovian
  dynamics of open quantum systems},\ }\href
  {https://doi.org/10.1103/PhysRevLett.120.030402} {\bibfield  {journal}
  {\bibinfo  {journal} {Physical Review Letters}\ }\textbf {\bibinfo {volume}
  {120}},\ \bibinfo {pages} {030402} (\bibinfo {year} {2018})}\BibitemShut
  {NoStop}%
\bibitem [{\citenamefont {Denning}\ \emph {et~al.}(2019)\citenamefont
  {Denning}, \citenamefont {Iles-Smith},\ and\ \citenamefont {Mork}}]{PRBjm}%
  \BibitemOpen
  \bibfield  {author} {\bibinfo {author} {\bibfnamefont {E.~V.}\ \bibnamefont
  {Denning}}, \bibinfo {author} {\bibfnamefont {J.}~\bibnamefont
  {Iles-Smith}},\ and\ \bibinfo {author} {\bibfnamefont {J.}~\bibnamefont
  {Mork}},\ }\bibfield  {title} {\bibinfo {title} {Quantum light-matter
  interaction and controlled phonon scattering in a photonic fano cavity},\
  }\href {https://doi.org/10.1103/PhysRevB.100.214306} {\bibfield  {journal}
  {\bibinfo  {journal} {Physical Review B}\ }\textbf {\bibinfo {volume}
  {100}},\ \bibinfo {pages} {214306} (\bibinfo {year} {2019})}\BibitemShut
  {NoStop}%
\bibitem [{\citenamefont {Lu}\ \emph {et~al.}(2022)\citenamefont {Lu},
  \citenamefont {Zhou}, \citenamefont {Li}, \citenamefont {Li}, \citenamefont
  {Liu}, \citenamefont {Wu},\ and\ \citenamefont {Tan}}]{np2022}%
  \BibitemOpen
  \bibfield  {author} {\bibinfo {author} {\bibfnamefont {Y.-W.}\ \bibnamefont
  {Lu}}, \bibinfo {author} {\bibfnamefont {W.-J.}\ \bibnamefont {Zhou}},
  \bibinfo {author} {\bibfnamefont {Y.}~\bibnamefont {Li}}, \bibinfo {author}
  {\bibfnamefont {R.}~\bibnamefont {Li}}, \bibinfo {author} {\bibfnamefont
  {J.-F.}\ \bibnamefont {Liu}}, \bibinfo {author} {\bibfnamefont
  {L.}~\bibnamefont {Wu}},\ and\ \bibinfo {author} {\bibfnamefont
  {H.}~\bibnamefont {Tan}},\ }\bibfield  {title} {\bibinfo {title} {Unveiling
  atom-photon quasi-bound states in hybrid plasmonic-photonic cavity},\ }\href
  {https://doi.org/doi:10.1515/nanoph-2022-0162} {\bibfield  {journal}
  {\bibinfo  {journal} {Nanophotonics}\ }\textbf {\bibinfo {volume} {11}},\
  \bibinfo {pages} {3307} (\bibinfo {year} {2022})}\BibitemShut {NoStop}%
\bibitem [{\citenamefont {Scully}\ and\ \citenamefont {Zubairy}(1999)}]{RN72}%
  \BibitemOpen
  \bibfield  {author} {\bibinfo {author} {\bibfnamefont {M.~O.}\ \bibnamefont
  {Scully}}\ and\ \bibinfo {author} {\bibfnamefont {M.~S.}\ \bibnamefont
  {Zubairy}},\ }\href@noop {} {\emph {\bibinfo {title} {Quantum optics}}}\
  (\bibinfo  {publisher} {Cambridge University Press},\ \bibinfo {year}
  {1999})\BibitemShut {NoStop}%
\bibitem [{\citenamefont {Shen}\ and\ \citenamefont {Shen}(2012)}]{RN63}%
  \BibitemOpen
  \bibfield  {author} {\bibinfo {author} {\bibfnamefont {Y.}~\bibnamefont
  {Shen}}\ and\ \bibinfo {author} {\bibfnamefont {J.-T.}\ \bibnamefont
  {Shen}},\ }\bibfield  {title} {\bibinfo {title} {Nanoparticle sensing using
  whispering-gallery-mode resonators: Plasmonic and rayleigh scatterers},\
  }\href {https://doi.org/10.1103/PhysRevA.85.013801} {\bibfield  {journal}
  {\bibinfo  {journal} {Physical Review A}\ }\textbf {\bibinfo {volume} {85}},\
  \bibinfo {pages} {013801} (\bibinfo {year} {2012})}\BibitemShut {NoStop}%
\bibitem [{\citenamefont {Doeleman}\ \emph {et~al.}(2016)\citenamefont
  {Doeleman}, \citenamefont {Verhagen},\ and\ \citenamefont
  {Koenderink}}]{RN73}%
  \BibitemOpen
  \bibfield  {author} {\bibinfo {author} {\bibfnamefont {H.~M.}\ \bibnamefont
  {Doeleman}}, \bibinfo {author} {\bibfnamefont {E.}~\bibnamefont {Verhagen}},\
  and\ \bibinfo {author} {\bibfnamefont {A.~F.}\ \bibnamefont {Koenderink}},\
  }\bibfield  {title} {\bibinfo {title} {Antenna–cavity hybrids: Matching
  polar opposites for purcell enhancements at any linewidth},\ }\href
  {https://doi.org/10.1021/acsphotonics.6b00453} {\bibfield  {journal}
  {\bibinfo  {journal} {ACS Photonics}\ }\textbf {\bibinfo {volume} {3}},\
  \bibinfo {pages} {1943} (\bibinfo {year} {2016})}\BibitemShut {NoStop}%
\bibitem [{\citenamefont {Medina}\ \emph {et~al.}(2021)\citenamefont {Medina},
  \citenamefont {García-Vidal}, \citenamefont {Fernández-Domínguez},\ and\
  \citenamefont {Feist}}]{RN74}%
  \BibitemOpen
  \bibfield  {author} {\bibinfo {author} {\bibfnamefont {I.}~\bibnamefont
  {Medina}}, \bibinfo {author} {\bibfnamefont {F.~J.}\ \bibnamefont
  {García-Vidal}}, \bibinfo {author} {\bibfnamefont {A.~I.}\ \bibnamefont
  {Fernández-Domínguez}},\ and\ \bibinfo {author} {\bibfnamefont
  {J.}~\bibnamefont {Feist}},\ }\bibfield  {title} {\bibinfo {title} {Few-mode
  field quantization of arbitrary electromagnetic spectral densities},\ }\href
  {https://doi.org/10.1103/PhysRevLett.126.093601} {\bibfield  {journal}
  {\bibinfo  {journal} {Physical Review Letters}\ }\textbf {\bibinfo {volume}
  {126}},\ \bibinfo {pages} {093601} (\bibinfo {year} {2021})}\BibitemShut
  {NoStop}%
\bibitem [{\citenamefont {Franke}\ \emph {et~al.}(2019)\citenamefont {Franke},
  \citenamefont {Hughes}, \citenamefont {Dezfouli}, \citenamefont {Kristensen},
  \citenamefont {Busch}, \citenamefont {Knorr},\ and\ \citenamefont
  {Richter}}]{RN75}%
  \BibitemOpen
  \bibfield  {author} {\bibinfo {author} {\bibfnamefont {S.}~\bibnamefont
  {Franke}}, \bibinfo {author} {\bibfnamefont {S.}~\bibnamefont {Hughes}},
  \bibinfo {author} {\bibfnamefont {M.~K.}\ \bibnamefont {Dezfouli}}, \bibinfo
  {author} {\bibfnamefont {P.~T.}\ \bibnamefont {Kristensen}}, \bibinfo
  {author} {\bibfnamefont {K.}~\bibnamefont {Busch}}, \bibinfo {author}
  {\bibfnamefont {A.}~\bibnamefont {Knorr}},\ and\ \bibinfo {author}
  {\bibfnamefont {M.}~\bibnamefont {Richter}},\ }\bibfield  {title} {\bibinfo
  {title} {Quantization of quasinormal modes for open cavities and plasmonic
  cavity quantum electrodynamics},\ }\href
  {https://doi.org/10.1103/PhysRevLett.122.213901} {\bibfield  {journal}
  {\bibinfo  {journal} {Physical Review Letters}\ }\textbf {\bibinfo {volume}
  {122}},\ \bibinfo {pages} {213901} (\bibinfo {year} {2019})}\BibitemShut
  {NoStop}%
\bibitem [{\citenamefont {Lu}\ \emph {et~al.}(2021{\natexlab{a}})\citenamefont
  {Lu}, \citenamefont {Liu}, \citenamefont {Liao},\ and\ \citenamefont
  {Wang}}]{RN76}%
  \BibitemOpen
  \bibfield  {author} {\bibinfo {author} {\bibfnamefont {Y.-W.}\ \bibnamefont
  {Lu}}, \bibinfo {author} {\bibfnamefont {J.-F.}\ \bibnamefont {Liu}},
  \bibinfo {author} {\bibfnamefont {Z.}~\bibnamefont {Liao}},\ and\ \bibinfo
  {author} {\bibfnamefont {X.-H.}\ \bibnamefont {Wang}},\ }\bibfield  {title}
  {\bibinfo {title} {Plasmonic-photonic cavity for high-efficiency
  single-photon blockade},\ }\href {https://doi.org/10.1007/s11433-021-1712-2}
  {\bibfield  {journal} {\bibinfo  {journal} {Science China Physics, Mechanics
  \& Astronomy}\ }\textbf {\bibinfo {volume} {64}},\ \bibinfo {pages} {274212}
  (\bibinfo {year} {2021}{\natexlab{a}})}\BibitemShut {NoStop}%
\bibitem [{\citenamefont {Van~Vlack}\ \emph {et~al.}(2012)\citenamefont
  {Van~Vlack}, \citenamefont {Kristensen},\ and\ \citenamefont
  {Hughes}}]{RN77}%
  \BibitemOpen
  \bibfield  {author} {\bibinfo {author} {\bibfnamefont {C.}~\bibnamefont
  {Van~Vlack}}, \bibinfo {author} {\bibfnamefont {P.~T.}\ \bibnamefont
  {Kristensen}},\ and\ \bibinfo {author} {\bibfnamefont {S.}~\bibnamefont
  {Hughes}},\ }\bibfield  {title} {\bibinfo {title} {Spontaneous emission
  spectra and quantum light-matter interactions from a strongly coupled quantum
  dot metal-nanoparticle system},\ }\href
  {https://doi.org/10.1103/PhysRevB.85.075303} {\bibfield  {journal} {\bibinfo
  {journal} {Physical Review B}\ }\textbf {\bibinfo {volume} {85}},\ \bibinfo
  {pages} {025303} (\bibinfo {year} {2012})}\BibitemShut {NoStop}%
\bibitem [{\citenamefont {Lu}\ \emph {et~al.}(2021{\natexlab{b}})\citenamefont
  {Lu}, \citenamefont {Liu}, \citenamefont {Liu}, \citenamefont {Su},\ and\
  \citenamefont {Wang}}]{RN78}%
  \BibitemOpen
  \bibfield  {author} {\bibinfo {author} {\bibfnamefont {Y.-W.}\ \bibnamefont
  {Lu}}, \bibinfo {author} {\bibfnamefont {J.-F.}\ \bibnamefont {Liu}},
  \bibinfo {author} {\bibfnamefont {R.}~\bibnamefont {Liu}}, \bibinfo {author}
  {\bibfnamefont {R.}~\bibnamefont {Su}},\ and\ \bibinfo {author}
  {\bibfnamefont {X.-H.}\ \bibnamefont {Wang}},\ }\bibfield  {title} {\bibinfo
  {title} {Quantum exceptional chamber induced by large nondipole effect of a
  quantum dot coupled to a nano-plasmonic resonator},\ }\href
  {https://doi.org/doi:10.1515/nanoph-2021-0088} {\bibfield  {journal}
  {\bibinfo  {journal} {Nanophotonics}\ }\textbf {\bibinfo {volume} {10}},\
  \bibinfo {pages} {2431} (\bibinfo {year} {2021}{\natexlab{b}})}\BibitemShut
  {NoStop}%
\bibitem [{\citenamefont {Johansson}\ \emph {et~al.}(2013)\citenamefont
  {Johansson}, \citenamefont {Nation},\ and\ \citenamefont {Nori}}]{RN79}%
  \BibitemOpen
  \bibfield  {author} {\bibinfo {author} {\bibfnamefont {J.~R.}\ \bibnamefont
  {Johansson}}, \bibinfo {author} {\bibfnamefont {P.~D.}\ \bibnamefont
  {Nation}},\ and\ \bibinfo {author} {\bibfnamefont {F.}~\bibnamefont {Nori}},\
  }\bibfield  {title} {\bibinfo {title} {Qutip 2: A python framework for the
  dynamics of open quantum systems},\ }\href
  {https://doi.org/https://doi.org/10.1016/j.cpc.2012.11.019} {\bibfield
  {journal} {\bibinfo  {journal} {Computer Physics Communications}\ }\textbf
  {\bibinfo {volume} {184}},\ \bibinfo {pages} {1234} (\bibinfo {year}
  {2013})}\BibitemShut {NoStop}%
\bibitem [{\citenamefont {Wootters}(1998)}]{RN86}%
  \BibitemOpen
  \bibfield  {author} {\bibinfo {author} {\bibfnamefont {W.~K.}\ \bibnamefont
  {Wootters}},\ }\bibfield  {title} {\bibinfo {title} {Entanglement of
  formation of an arbitrary state of two qubits},\ }\href
  {https://doi.org/10.1103/PhysRevLett.80.2245} {\bibfield  {journal} {\bibinfo
   {journal} {Physical Review Letters}\ }\textbf {\bibinfo {volume} {80}},\
  \bibinfo {pages} {2245} (\bibinfo {year} {1998})}\BibitemShut {NoStop}%
\bibitem [{\citenamefont {Hsu}\ \emph {et~al.}(2016)\citenamefont {Hsu},
  \citenamefont {Zhen}, \citenamefont {Stone}, \citenamefont {Joannopoulos},\
  and\ \citenamefont {Soljačić}}]{RN83}%
  \BibitemOpen
  \bibfield  {author} {\bibinfo {author} {\bibfnamefont {C.~W.}\ \bibnamefont
  {Hsu}}, \bibinfo {author} {\bibfnamefont {B.}~\bibnamefont {Zhen}}, \bibinfo
  {author} {\bibfnamefont {A.~D.}\ \bibnamefont {Stone}}, \bibinfo {author}
  {\bibfnamefont {J.~D.}\ \bibnamefont {Joannopoulos}},\ and\ \bibinfo {author}
  {\bibfnamefont {M.}~\bibnamefont {Soljačić}},\ }\bibfield  {title}
  {\bibinfo {title} {Bound states in the continuum},\ }\href
  {https://doi.org/10.1038/natrevmats.2016.48} {\bibfield  {journal} {\bibinfo
  {journal} {Nature Reviews Materials}\ }\textbf {\bibinfo {volume} {1}},\
  \bibinfo {pages} {16048} (\bibinfo {year} {2016})}\BibitemShut {NoStop}%
\bibitem [{\citenamefont {Cotrufo}\ and\ \citenamefont {Alù}(2019)}]{RN84}%
  \BibitemOpen
  \bibfield  {author} {\bibinfo {author} {\bibfnamefont {M.}~\bibnamefont
  {Cotrufo}}\ and\ \bibinfo {author} {\bibfnamefont {A.}~\bibnamefont {Alù}},\
  }\bibfield  {title} {\bibinfo {title} {Excitation of single-photon embedded
  eigenstates in coupled cavity–atom systems},\ }\href
  {https://doi.org/10.1364/optica.6.000799} {\bibfield  {journal} {\bibinfo
  {journal} {Optica}\ }\textbf {\bibinfo {volume} {6}},\ \bibinfo {pages} {799}
  (\bibinfo {year} {2019})}\BibitemShut {NoStop}%
\bibitem [{\citenamefont {Friedrich}\ and\ \citenamefont
  {Wintgen}(1985)}]{RN85}%
  \BibitemOpen
  \bibfield  {author} {\bibinfo {author} {\bibfnamefont {H.}~\bibnamefont
  {Friedrich}}\ and\ \bibinfo {author} {\bibfnamefont {D.}~\bibnamefont
  {Wintgen}},\ }\bibfield  {title} {\bibinfo {title} {Interfering resonances
  and bound states in the continuum},\ }\href
  {https://doi.org/10.1103/PhysRevA.32.3231} {\bibfield  {journal} {\bibinfo
  {journal} {Physical Review A}\ }\textbf {\bibinfo {volume} {32}},\ \bibinfo
  {pages} {3231} (\bibinfo {year} {1985})}\BibitemShut {NoStop}%
\bibitem [{\citenamefont {Imamoḡlu}\ \emph {et~al.}(1997)\citenamefont
  {Imamoḡlu}, \citenamefont {Schmidt}, \citenamefont {Woods},\ and\
  \citenamefont {Deutsch}}]{RN80}%
  \BibitemOpen
  \bibfield  {author} {\bibinfo {author} {\bibfnamefont {A.}~\bibnamefont
  {Imamoḡlu}}, \bibinfo {author} {\bibfnamefont {H.}~\bibnamefont {Schmidt}},
  \bibinfo {author} {\bibfnamefont {G.}~\bibnamefont {Woods}},\ and\ \bibinfo
  {author} {\bibfnamefont {M.}~\bibnamefont {Deutsch}},\ }\bibfield  {title}
  {\bibinfo {title} {Strongly interacting photons in a nonlinear cavity},\
  }\href {https://doi.org/10.1103/PhysRevLett.79.1467} {\bibfield  {journal}
  {\bibinfo  {journal} {Physical Review Letters}\ }\textbf {\bibinfo {volume}
  {79}},\ \bibinfo {pages} {1467} (\bibinfo {year} {1997})}\BibitemShut
  {NoStop}%
\bibitem [{\citenamefont {Zubizarreta~Casalengua}\ \emph
  {et~al.}(2020)\citenamefont {Zubizarreta~Casalengua}, \citenamefont
  {López~Carreño}, \citenamefont {Laussy},\ and\ \citenamefont
  {Valle}}]{RN81}%
  \BibitemOpen
  \bibfield  {author} {\bibinfo {author} {\bibfnamefont {E.}~\bibnamefont
  {Zubizarreta~Casalengua}}, \bibinfo {author} {\bibfnamefont {J.~C.}\
  \bibnamefont {López~Carreño}}, \bibinfo {author} {\bibfnamefont {F.~P.}\
  \bibnamefont {Laussy}},\ and\ \bibinfo {author} {\bibfnamefont {E.~d.}\
  \bibnamefont {Valle}},\ }\bibfield  {title} {\bibinfo {title} {Conventional
  and unconventional photon statistics},\ }\href
  {https://doi.org/10.1002/lpor.201900279} {\bibfield  {journal} {\bibinfo
  {journal} {Laser \& Photonics Reviews}\ }\textbf {\bibinfo {volume} {14}},\
  \bibinfo {pages} {1900279} (\bibinfo {year} {2020})}\BibitemShut {NoStop}%
\bibitem [{\citenamefont {Liu}\ \emph {et~al.}(2017)\citenamefont {Liu},
  \citenamefont {Zhou}, \citenamefont {Yu}, \citenamefont {Zhang},
  \citenamefont {Wang}, \citenamefont {Liu}, \citenamefont {Wei}, \citenamefont
  {Chen},\ and\ \citenamefont {Wang}}]{RN82}%
  \BibitemOpen
  \bibfield  {author} {\bibinfo {author} {\bibfnamefont {R.}~\bibnamefont
  {Liu}}, \bibinfo {author} {\bibfnamefont {Z.~K.}\ \bibnamefont {Zhou}},
  \bibinfo {author} {\bibfnamefont {Y.~C.}\ \bibnamefont {Yu}}, \bibinfo
  {author} {\bibfnamefont {T.}~\bibnamefont {Zhang}}, \bibinfo {author}
  {\bibfnamefont {H.}~\bibnamefont {Wang}}, \bibinfo {author} {\bibfnamefont
  {G.}~\bibnamefont {Liu}}, \bibinfo {author} {\bibfnamefont {Y.}~\bibnamefont
  {Wei}}, \bibinfo {author} {\bibfnamefont {H.}~\bibnamefont {Chen}},\ and\
  \bibinfo {author} {\bibfnamefont {X.~H.}\ \bibnamefont {Wang}},\ }\bibfield
  {title} {\bibinfo {title} {Strong light-matter interactions in single open
  plasmonic nanocavities at the quantum optics limit},\ }\href
  {https://doi.org/10.1103/PhysRevLett.118.237401} {\bibfield  {journal}
  {\bibinfo  {journal} {Physical Review Letters}\ }\textbf {\bibinfo {volume}
  {118}},\ \bibinfo {pages} {237401} (\bibinfo {year} {2017})}\BibitemShut
  {NoStop}%
\bibitem [{\citenamefont {Wang}\ \emph
  {et~al.}(2019{\natexlab{a}})\citenamefont {Wang}, \citenamefont {He},
  \citenamefont {Chung}, \citenamefont {Hu}, \citenamefont {Yu}, \citenamefont
  {Chen}, \citenamefont {Ding}, \citenamefont {Chen}, \citenamefont {Qin},
  \citenamefont {Yang}, \citenamefont {Liu}, \citenamefont {Duan},
  \citenamefont {Li}, \citenamefont {Gerhardt}, \citenamefont {Winkler},
  \citenamefont {Jurkat}, \citenamefont {Wang}, \citenamefont {Gregersen},
  \citenamefont {Huo}, \citenamefont {Dai}, \citenamefont {Yu}, \citenamefont
  {Höfling}, \citenamefont {Lu},\ and\ \citenamefont {Pan}}]{RN87}%
  \BibitemOpen
  \bibfield  {author} {\bibinfo {author} {\bibfnamefont {H.}~\bibnamefont
  {Wang}}, \bibinfo {author} {\bibfnamefont {Y.-M.}\ \bibnamefont {He}},
  \bibinfo {author} {\bibfnamefont {T.~H.}\ \bibnamefont {Chung}}, \bibinfo
  {author} {\bibfnamefont {H.}~\bibnamefont {Hu}}, \bibinfo {author}
  {\bibfnamefont {Y.}~\bibnamefont {Yu}}, \bibinfo {author} {\bibfnamefont
  {S.}~\bibnamefont {Chen}}, \bibinfo {author} {\bibfnamefont {X.}~\bibnamefont
  {Ding}}, \bibinfo {author} {\bibfnamefont {M.~C.}\ \bibnamefont {Chen}},
  \bibinfo {author} {\bibfnamefont {J.}~\bibnamefont {Qin}}, \bibinfo {author}
  {\bibfnamefont {X.}~\bibnamefont {Yang}}, \bibinfo {author} {\bibfnamefont
  {R.-Z.}\ \bibnamefont {Liu}}, \bibinfo {author} {\bibfnamefont {Z.~C.}\
  \bibnamefont {Duan}}, \bibinfo {author} {\bibfnamefont {J.~P.}\ \bibnamefont
  {Li}}, \bibinfo {author} {\bibfnamefont {S.}~\bibnamefont {Gerhardt}},
  \bibinfo {author} {\bibfnamefont {K.}~\bibnamefont {Winkler}}, \bibinfo
  {author} {\bibfnamefont {J.}~\bibnamefont {Jurkat}}, \bibinfo {author}
  {\bibfnamefont {L.-J.}\ \bibnamefont {Wang}}, \bibinfo {author}
  {\bibfnamefont {N.}~\bibnamefont {Gregersen}}, \bibinfo {author}
  {\bibfnamefont {Y.-H.}\ \bibnamefont {Huo}}, \bibinfo {author} {\bibfnamefont
  {Q.}~\bibnamefont {Dai}}, \bibinfo {author} {\bibfnamefont {S.}~\bibnamefont
  {Yu}}, \bibinfo {author} {\bibfnamefont {S.}~\bibnamefont {Höfling}},
  \bibinfo {author} {\bibfnamefont {C.-Y.}\ \bibnamefont {Lu}},\ and\ \bibinfo
  {author} {\bibfnamefont {J.-W.}\ \bibnamefont {Pan}},\ }\bibfield  {title}
  {\bibinfo {title} {Towards optimal single-photon sources from polarized
  microcavities},\ }\href {https://doi.org/10.1038/s41566-019-0494-3}
  {\bibfield  {journal} {\bibinfo  {journal} {Nature Photonics}\ }\textbf
  {\bibinfo {volume} {13}},\ \bibinfo {pages} {770} (\bibinfo {year}
  {2019}{\natexlab{a}})}\BibitemShut {NoStop}%
\bibitem [{\citenamefont {Iles-Smith}\ \emph {et~al.}(2017)\citenamefont
  {Iles-Smith}, \citenamefont {McCutcheon}, \citenamefont {Nazir},\ and\
  \citenamefont {Mørk}}]{RN88}%
  \BibitemOpen
  \bibfield  {author} {\bibinfo {author} {\bibfnamefont {J.}~\bibnamefont
  {Iles-Smith}}, \bibinfo {author} {\bibfnamefont {D.~P.~S.}\ \bibnamefont
  {McCutcheon}}, \bibinfo {author} {\bibfnamefont {A.}~\bibnamefont {Nazir}},\
  and\ \bibinfo {author} {\bibfnamefont {J.}~\bibnamefont {Mørk}},\ }\bibfield
   {title} {\bibinfo {title} {Phonon scattering inhibits simultaneous
  near-unity efficiency and indistinguishability in semiconductor single-photon
  sources},\ }\href {https://doi.org/10.1038/nphoton.2017.101} {\bibfield
  {journal} {\bibinfo  {journal} {Nature Photonics}\ }\textbf {\bibinfo
  {volume} {11}},\ \bibinfo {pages} {521} (\bibinfo {year} {2017})}\BibitemShut
  {NoStop}%
\bibitem [{\citenamefont {Choi}\ \emph {et~al.}(2017)\citenamefont {Choi},
  \citenamefont {Heuck},\ and\ \citenamefont {Englund}}]{RN89}%
  \BibitemOpen
  \bibfield  {author} {\bibinfo {author} {\bibfnamefont {H.}~\bibnamefont
  {Choi}}, \bibinfo {author} {\bibfnamefont {M.}~\bibnamefont {Heuck}},\ and\
  \bibinfo {author} {\bibfnamefont {D.}~\bibnamefont {Englund}},\ }\bibfield
  {title} {\bibinfo {title} {Self-similar nanocavity design with ultrasmall
  mode volume for single-photon nonlinearities},\ }\href
  {https://doi.org/10.1103/PhysRevLett.118.223605} {\bibfield  {journal}
  {\bibinfo  {journal} {Physical Review Letters}\ }\textbf {\bibinfo {volume}
  {118}},\ \bibinfo {pages} {223605} (\bibinfo {year} {2017})}\BibitemShut
  {NoStop}%
\bibitem [{\citenamefont {Mok}\ \emph {et~al.}(2020)\citenamefont {Mok},
  \citenamefont {Aghamalyan}, \citenamefont {You}, \citenamefont {Haug},
  \citenamefont {Zhang}, \citenamefont {Png},\ and\ \citenamefont
  {Kwek}}]{RN90}%
  \BibitemOpen
  \bibfield  {author} {\bibinfo {author} {\bibfnamefont {W.-K.}\ \bibnamefont
  {Mok}}, \bibinfo {author} {\bibfnamefont {D.}~\bibnamefont {Aghamalyan}},
  \bibinfo {author} {\bibfnamefont {J.-B.}\ \bibnamefont {You}}, \bibinfo
  {author} {\bibfnamefont {T.}~\bibnamefont {Haug}}, \bibinfo {author}
  {\bibfnamefont {W.}~\bibnamefont {Zhang}}, \bibinfo {author} {\bibfnamefont
  {C.~E.}\ \bibnamefont {Png}},\ and\ \bibinfo {author} {\bibfnamefont {L.-C.}\
  \bibnamefont {Kwek}},\ }\bibfield  {title} {\bibinfo {title} {Long-distance
  dissipation-assisted transport of entangled states via a chiral waveguide},\
  }\href {https://doi.org/10.1103/PhysRevResearch.2.013369} {\bibfield
  {journal} {\bibinfo  {journal} {Physical Review Research}\ }\textbf {\bibinfo
  {volume} {2}},\ \bibinfo {pages} {013369} (\bibinfo {year}
  {2020})}\BibitemShut {NoStop}%
\bibitem [{\citenamefont {Gonzalez-Ballestero}\ \emph
  {et~al.}(2015)\citenamefont {Gonzalez-Ballestero}, \citenamefont
  {Gonzalez-Tudela}, \citenamefont {Garcia-Vidal},\ and\ \citenamefont
  {Moreno}}]{RN91}%
  \BibitemOpen
  \bibfield  {author} {\bibinfo {author} {\bibfnamefont {C.}~\bibnamefont
  {Gonzalez-Ballestero}}, \bibinfo {author} {\bibfnamefont {A.}~\bibnamefont
  {Gonzalez-Tudela}}, \bibinfo {author} {\bibfnamefont {F.~J.}\ \bibnamefont
  {Garcia-Vidal}},\ and\ \bibinfo {author} {\bibfnamefont {E.}~\bibnamefont
  {Moreno}},\ }\bibfield  {title} {\bibinfo {title} {Chiral route to
  spontaneous entanglement generation},\ }\href
  {https://doi.org/10.1103/PhysRevB.92.155304} {\bibfield  {journal} {\bibinfo
  {journal} {Physical Review B}\ }\textbf {\bibinfo {volume} {92}},\ \bibinfo
  {pages} {155304} (\bibinfo {year} {2015})}\BibitemShut {NoStop}%
\bibitem [{\citenamefont {Chen}\ \emph {et~al.}(2022)\citenamefont {Chen},
  \citenamefont {Abbasi}, \citenamefont {Ha}, \citenamefont {Erdamar},
  \citenamefont {Joglekar},\ and\ \citenamefont {Murch}}]{RN23}%
  \BibitemOpen
  \bibfield  {author} {\bibinfo {author} {\bibfnamefont {W.}~\bibnamefont
  {Chen}}, \bibinfo {author} {\bibfnamefont {M.}~\bibnamefont {Abbasi}},
  \bibinfo {author} {\bibfnamefont {B.}~\bibnamefont {Ha}}, \bibinfo {author}
  {\bibfnamefont {S.}~\bibnamefont {Erdamar}}, \bibinfo {author} {\bibfnamefont
  {Y.~N.}\ \bibnamefont {Joglekar}},\ and\ \bibinfo {author} {\bibfnamefont
  {K.~W.}\ \bibnamefont {Murch}},\ }\bibfield  {title} {\bibinfo {title}
  {Decoherence-induced exceptional points in a dissipative superconducting
  qubit},\ }\href {https://doi.org/10.1103/PhysRevLett.128.110402} {\bibfield
  {journal} {\bibinfo  {journal} {Physical Review Letters}\ }\textbf {\bibinfo
  {volume} {128}},\ \bibinfo {pages} {110402} (\bibinfo {year}
  {2022})}\BibitemShut {NoStop}%
\bibitem [{\citenamefont {Nie}\ \emph {et~al.}(2020)\citenamefont {Nie},
  \citenamefont {Peng}, \citenamefont {Nori},\ and\ \citenamefont
  {Liu}}]{RN92}%
  \BibitemOpen
  \bibfield  {author} {\bibinfo {author} {\bibfnamefont {W.}~\bibnamefont
  {Nie}}, \bibinfo {author} {\bibfnamefont {Z.}~\bibnamefont {Peng}}, \bibinfo
  {author} {\bibfnamefont {F.}~\bibnamefont {Nori}},\ and\ \bibinfo {author}
  {\bibfnamefont {Y.-x.}\ \bibnamefont {Liu}},\ }\bibfield  {title} {\bibinfo
  {title} {Topologically protected quantum coherence in a superatom},\ }\href
  {https://doi.org/10.1103/PhysRevLett.124.023603} {\bibfield  {journal}
  {\bibinfo  {journal} {Physical Review Letters}\ }\textbf {\bibinfo {volume}
  {124}},\ \bibinfo {pages} {023603} (\bibinfo {year} {2020})}\BibitemShut
  {NoStop}%
\bibitem [{\citenamefont {Iorsh}\ \emph {et~al.}(2020)\citenamefont {Iorsh},
  \citenamefont {Poshakinskiy},\ and\ \citenamefont {Poddubny}}]{RN93}%
  \BibitemOpen
  \bibfield  {author} {\bibinfo {author} {\bibfnamefont {I.}~\bibnamefont
  {Iorsh}}, \bibinfo {author} {\bibfnamefont {A.}~\bibnamefont
  {Poshakinskiy}},\ and\ \bibinfo {author} {\bibfnamefont {A.}~\bibnamefont
  {Poddubny}},\ }\bibfield  {title} {\bibinfo {title} {Waveguide quantum
  optomechanics: Parity-time phase transitions in ultrastrong coupling
  regime},\ }\href {https://doi.org/10.1103/PhysRevLett.125.183601} {\bibfield
  {journal} {\bibinfo  {journal} {Physical Review Letters}\ }\textbf {\bibinfo
  {volume} {125}},\ \bibinfo {pages} {183601} (\bibinfo {year}
  {2020})}\BibitemShut {NoStop}%
\bibitem [{\citenamefont {Primo}\ \emph {et~al.}(2020)\citenamefont {Primo},
  \citenamefont {Carvalho}, \citenamefont {Kersul}, \citenamefont {Frateschi},
  \citenamefont {Wiederhecker},\ and\ \citenamefont {Alegre}}]{RN94}%
  \BibitemOpen
  \bibfield  {author} {\bibinfo {author} {\bibfnamefont {A.~G.}\ \bibnamefont
  {Primo}}, \bibinfo {author} {\bibfnamefont {N.~C.}\ \bibnamefont {Carvalho}},
  \bibinfo {author} {\bibfnamefont {C.~M.}\ \bibnamefont {Kersul}}, \bibinfo
  {author} {\bibfnamefont {N.~C.}\ \bibnamefont {Frateschi}}, \bibinfo {author}
  {\bibfnamefont {G.~S.}\ \bibnamefont {Wiederhecker}},\ and\ \bibinfo {author}
  {\bibfnamefont {T.~P.}\ \bibnamefont {Alegre}},\ }\bibfield  {title}
  {\bibinfo {title} {Quasinormal-mode perturbation theory for dissipative and
  dispersive optomechanics},\ }\href
  {https://doi.org/10.1103/PhysRevLett.125.233601} {\bibfield  {journal}
  {\bibinfo  {journal} {Physical Review Letters}\ }\textbf {\bibinfo {volume}
  {125}},\ \bibinfo {pages} {233601} (\bibinfo {year} {2020})}\BibitemShut
  {NoStop}%
\bibitem [{\citenamefont {Wang}\ \emph
  {et~al.}(2019{\natexlab{b}})\citenamefont {Wang}, \citenamefont {Rao},
  \citenamefont {Yang}, \citenamefont {Xu}, \citenamefont {Gui}, \citenamefont
  {Yao}, \citenamefont {You},\ and\ \citenamefont {Hu}}]{RN95}%
  \BibitemOpen
  \bibfield  {author} {\bibinfo {author} {\bibfnamefont {Y.~P.}\ \bibnamefont
  {Wang}}, \bibinfo {author} {\bibfnamefont {J.~W.}\ \bibnamefont {Rao}},
  \bibinfo {author} {\bibfnamefont {Y.}~\bibnamefont {Yang}}, \bibinfo {author}
  {\bibfnamefont {P.~C.}\ \bibnamefont {Xu}}, \bibinfo {author} {\bibfnamefont
  {Y.~S.}\ \bibnamefont {Gui}}, \bibinfo {author} {\bibfnamefont {B.~M.}\
  \bibnamefont {Yao}}, \bibinfo {author} {\bibfnamefont {J.~Q.}\ \bibnamefont
  {You}},\ and\ \bibinfo {author} {\bibfnamefont {C.~M.}\ \bibnamefont {Hu}},\
  }\bibfield  {title} {\bibinfo {title} {Nonreciprocity and unidirectional
  invisibility in cavity magnonics},\ }\href
  {https://doi.org/10.1103/PhysRevLett.123.127202} {\bibfield  {journal}
  {\bibinfo  {journal} {Physical Review Letters}\ }\textbf {\bibinfo {volume}
  {123}},\ \bibinfo {pages} {127202} (\bibinfo {year}
  {2019}{\natexlab{b}})}\BibitemShut {NoStop}%
\end{thebibliography}%

\end{document}